\documentclass[%
 reprint,
superscriptaddress,
groupedaddress,
nofootinbib,
 amsmath,amssymb,
 aps,
floatfix,
]{revtex4-2}

\usepackage{amsmath}
\usepackage{mathrsfs}
\usepackage{url}
\usepackage{hyperref}
\usepackage{amssymb}
\usepackage[export]{adjustbox}
\usepackage[english]{babel}
\usepackage{float}
\usepackage{silence}
\usepackage{siunitx}
\usepackage{graphicx}
\usepackage{dcolumn}
\usepackage{bm}
\usepackage{hyperref}
\usepackage[mathlines]{lineno}
\usepackage{mathtools}
\usepackage[capitalize]{cleveref}
\usepackage[labelformat=simple]{subcaption}
\usepackage{bigints}
\usepackage[italicdiff]{physics} 
\usepackage{slashed}
\newcommand{\epkl}{\epsilon_{\text{KL}}}
\newcommand{\eprr}{\epsilon_{\text{RR}}}
\newcommand{\ellkl}{\ell_{\text{KL}}}
\newcommand{\ellrr}{\ell_{\text{RR}}}
\newcommand{\emax}{e_{\text{max}}}
\newcommand{\emin}{e_{\text{min}}}
\newcommand{\ftilde}{\tilde{F}}

\newcommand{\mczeta}{\mathcal{\zeta}}
\newcommand{\mcgamma}{\mathcal{\gamma}}
\newcommand{\cn}{\text{cn}}
\newcommand{\sn}{\text{sn}}
\newcommand{\dn}{\text{dn}}
\newcommand{\ellout}{\ell_{\text{out}}}
\newcommand{\ftb}{\Vec{f}_{\text{3b}}}
\newcommand{\ellmean}{\tilde{\ell}}
\newcommand{\ellmeankl}{\tilde{\ell}_{\text{KL}}}
\newcommand{\ellmeanrr}{\tilde{\ell}_{\text{RR}}}

\usepackage[dvipsnames, usenames]{xcolor}
\definecolor{linkcolor}{rgb}{0.0,0.3,0.5}
\definecolor{darkgreen}{rgb}{0.0, 0.5, 0.0}
\definecolor{darkcyan}{rgb}{0.0, 0.5,0.5}

\newcommand{\EDIT}[1]{{\color{black}{{#1}} }}
\usepackage{etoolbox}
\makeatletter
\patchcmd{\math@cr@@@align}
  {\place@tag}
  {\bgroup\color{black}\place@tag\egroup}
  {}{}
\makeatother
\begin{document}


\title{Ready-to-use analytic model for gravitational waves from a hierarchical triple with Kozai-Lidov oscillations}
\author{Rohit S. Chandramouli}%
 \email{rsc4@illinois.edu}%
   \affiliation{Illinois Center for Advanced Studies of the Universe and Department of Physics, \\
   University of Illinois at Urbana-Champaign, Urbana, IL 61801, USA 
   }%
\author{Nicol\'as Yunes}%
 \email{nyunes@illinois.edu}%
  \affiliation{Illinois Center for Advanced Studies of the Universe and Department of Physics, \\
   University of Illinois at Urbana-Champaign, Urbana, IL 61801, USA }
\date{\today}

\begin{abstract}
     Gravitational waves emitted by inner binaries in hierarchical triple systems are interesting astrophysical candidates for space-based detectors like the Laser Interferometer Space Antenna, LISA. In the presence of a third body, such as a supermassive black hole, an inner binary consisting of intermediate mass black holes can undergo oscillations in eccentricity and inclination angle due to the Kozai-Lidov mechanism. In this work, we construct analytic gravitational waveforms in the Fourier domain, taking into account the Kozai-Lidov effect at Newtonian (leading) order. Using multiple-scale analysis, we make use of the separability of timescales to combine the effects of both Kozai-Lidov oscillations and radiation reaction. We assume small eccentricity and present analytic solutions to the evolution of the other orbital elements. Our analytic calculation can be systematically extended to higher orders in eccentricity, and can be used to construct inspiral-merger-ringdown models. The imprint on the waveform, due to this combined evolution, is computed under the stationary-phase approximation. We find that the oscillations leave a clear signature on the Fourier amplitude of the waveform while leaving a measurable imprint on the gravitational wave phase, and that our analytic results are consistent with numerics. Further, with our study of the astrophysical parameters of the hierarchical triple, we outline potential source candidates, along with potential implications for gravitational wave data analysis.
\end{abstract}

\maketitle


\section{Introduction} \label{sec:intro}


The detection of gravitational waves (GWs) from compact objects by the Laser Interferometer Gravitational Wave Observatory (LIGO)/Virgo Collaboration~\cite{gwtc-1,gwtc-2} has opened new research directions in relativistic astrophysics and gravitational physics~\cite{Sathyaprakash:2009ws,Vitale:2020rid,Berti:2015itd,Will:2014kxa,Sedda:2021hpg,Sedda:2019uro,Barausse:2020rsu,Gair:2012nm}. The GW sources thus far have been mergers of binary systems composed of black holes (BHs) and neutron stars. The BH masses inferred from these observations have largely been in the stellar range, with the exception of GW190521 that resulted in the formation of an intermediate mass BH (IMBH) of mass $\sim 142 M_{\odot}$~\cite{GW190521_ligo_virgo,prop_GW190521_ligo_virgo}. As such, this event provided the first direct evidence of the existence of BHs in this intermediate mass range. The astrophysical origin of the GW190521 event is still contested~\cite{Palmese:2021wcv,Farrell:2020zju,Gayathri:2020coq,Romero-Shaw:2020thy,Fishbach:2020qag,Fragione:2020han,Kimball:2020qyd}, particularly after the observation of what is potentially an electromagnetic counterpart~\cite{Graham:2020gwr}. Provided the counterpart is genuine, its origin could be explained by the presence of a supermassive BH (SMBH)~\cite{Liu:2020gif}. One possibility to explain the origins of the merging BHs of the GW190521 event is that they formed through a hierarchical merger. Such mergers are expected to occur in dense globular clusters and galactic nuclei, and serve as exciting potential sources for future detectors, like the Laser Interferometer Space Antenna (LISA)~\cite{Toubiana:2020drf}. A crucial aspect of mergers in such clusters or nuclei is the presence of a third body, such as the aforementioned SMBH. 

In a hierarchical triple, one in which the inner binary's center of mass is far from the third body, the Kozai-Lidov (KL) mechanism is particularly interesting. The KL mechanism occurs due to the torquing of the inner binary's orbit by the outer third-body's orbit. Due to this mechanism, the eccentricity and inclination of the inner binary's orbit undergo \textit{oscillations}, as first studied (to quadrupole-order in the third-body's perturbation) by Kozai~\cite{kozai} and Lidov~\cite{lidov} independently. For highly inclined orbits, the eccentricity can grow close to unity due to the KL mechanism, making it a promising channel for the production of eccentric binaries observable by future GW detectors. We refer the interested reader to~\cite{smadar_review} for a review of the KL mechanism and its myriad of implications for astrophysics.

The KL mechanism is expected to play a particularly important role in GW astrophysics for a variety of reasons. 
Numerical studies~\cite{wen,Samsing:2013kua,VanLandingham1,Miller_2002,Blaes:2002cs} have showed that a few tens of percent of binaries, due to the third-body's perturbation (and amplified by four body effects~\cite{Miller_2002}), will retain eccentricity greater than a value of $0.1$ when they enter the LIGO band. Moreover, the KL mechanism is expected to reduce the merger time due to the larger eccentricity induced by the KL oscillations. Estimating the event rate of coalescing eccentric binaries, particularly those in dense environments, calls for accurate modeling of these third-body effects, in tandem with relativistic post-Newtonian (PN) effects. For these reasons, the study of the KL mechanism, with higher order effects from the third-body and PN effects due to the inner binary included has become an active research front in recent years~\cite{VanLandingham1,Miller_2002,Blaes:2002cs,Lithwick:2011hh,benjamin,Naoz:2011mb,Naoz:2012bx,Randall:2018qna,Randall:2017jop,Antognini:2013lpa,Antognini:2016,Kimpson:2016,Trani:2019gqh,Fragione:2019poq,Yu:2020,Liu:2019a,Yamada:2011xy,stephan:2019,Liu:2019b,Will:2017vjc,Lim:2020cvm,Kuntz:2021ohi,Martinez:2020lzt,Li:2014_EKL}. While it is possible to infer the formation channel from GW observations alone, it may be desirable to also infer it through \textit{direct} imprints on the waveform. To this end, there has been a lot of recent work in modeling GWs emitted by binaries that are influenced by a third body~\cite{Deme:2020ewx,Hoang:2019kye,Li:2015lva,Antonini:2012,Yunes:2010sm,Inayoshi:2017hgw,Randall:2018lnh,Gupta:2019unn,Yang:2017aht,Yu:2020dlm,Bonga:2019ycj,Gupta:2021cno,Bonetti:2018tpf}. 

This paper is concerned with the KL imprint that a third body could have on the GWs emitted from a binary system in a hierarchical triple. Specifically, we focus on the KL effect induced by a SMBH on the GWs emitted by an IMBH binary. We concentrate on such a subset of hierarchical triple systems so that inner IMBH binary emits GWs that would be detectable by LISA~\cite{Miller:2008fi,AmaroSeoane:2009yr}. The main objective of our work is to construct analytic GW models for the waves emitted by the inner binary in the Fourier domain. As a first pass we only include the KL effect and radiation-reaction (RR) effect at \textit{leading order} (LO) in the ratio of the KL timescale to the RR timescale; with this under control, higher PN-order corrections can be investigated in future work. More specifically, we consider the quadrupole-order contribution to the KL effect and the $2.5$PN-order contribution to the RR effect. Our analytic results are useful to establish how the KL effect manifests in both the amplitude and phase of the gravitational waveform and also serve as the first step in developing more sophisticated inspiral-merger-ringdown models. 
 
In order to construct a waveform model that accounts for the KL effect, we first make use of the separability of timescales in our hierarchical triple system to extract the behavior of the orbital elements over both the KL and RR timescales. We perform a thorough analysis of the timescales and other astrophysical constraints involved, to determine the appropriate window in parameter space that we can study analytically. \EDIT{We find that for systems consisting of a comparable-mass IMBH inner binary with total masses $m \sim (10^4  - 10^5) M_{\odot}$, and a SMBH third body of mass $m_3 \sim (10^6 - 10^7) M_{\odot}$ at a separation of $R \sim (1 - 4)$ Astronomical Units (AU), are of most interest for our work. The formation of such systems is a field of study in itself and we point the reader to~\cite{Miller:2003sc} (and references therein)  for the formation of IMBHs, and to~\cite{Miller_2002,Antognini:2016} (and references therein) for the formation of triple systems. For masses and separations of interest to us, the parameter space to probe KL oscillations is appreciable when the inner binary is in the LISA band, emitting GW frequencies in the range $(10^{-4} - 10^{-3})$ Hz. Such a regime in parameter space corresponds to the early inspiral of the inner binary, where the KL effect would be, for the most part, more important than relativistic PN effects. Furthermore, the dynamics of the outer binary would occur on a longer timescale compared to that of the inner binary, and therefore in this work, we limit attention to the case where the outer binary is stationary. }

The separability of timescales further allows us to take advantage of analytic tools such as the \textit{osculating orbital formalism} and \textit{multiple-scale analysis} (MSA). Such tools \EDIT{allow us to directly determine} the \EDIT{secular} evolution of the orbital elements over the (quadrupolar) KL timescale and the (2.5PN) RR timescale. We obtain the evolution of the eccentricity $e$ in terms of the mean orbital frequency $F$, aided by an expansion in small eccentricity, and the long-RR timescale behavior is obtained by averaging over the KL cycles. Using this $e(F)$, we obtain the evolution of the inclination angle $\iota (F)$ and the pericenter angle $\omega (F)$, exploiting two of the ``KL constants" (constants over the KL timescale, but slowly varying over the RR timescale). The analytic result for the evolution of the orbital elements over both timescales, and particularly in terms of the orbital frequency, is our first result, and we validate it using numerical evolution of the PN orbital equations. To the best of our knowledge, this is the first explicit analytic calculation done in this context.

With that in hand, we then move on to computing an analytic GW model in the time and frequency domains. Assuming general relativity is valid, we make use of the Quadrupole Formula, following e.g.~\cite{martel-poisson,Wahlquist:1987rx,moreno-garrido,NicoBertiArunWill}. The time-domain polarizations resulting from the Quadrupole Formula are Fourier decomposed into a sum of harmonics that depend on the mean anomaly $\ellmean$. The waveform in the Fourier domain is computed under the stationary-phase approximation (SPA), and our calculation avoids mathematical catastrophes~\cite{Klein:2013qda} since we perform a LO approximation using MSA. We find that the Fourier GW phase takes on the form 
\begin{align}
    \psi_{n} (f) \sim \psi_{n}^{\text{PC}} (f) + \psi_{n}^{\text{KL}} (f) \label{eqn:psi_intro_result},
\end{align}
where $ \psi_{n}^{\text{PC}} $ is the ``postcircular" phase~\cite{NicoBertiArunWill}, while $\psi_{n}^{\text{KL}}$ encodes the KL corrections to the Fourier phase. We find that by ``turning off" the KL effect, the $ \psi_{n}^{\text{KL}}$ term vanishes and the ``postcircular" result is recovered exactly. In addition to the chirp mass $\mathcal{M}$, we find that the GW phase depends on 3 parameters denoted by $\overline{e}_0, \delta e_0,$ and $k_0$, that are induced by the KL effect. Mathematically, these parameters are combinations of the initial values of the eccentricity, inclination angle, and pericenter angle $\{ e_0, \iota_0, \omega_0 \}$. Physically, $\overline{e}_0$ represents the initial average value of the eccentricity oscillations, $ \delta e_0 $ represents the initial difference between the average value and the minimum value of the eccentricity oscillations, and $k_0$ controls the `shape' as well as the time period associated with the eccentricity oscillations. We validate our analytic GW phase by comparing it to its numerical counterpart, which is obtained by evaluating the SPA result using the numerical solution to the orbital evolution. We also validate our ``PCKL" phase by comparing it to the PC phase, to determine in which region of parameter space the PCKL model would be most relevant. These calculations of the GW phase constitute our second main result, and once more it is the first time it appears in the literature. 

Our analysis shows that the amplitude of the $n$th harmonic of the GW polarization in the Fourier domain, up to normalization, takes the form 
\begin{align}
\tilde{h}_{+, \times }^{(n)} (f) \propto f^{-7/6} \Big [  C_{+, \times}^{(n)} (f) + i S_{+, \times}^{(n)} (f) \Big ] e^{-i \psi_{n} (f) } \label{eqn:polarizations_intro_result},
\end{align} 
where the coefficients $C_{+,\times}^{(n)}$ and $S_{+, \times}^{(n)}$ are functions of the eccentricity $e$, inclination angle $\iota$, and polarization angle $\beta$ and are identical to the ones computed in~\cite{NicoBertiArunWill,blake_2018,moreno-garrido} (see Sec~\ref{subsec:Fourier_analysis_GW}). Since the eccentricity and inclination angle undergo oscillations due to the KL mechanism, the latter leave a direct imprint on the frequency-evolution of the amplitudes of the gravitational waveform, through  the coefficients $C_{+,\times}^{(n)}, S_{+, \times}^{(n)}$. This explicit KL imprint on the amplitude of the Fourier-domain GW model is our third result, and once more we validate it against a numerical evaluation of the SPA amplitude. Further, the GW amplitude depends on $m_3$ and $R$, in addition to $ \{ \overline{e}_0, \delta e_0, k_0 \}$ and hence it offers complementary information to the GW phase. Therefore, a prediction of our model is that the family of waveform parameters is to be extended by $ \{ m_3, R, \overline{e}_0, \delta e_0, k_0 \}$  to capture the KL effect to leading order in perturbation theory.

The remainder of this paper presents the details of the results summarized above and it is organized as follows. In Sec.~\ref{sec:Modeling}, we discuss the modeling approach we use in this paper. We go into details on the osculating orbit formalism in Sec.~\ref{subsec:Osculating_Orbit_Formalism}, followed by an analysis of the physical timescales in Sec.~\ref{subsec:timescales}. We then go into details in Sec.~\ref{sec:MSA} on how to apply MSA, first to the KL perturbation in Sec.~\ref{subsec:2timescale}, and subsequently extend it to the KL$+$RR perturbations in Sec.~\ref{subsec:3timescale}. We discuss the details of the waveform modeling in Sec.~\ref{sec:waveforms}. In Sec.~\ref{sec:validation} we present the validation of our analytic results with numerics. We finally conclude and outline future directions in Sec.~\ref{sec:conclusion}. Henceforth we follow the conventions of~\cite{poisson_will_2014} and use geometric units in which $G = 1 = c$.


\section{Modeling third-body and radiation-reaction effects} \label{sec:Modeling}

\begin{figure}[t]
\centering
\includegraphics[width=\linewidth]{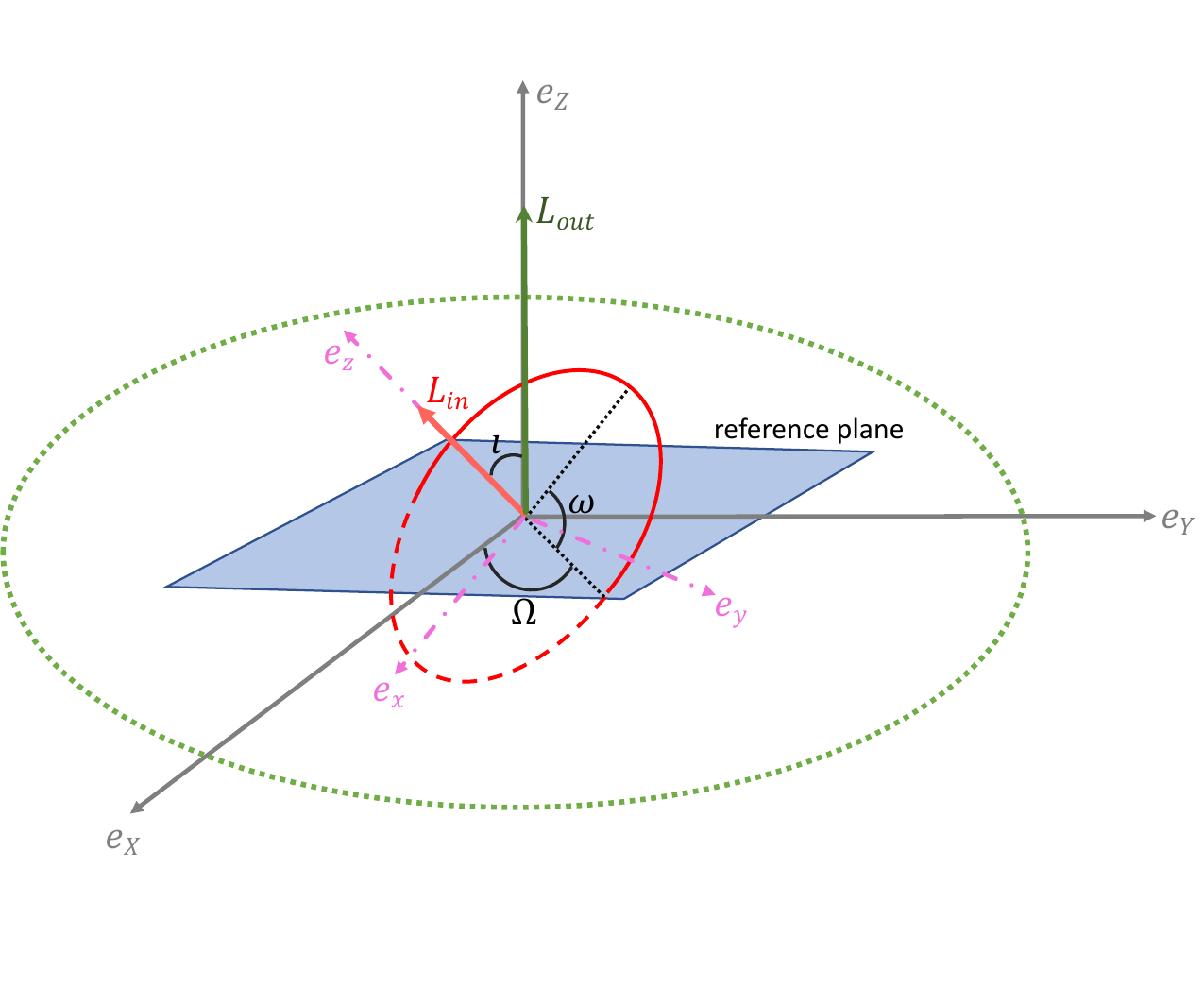}
\caption{Cartoon depiction of the hierarchical triple system. Note the two coordinate systems associated with the inner and outer orbits. }
\label{fig:fundamental_frame_PW}
\end{figure}

The evolution of the inner binary under the influence of the perturbing gravitational force of the third body (a cartoon depiction of the hierarchical triple system is shown in~\cref{fig:fundamental_frame_PW}) and that due to the RR force can be simultaneously treated under the osculating orbit formalism with the help of MSA. Using the former technique, one obtains the evolution equations for the orbital elements under both perturbations. The separation of timescales in the system further allows us to deconstruct the orbital equations into two sets of equations -- one that varies over the KL timescale and another that varies over the RR timescale. The perturbations on the inner orbit are modeled in the manner of a \textit{perturbed Kepler problem}. On the orbital timescale, the description of the orbit follows Kepler's laws and the orbital elements are constants. On the longer timescales, the effect of the perturbing forces is to make the orbital elements vary \textit{slowly}. Over the KL timescale, the eccentricity and inclination angle will oscillate, while on the RR timescale, the orbit circularizes and shrinks.\\

Let us now describe in more detail how each of the perturbations are modeled. The third body of mass $m_3$ exerts a gravitational acceleration on each of the masses $m_1$ and $m_2$ in the inner binary. What matters however is the relative acceleration induced by the third body, which is nothing but the tidal acceleration. One can perform a simple Fermi estimate of this to get a sense of how the perturbation scales with respect to the Newtonian 2-body acceleration. Let the distance of the third body be $R$, and the orbital separation of the inner binary be $a$. To LO, i.\,e. quadrupole-order, the tidal acceleration due to the third body on the inner binary scales as
\begin{align}
   | \Vec{a}_{3b} | \sim \dfrac{m_3 a}{R^3}.
\end{align}
Relative to the Newtonian two-body acceleration of the inner binary $| \Vec{a}_{\text{in}} | = m/a^2$, the third-body's tidal acceleration is then smaller by a factor of 
\begin{align}
    \dfrac{| \Vec{a}_{\text{3b}} |}{| \Vec{a}_{\text{in}} |} \sim \dfrac{m_3}{m} \Big ( \dfrac{a}{R} \Big )^3.
\end{align}
With this estimate of the ratio of accelerations, we immediately have an estimate of the KL timescale, namely
\begin{align}
    P_{\text{KL}} \sim P_{\text{orb,in}} \dfrac{m}{m_3} \Big ( \dfrac{R}{a} \Big )^3, \label{eqn:fermi_P_KL}
\end{align}
where $P_{\text{orb,in}}$ is the orbital period of the inner binary. We will rederive this in more detail by exploiting the osculating orbit formalism and MSA in Sec~\ref{subsec:2timescale}. The point to note here is that we have only treated the effect of the third-body's perturbation to quadrupolar order. One can see that the higher order perturbations from the third body will scale with powers of $(a/R)$. Therefore, the next-order correction, the octupole perturbation relative to the Newtonian acceleration, will be of $\mathcal{O}(a^4/R^4)$. This term, however, happens to also be proportional to the mass difference $(m_2-m_1)$ of the inner binary. Since we focus on an inner binary composed of comparable mass BHs, the octupole term can be neglected. The next contribution is of hexadecapole order, which is much smaller than the LO term, and we are thus justified in ignoring it for the kind of systems we are considering. Our analysis, nonetheless, can be systematically extended to higher order if one desires to model more generic binaries. We refer the reader to~\cite{poisson_will_2014} as well as~\cite{Will:2017vjc} for the higher order contributions due to the perturbation of the third body. 

The RR perturbation can similarly be Fermi estimated to make the associated timescale tangible. The luminosity $ \mathscr{L}$ of the GWs emitted by the inner binary scales, to LO, as
\begin{align}
    \mathscr{L} \sim \dfrac{32}{5} \eta^2 v^{10},
\end{align} 
where $v$ is its orbital velocity. Qualitatively, we can express the luminosity as the product of the RR force and the orbital velocity, and from that we can obtain the scaling of the acceleration due to RR, namely
\begin{align}
    \Big | \Vec{a}_{\text{RR}} \Big | \sim \frac{\Big | \Vec{f}_{\text{RR}} \Big |}{\mu} 
    \sim \frac{{\mathscr{L}}}{\mu \, v}  
    \sim \dfrac{32}{5} \dfrac{\eta}{m} v^9,
\end{align}
where we have used the virial relation $v^2 \sim m/a$, and $\eta = m_1 m_2/(m_1+m_2)^2$ is the symmetric mass ratio. Relative to the Newtonian two-body acceleration, which can also be expressed as $| \Vec{a}_{\text{in}} | = v^4/m$,  we find
\begin{align}
 \dfrac{| \Vec{a}_{\text{RR}} |}{| \Vec{a}_{\text{in}} |} \sim \dfrac{32}{5} \eta \ v^5.
\end{align}
Since $| \Vec{a}_{\text{RR}} |$ is smaller than the Newtonian two-body acceleration by a factor of $v^5$, this is referred to as a 2.5PN term in the PN framework, where the order counting is done in powers of $v^{2}$. Furthermore, one can perform a sanity check by taking the test-particle limit $\eta \rightarrow 0$ and find that the RR acceleration vanishes, which is consistent with the fact that to LO in perturbation theory, a test-particle will move on a geodesic. With this estimate in hand, the RR \EDIT{timescale} is 
\begin{align}
    P_{\text{RR}} \sim P_{\text{orb,in}} \dfrac{5}{32} \eta^{-1} v^{-5}. \label{eqn:fermi_P_RR}
\end{align}

In what follows, we first describe the formalism of osculating orbits, following~\cite{poisson_will_2014}. We then discuss in detail the timescales involved in the system. 

\subsection{Osculating orbit formalism} \label{subsec:Osculating_Orbit_Formalism}

In the \textit{perturbed Kepler problem}, on short timescales, Kepler's third law will hold. In the absence of perturbations, there will be six orbital constants of the motion, but due to the perturbing forces, these orbital elements will vary ``slowly." This allows one to approximate the dynamics of the inner binary by looking at how these orbital elements change, as opposed to looking at the changes in coordinates of the inner binary. In other words, the dynamics is simply a sequence of Keplerian orbits, which are varying \textit{secularly} due to the perturbations. This is the method of \textit{osculating orbital elements} and the reader is directed to~\cite{poisson_will_2014} for a textbook treatment. We are interested in the \textit{secular} evolution of the orbital elements of the inner binary, and hence the perturbations due to the third body and RR force can be orbit averaged and studied under a LO analysis. Qualitatively, to LO, the combined effects of the perturbation from the third body and that due to RR is simply the addition of the secular contributions of the individual perturbations, which leads to evolution over both the KL and RR timescales of~\cref{eqn:fermi_P_KL,eqn:fermi_P_RR}. 

We now review the osculating orbit formalism~\cite{poisson_will_2014}, suited to treat both the third body and RR perturbations. To do so, we use two sets of coordinate systems associated with the hierarchical triple -- the \textit{orbital frame} (OF) and the \textit{fundamental frame} (FF). The OF is tied to the inner binary, and it comes with the orthonormal basis vectors $\{ \Vec{e}_x ,\Vec{e}_y , \Vec{e}_z \}$, where $\Vec{e}_z$ is in the direction of $\Vec{L}_{\text{in}}$. The origin of the OF is the center of mass of the inner binary, and it is in this frame that the RR perturbation is computed. The FF on the other hand is better suited to compute the third-body's perturbation. The orthonormal basis vectors of the FF are denoted by $\{ \Vec{e}_X ,\Vec{e}_Y , \Vec{e}_Z \}$, where $\Vec{e}_Z$ is in the direction of $\Vec{L}_{\text{out}}$, and this choice is convenient because we assume that the motion of the outer orbit is stationary. Furthermore, the FF is chosen so that it shares the same origin as the OF. The transformation between the OF and FF can be found in~\cite{poisson_will_2014}. 

\allowdisplaybreaks[4]

Figure~\ref{fig:fundamental_frame_PW} shows the orbital setup in more detail. Observe the difference between the basis vectors of the OF and that of the FF. Furthermore, observe that the outer orbital plane is the same as the reference plane. From elementary classical mechanics~\cite{poisson_will_2014}, it is well known that the two-body orbital motion for the inner binary is effectively described by a test-particle moving under the influence of the gravitational potential induced by the total mass $m$ located at the center of mass. The orbital phase associated with this test-particle motion is denoted by $\phi$, and the radial distance of the test-particle is denoted as $r$. Furthermore, we have adopted the following definitions and notations for the orbital elements: the angle of pericenter is denoted as $\omega$; the angle of the ascending node is $\Omega$; the eccentricity is $e$; the semimajor axis is $a$; the semilatus rectum is $p$; and the inclination angle is $\iota$. We also have that the true anomaly $\ell$ of the inner binary is defined through $\ell = \phi - \omega$. 

With this picture in our minds, let us now introduce the perturbing force $\Vec{f}= \mathcal{R} \Vec{n} + \mathcal{S} \Vec{\lambda} + \mathcal{W} \Vec{e_z}$. Its components are then given by 
\begin{align}
\begin{split}
    \mathcal{R} &= \Vec{f} \cdot \Vec{n}, \\ 
\mathcal{S} &= \Vec{f} \cdot \Vec{\lambda}, \\
\mathcal{W} &= \Vec{f} \cdot \Vec{e_z},
\end{split}
\end{align}
where 
\begin{align}
\begin{split}
\label{eq:basis}
\Vec{n} &= \cos (\omega + \ell) \, \Vec{e}_X + \cos \iota \sin (\omega + \ell) \, \Vec{e}_Y \\ &+ \sin \iota \sin (\omega + \ell) \, \Vec{e}_Z, \\
\Vec{\lambda} &= -\sin (\omega + \ell) \, \Vec{e}_X + \cos \iota \cos (\omega + \ell) \, \Vec{e}_Y \\ &+ \sin \iota \cos (\omega + \ell) \, \Vec{e}_Z, \\
\Vec{e_z} &= -\sin \iota \, \Vec{e}_Y + \cos \iota \, \Vec{e}_Z. 
\end{split}
\end{align}
The explicit expressions for $\mathcal{R},\mathcal{S}, \mathcal{W}$ corresponding to the third body and RR perturbations can be found in~\cref{app:perturbations}. We use the true anomaly $\ell$ as our independent variable, and consequently the time variable $t$ becomes another dependent quantity whose equation adds to the set of osculating orbit equations, namely
\begin{align}
    \dv{p}{\ell}  & \simeq 2 \dfrac{p^3}{G m} \dfrac{1}{(1+e \cos \ell)} \mathcal{S}, 
    \nonumber \\
\dv{e}{\ell} & \simeq \dfrac{p^2}{G m} \big [ \dfrac{\sin \ell}{(1+e \cos \ell)^2} \mathcal{R}  + \dfrac{2 \cos \ell + e (1+\cos^2 \ell )}{(1+e \cos \ell)^3} \mathcal{S} \big ], 
\nonumber \\
\dv{\iota}{\ell} & \simeq \dfrac{p^2}{G m} \dfrac{\cos (\omega + \ell)}{(1+e \cos \ell)^3} \mathcal{W},
\nonumber \\
\sin \iota \dv{\Omega}{\ell} & \simeq \dfrac{p^2}{G m} \dfrac{\sin (\omega+\ell)}{(1+e \cos \ell)^3} \mathcal{W},
\nonumber \\
\dv{\omega}{\ell} & \simeq \dfrac{p^2}{e \, G m } \big [ - \dfrac{\cos \ell}{(1+e \cos \ell)^2} \mathcal{R} 
\nonumber \\ & + \dfrac{2+e \cos \ell}{(1+e \cos \ell)^3} \sin \ell \ \mathcal{S} - e \cot \iota \dfrac{\sin (\omega+\ell)}{(1+e\cos \ell)^3}\mathcal{W} \big ], 
\nonumber \\
\dv{t}{\ell} & \simeq \big (\frac{p}{Gm} \big )^{3/2} \dfrac{1}{(1+e\cos \ell)^2}  \Big  [ 1- \frac{1}{e} \frac{p^2}{G m } 
\nonumber \\ & \times \Big ( \dfrac{\cos \ell}{(1+e\cos \ell )^2}\mathcal{R} - \dfrac{2+e \cos \ell}{(1+e \cos \ell)^3} \sin \ell \; \mathcal{S} \Big ) \Big ].
\label{eq:osculating}
\end{align}
The osculating orbit equations constitute a system of first-order ordinary differential equations (ODEs) which require initial conditions on the orbital elements. Once the initial values, given by the set $\{p_0, e_0, \iota_0, \Omega_0, \omega_0, t_0 \}$, are specified, the system of equations can be integrated for a choice of system parameters $\{ m,m_3 , \eta, R \}$.

\subsection{Timescales of the triple system} \label{subsec:timescales}
Understanding the different timescales provides a clear picture of when and how the different physical effects that are relevant to the system manifest themselves. The physical effects that the system undergoes can be listed as follows:
\begin{enumerate}
\setlength\parskip{-0.1cm}
\setlength\parsep{-0.1cm}
    \item Orbital evolution of inner and outer orbits of the hierarchical triple, governed by Kepler's law,
    \item Mutual torquing of the inner and outer orbits that produces the KL oscillations, when we can treat the effect of the third body as a perturbation on the inner orbit's evolution,
    \item Relativistic effects of the inner orbit as it undergoes orbital evolution - this includes conservative PN effects, such as the 1PN pericenter precession, and dissipative PN effects, such as the 2.5PN RR, among other high PN-order effects.
\end{enumerate}

We first list the relevant timescales and discuss which ones are most relevant for us, and focus on those for our modeling. 

\subsubsection{Orbital Timescales}
Since we are interested in GWs from the inner binary, the orbital timescale is relevant for understanding the frequency window in which the inner binary will be observable. The orbital timescale is simply given by Kepler's law and for the inner orbit it is
\begin{equation}
    P_{\text{orb,in}} = 2\pi \Big ( \dfrac{a^3}{m} \Big )^{1/2},
\end{equation}
while for the outer orbit it is 
\begin{align}
\begin{split}
    P_{\text{orb,out}} &= 2\pi \Big ( \dfrac{R^3}{m_3+m} \Big )^{1/2} \\
    & \sim 2\pi \Big ( \dfrac{R^3}{m_3} \Big )^{1/2} ,
\end{split}
    \end{align}
where we have assumed that $m_3 \gg m$. 

\subsubsection{KL timescale}
The oscillations of eccentricity and inclination angle vary over what is known as the KL timescale, and using an order of magnitude estimate, it scales as
\begin{align}
    P_{\text{KL}} \sim P_{\text{orb, in}} \Big ( \dfrac{R}{a} \Big )^3 \Big ( \dfrac{m}{m_3} \Big ),
\end{align}
treating the tidal perturbation at quadrupole-order. It is clear that a more massive third body (with other parameters fixed) will induce oscillations that vary more rapidly, while a more distant third body (with other parameters fixed) will induce oscillations that vary less rapidly. Since we require the KL effect to be perturbative and for the timescales to separate, we must have that $P_{\text{orb, in}} \ll P_{\text{KL}} $ which can be recast into
\begin{align}
\begin{split}
 \epkl \equiv \Big ( \dfrac{a}{R} \Big )^3 \Big ( \dfrac{m_3}{m} \Big ) & \ll 1.
\end{split} \label{eqn:epkl_def}
\end{align}
This gives us a dimensionless small parameter $\epkl$ to work with when we model the perturbation from the third body. 

There are three approximations we will make regarding the quadrupole perturbation. First, we will average over the inner and outer orbits, where the latter amounts to considering the third body to be smeared over its orbit. The inner orbit averaging is justified by~\cref{eqn:epkl_def}, and requiring $P_{\text{orb, out}} \ll P_{\text{KL}} $ justifies the outer orbit averaging. Second, we assume that the outer orbit is circular, which means that the ``eccentric KL"~\cite{smadar_review} effect is absent in our model. Third, we assume that the orbit is stationary, meaning that the mutual torquing between the inner and outer orbits does not budge the latter. We can assume this provided the ratio $L_{\text{in}}/L_{\text{out}}$ is small, implying that
\begin{align}
\eta \dfrac{m}{m_3} \sqrt{ \dfrac{a}{R} \Big ( 1 + \dfrac{m_3}{m} \Big )} \ll 1.
\end{align}
As one can check by looking at the complete set of orbital equations at quadrupole-order~\cite{Will:2017vjc}, the variation of the outer orbit, relative to the inner orbit, is precisely the factor $L_{\text{in}}/L_{\text{out}}$ we just computed. To LO in $m/m_3$ and in $a/R$, this condition becomes
\begin{align}
\eta \sqrt{ \dfrac{a}{R} \dfrac{m}{m_3}} \ll 1.
\end{align}
Since we are interested in systems where $m_3 \gg m$ and when $R \gg a$, this condition is guaranteed to be satisfied. In passing, note that the requirement above can be satisfied either by requiring \EDIT{$m/m_3 \ll 1$} and  $a/R \ll 1$, or by requiring that $\eta \ll 1$, the latter of which leads to the classic ``test-particle KL" effect~\cite{smadar_review}. Here, however, we will keep $\eta \sim 1/4$, since we are considering comparable-mass binaries. 

Before proceeding, let us make one final observation. In order to find the cumulative effect of the KL oscillations on the GW phase, we must require that during the observing time period $P_{\text{obs}}$ there are sufficient number of KL cycles. In other words, we must require that $P_{\text{KL}} \ll P_{\text{obs}}$.

\subsubsection{Relativistic PN effects}
Relativistic PN effects in a binary's evolution occur due to the effect of the curvature induced by the masses in the binary. Broadly speaking, PN effects can be classified as conservative (even powers of $v$) and dissipative (odd powers of $v$). The LO dissipative PN effect occurs at 2.5PN-order in the form of RR and the associated timescale is
\begin{align}
\begin{split}
    P_{\text{RR,in}} & \sim P_{\text{orb,in}} \Big ( \dfrac{m}{a} \Big )^{-5/2} \sim  P_{\text{orb,in}} \ \eta^{-1} v^{-5},
\end{split}
\end{align}
and to treat this effect perturbatively, we require that the dimensionless quantity $\eprr \ll 1$ where we have defined
\begin{equation}
    \eprr \equiv v^5.
\end{equation}
In the PN framework, $v \ll 1$ so that means $\eprr \ll 1$ as well. 

At 1PN-order, conservative effects modify the orbit, for example introducing pericenter precession. The associated timescale of this precession effect scales as
\EDIT{\begin{align}
 P_{\text{1PN,in}} \sim P_{\text{orb,in}} \dfrac{a (1-e^2)}{m},
\end{align}}
which for small eccentricity is just \EDIT{$ P_{\text{1PN,in}} \sim P_{\text{orb,in}} v^{-2}$}. We can again define a dimensionless parameter $\epsilon_{\text{1PN}} \equiv v^2 \ll 1$. The 1PN precession of the pericenter can cause the KL effect to be subdued (meaning smaller amplitude compared to when the 1PN effect is  absent) and if strong enough, it can completely quench the KL oscillations~\cite{Blaes:2002cs,smadar_review,Gupta:2019unn}. The quenching happens when the 1PN timescale is faster than the KL timescale. Thus, to observe KL oscillations in the inner binary, one needs to require that  $P_{\text{KL}} \ll P_{\text{1PN,in}}$, and that implies that $\epkl \gg \epsilon_{\text{1PN}}$. Recall that $\epsilon_{\text{1PN}} \sim v^2$ is larger than the 2.5PN RR term (which scales as $v^5$). Therefore, we have that $\epkl \gg \eprr$, and requiring that the KL oscillations not be quenched by the 1PN precession means that the KL effect must be more dominant than the RR effect. Qualitatively this tells us that the KL effect is going to be significant in the early inspiral of the inner binary. 

\subsubsection{Separation of timescales}
From our discussion of the timescales thus far, we have the following separation of timescales for probing the KL effect
\begin{align}
    P_{\text{orb,in}} \ll P_{\text{orb,out}} \ll P_{\text{KL}} \ll P_{\text{1PN,in}} \ll P_{\text{RR,in}} \label{eqn:sep_time},
\end{align}
and it is this separation of timescales that allows us to use analytical tools such as MSA and the osculating orbit formalism. It is apparent from the expression for the timescales, that as the inner binary inspirals, the PN effects will become stronger and the KL effect will become weaker. During the course of the inspiral, this separation also implies that the ratio $P_{\text{KL}} / P_{\text{1PN,in}}$ will flip from being much larger than 1 to being smaller than 1, signaling that the KL effect is quenched at that stage. Although such PN effects are important in hierarchical triple systems, in this work we restrict attention to a LO analysis, and therefore the flipping of timescales is not an issue when it comes to using MSA. Furthermore, for a significant portion of the inspiral, $\epkl$ is going to be much larger than $\eprr$, which means that the separation of scales between the KL and RR timescales is guaranteed, and one can take advantage of MSA.  

\section{multiple-scale analysis applied to osculating orbits} \label{sec:MSA}
In this section, we apply the separability of timescales to solve the osculating orbit equations, using MSA. The goal of MSA is to make explicit use of the \textit{separability of timescales} of the system to extract its \textit{long-timescale} behavior. We review in Sec~\ref{subsec:2timescale} the application of MSA to the case when only the (quadrupole-order) perturbation from the third body is considered. Such an application of MSA recovers a well-known exact solution to the KL problem~\cite{Kinoshita:2007}. Then, in Sec~\ref{subsec:3timescale}, we apply MSA to the case when both the perturbation of the third body and RR effects are included, and solve the resulting equations perturbatively using a small-eccentricity approximation. 

\subsection{Review of two-timescale analysis of the Kozai-Lidov problem} \label{subsec:2timescale}
When the perturbing force is due to just the third body, at quadrupolar order it is given by $\ftb= \mathcal{R}_{\text{3b}} \Vec{n} + \mathcal{S}_{\text{3b}} \Vec{\lambda} + \mathcal{W}_{\text{3b}} \Vec{e_z}$. The osculating equations can be written as 
\begin{align}
\dv{\mu_{i}}{\ell} &= \epkl \, g^{(\text{KL})}_{i} (\mu_{\alpha}),\\
\dv{t}{\ell} &= \Big (\frac{p}{m} \Big )^{3/2} \dfrac{1}{(1+e\cos \ell)^2} + \epkl \,  g_t^{(\text{KL})} (\mu_{\alpha}), 
\end{align}
where the functions $g_{i}^{(\text{KL})}$ and $g_t^{(\text{KL})}$ can be computed explicitly using the components of the perturbing force, given in Appendix~\ref{app:perturbations} and in~\cite{poisson_will_2014}. Using MSA, we introduce a slowly varying timescale $\ellkl = \epkl \, \ell$ to extract the longer KL timescale behavior.  We introduce an explicit dependence of the orbital elements over the KL timescale through $\mu_i (\ell) \rightarrow \mu_i (\ell,\ellkl)$ and $t (\ell) \rightarrow t (\ell , \ellkl)$. Further, we decompose the orbital elements into  $ \mu_i (\ell,\ellkl) = \mu_i^{(0)}(\ell,\ellkl) + \epkl \, \mu_i^{(1)} (\ell,\ellkl)$ and $t(\ell,\ellkl) = \epkl^{-1} \ t^{(-1)}(\ell,\ellkl) + t^{0}(\ell,\ellkl) + \epkl \ t^{(1)}(\ell,\ellkl)$. We require the $t^{(-1)}$ term because $t$ has a secular growth in the absence of perturbations. At each order $n$, we can also decompose the orbital elements into an oscillatory piece and a secular piece, in the form $\mu_i^{(n)}(\ell,\ellkl) = \mu_i^{(n),\text{osc}}(\ell,\ellkl) + \mu_i^{(n),\text{sec}} (\ellkl)$. 

The evolution equations at LO can then be found to be
\begin{align}
\pdv{\mu_i^{0}}{\ell} &= 0, \label{eqn:2timescale_MSA_mu0_osc} \\ 
\pdv{ t^{(-1)}}{\ell} &= 0, \label{eqn:2timescale_MSA_t_osc}  \\
\pdv{\mu_i^{(0)} }{\ell_{KL}} + \pdv{\mu_i^{(1)}}{\ell} &= g_{i}^{(\text{KL})}(\mu_{\alpha}^{(0)}), \label{eqn:2timescale_MSA_mu0_sec}\\ 
\pdv{t^{(0)}}{\ell} + \pdv{t^{(-1)}}{\ell_{KL}} &= \Big (\frac{p}{m} \Big )^{3/2} \dfrac{1}{(1+e\cos \ell)^2}. \label{eqn:2timescale_MSA_t_sec}
\end{align}
At LO, there is only secular growth, which is expected because in the absence of a perturbation, the orbital elements are all constants of the motion and $t$ just keeps evolving secularly. We consider the adiabatic approximation where we average over the orbital motion of the inner and outer orbits to obtain the secular evolution of $\mu_i$. 

For the inner orbit, we average over $\ell$, and for the outer orbit, we average over the true anomaly associated with the outer orbit, $\ellout$. Doing this gives us
\begin{align}
 \Big ( \dv{\mu_i^{(0)}}{\ellkl} \Big )^{\text{sec}} &= \langle g_{i}^{(\text{KL})} (\mu_{\alpha}^0) \rangle_{\ell, \ellout}, \label{eqn:mu0_MSA_avg}\\
 \Big ( \dv{t^{(-1)}}{\ellkl} \Big )^{\text{sec}} &= \Big ( \dfrac{a^3}{m} \Big )^{3/2}, \label{eqn:t_MSA_avg}
\end{align}
where the notation $\langle \dots \rangle_{\ell, \ \ellout}$ means
\begin{align}
    \langle \dots \rangle_{\ell, \ellout} = \dfrac{1}{4 \pi^2} \int \limits_{0}^{2\pi} \int \limits_{0}^{2\pi} (\dots) d\ell \ d\ellout.
\end{align}
The double-orbit averaging would imply that the oscillatory contributions will vanish at LO because $\mu_i^{(1)}$ obeys the periodicity condition $\mu_i^{(1)} (\ell) = \mu_i^{(1)}(\ell+2\pi)$. Therefore we obtain a set of secular differential equations for $\mu_i^{(0),\text{sec}}$ and $t^{(-1),\text{sec}}$. The way to compute the post adiabatic corrections is outlined in~\cite{poisson_will_2014}, but in this work we restrict to a LO adiabatic analysis using MSA. 

Henceforth, we drop the $(\dots)^{\text{sec}}$ notation for LO secular terms and we simply use $\mu_i$ instead of $\mu_i^{(0)}$, with the understanding that we only compute the LO secular evolution. For example, the secular derivative $( d \mu_i^{(0)}/ d\ellkl )^{\text{sec}}$ can be simplified to
\begin{align}
    \Big (\dv{\mu_i^{(0)}}{\ellkl} \Big )^{\text{sec}} \equiv \dv{\mu_i}{\ellmeankl},
\end{align}
where $\ellmeankl$ is the mean anomaly that is defined by~\cref{eqn:t_MSA_avg}.  

After carrying out the double averaging, the orbital equations relevant for the KL mechanism~\cite{poisson_will_2014} are given below as a system of ODEs:
\begin{align}
    \begin{split}
       \dv{e}{\ellmeankl} &= \dfrac{15}{8}\Big ( \dfrac{a}{a_0} \Big )^3 e (1-e^2)^{1/2}  \sin^2 \iota \ \sin 2 \omega ,  \\
      \dv{a}{\ellmeankl} &= 0, \\
    \dv{\iota}{\ellmeankl}  &= -\dfrac{15}{16} \Big ( \dfrac{a}{a_0} \Big )^3 e^2 (1-e^2)^{-1/2} \sin 2 \iota \ \sin 2 \omega, \\
      \dv{\omega}{\ellmeankl} &= \dfrac{3}{4} \Big ( \dfrac{a}{a_0} \Big )^3  (1-e^2)^{-1/2} \Big [ 5 \cos^2 \iota  \sin^2 \omega \\ 
      & + (1-e^2)(5 \cos^2 \omega-3) \Big ],\\
     \dv{t}{\ellmeankl} & = \Big ( \dfrac{a^3}{m} \Big )^{1/2}.
    \end{split} \label{eqn:KL_equations}
\end{align}
We see that the semimajor axis remains constant $a=a_0$, which is a typical feature of the KL effect. The remaining equations can be integrated with the specification of the set of initial values $\{e_0, \iota_0, \omega_0, t_0 \}$. In fact, there are two constants of the motion associated with these equations -- one denoted by $\mcgamma$, and another denoted by $\zeta$. The constant $\mcgamma$ arises due to the interaction (with the third body) term of the conservative double averaged Hamiltonian of the system~\cite{smadar_review}. The constant $\mczeta$ arises due to the long-timescale conservation of $\Vec{L}_{\text{in}} \cdot \Vec{e}_{Z}$~\cite{smadar_review,poisson_will_2014}.

With these two constants, the system of equations is integrable and an exact solution to these equations was discovered by~\cite{Kinoshita:2007}. In the following, we review this exact solution, as it is instructive for what is to come; the reader familiar with this solution may wish to skip ahead to Sec.~\ref{subsec:3timescale}. Following~\cite{Kinoshita:2007}, we introduce 
\begin{equation}
x=1-e^2    
\end{equation}
and express the constants $\mczeta$ and $\mcgamma$ as
\begin{align}
\begin{split}
    \zeta & = x \cos^2 \iota, \\
    \mcgamma &= (5-3x) \Big ( 3\dfrac{\mczeta}{x}-1 \Big ) + 15 (1-x) \Big (1-\dfrac{\mczeta}{x} \Big ) \cos 2 \omega. \label{eqn:KL_constants}
\end{split}
\end{align}
Using these constants, we can write~\cref{eqn:KL_equations} entirely in terms of $x$, which is given by
\begin{equation}
\dv{x}{\ellmeankl} = - \dfrac{3 \sqrt{6}}{2} \sqrt{(x-x_0^*)(x-x_1^*)(x-x_2^*)} , \label{eqn:x_KL_exact}
\end{equation}
where $x_0^*, x_1^*$ and $x_2^*$, are the roots of the polynomial 
\begin{equation}
(12 x + \mcgamma-6\zeta-10)[(20+\mcgamma-18x)x+6 \mczeta (4x-5)]=0.
\end{equation}
and explicitly, they are given by (for $\iota \lesssim 40^{\circ}$)
\begin{align}
\begin{split}
    x_0^* &= \dfrac{\mczeta}{2} - \dfrac{\mcgamma}{12} + \dfrac{5}{6}, \\
    x_1^* &= \kappa - \sqrt{-5 \mczeta /3 + \kappa^2}, \\
    x_2^* &= \kappa + \sqrt{-5 \mczeta /3 + \kappa^2}, \\
    \kappa &= \dfrac{\mcgamma}{36} + \dfrac{5}{9} + \dfrac{2 \mczeta}{3}.
\end{split}
\end{align}
The solution to~\cref{eqn:x_KL_exact} is then given by
\begin{equation}
    x = \alpha_1 + (\alpha_0 - \alpha_1) \sn^2 (\theta,k^2) \label{eqn:KL_Exact_sol},
\end{equation}
where $\sn(\cdot)$ is one of the Jacobi elliptic functions~\cite{abramowitz_stegun} (for a physics inspired review of these functions, refer to~\cite{more_feshbach_hill,erdods}), and we also have the following definitions,
\begin{align}
\begin{split}
     \theta &=  \dfrac{3 \sqrt{6}}{8 \pi} \sqrt{(\alpha_2 - \alpha_0)} \epkl \ellmean + \theta_0, \\
  \theta_0 &= \sn^{-1} \Big [ \Big ( \dfrac{x_0-\alpha_1}{\alpha_0-\alpha_1} \Big )^{1/2} \Big ],\\
  k^2 &= \dfrac{\alpha_1-\alpha_0}{\alpha_2-\alpha_0}, \\
    \alpha_0 &= \text{min} ( \{ x_0^*,x_1^*,x_2^* \}), \\
    \alpha_1 &= \text{med} ( \{ x_0^*,x_1^*,x_2^* \}), \\
    \alpha_2 &= \text{max} ( \{ x_0^*,x_1^*,x_2^* \}),
\end{split}\label{eqn:params_KL_exact}
\end{align}
where we note that $k$ is the \emph{elliptic modulus}. With these definitions we can obtain an explicit expression for the KL time period that is given by
\begin{equation}
    P_{\text{KL}} = \dfrac{16 K }{3 \sqrt{6}} (\alpha_2 - \alpha_0)^{-1/2} \dfrac{m}{m_3} \left(\dfrac{R}{a}\right)^3 P_{\text{orb,in}}, \label{eqn:PKL_exact}
\end{equation}
where $K \equiv K (k^2)$ is the complete elliptic integral of the first kind. Note that $P_{\text{KL}}$ contains dependence on the initial conditions $e_0,\iota_0$, and $\omega_0$. Restricting the inclination to $\iota \lesssim \ang{40}$ implies a hierarchy in the roots: $x_1^* < x_0^*< x_2^*$. This hierarchy then means that $\alpha_0 = x_1^*$, $\alpha_1 = x_0^*$, and $\alpha_2 = x_2^*$, and it also ensures that the maximum eccentricity remains small. This is because the eccentricity can be written as
\begin{align}
    e^2 = \emin^2 + (\emax^2-\emin^2) \sn^2 (\theta,k^2), \label{eqn:2timescale_eccentricity_sol}
\end{align}
where $\emax^2 = 1-\alpha_1$ and $\emin^2=1-\alpha_0$ represent the maximum and minimum eccentricity respectively. For larger inclination angles, the hierarchy of roots changes depending on the value of $\omega_0$.  A more detailed analysis of the behavior of these roots, as well as the maximum eccentricity, can be found in~\cite{kozai,lidov} and we also present our own analysis in Appendix~\ref{app:roots_KL_problem}. In Sec.~\ref{subsubsec:2timescale_physical}, using two examples, we highlight key features of~\cref{eqn:2timescale_eccentricity_sol} that help provide a physical interpretation of the KL oscillations in the absence of RR; this helps understand the orbital dynamics when RR is included in Sec.~\ref{subsec:3timescale}, and will also shed light on the GW-waveform analysis discussed in Sec.~\ref{sec:waveforms}.

\subsubsection{Physical properties of the exact Kozai-Lidov solution} \label{subsubsec:2timescale_physical}

\begin{figure}[H]
    \centering
    \begin{subfigure}[t]{0.49\linewidth}
    \centering
    \includegraphics[width=\linewidth]{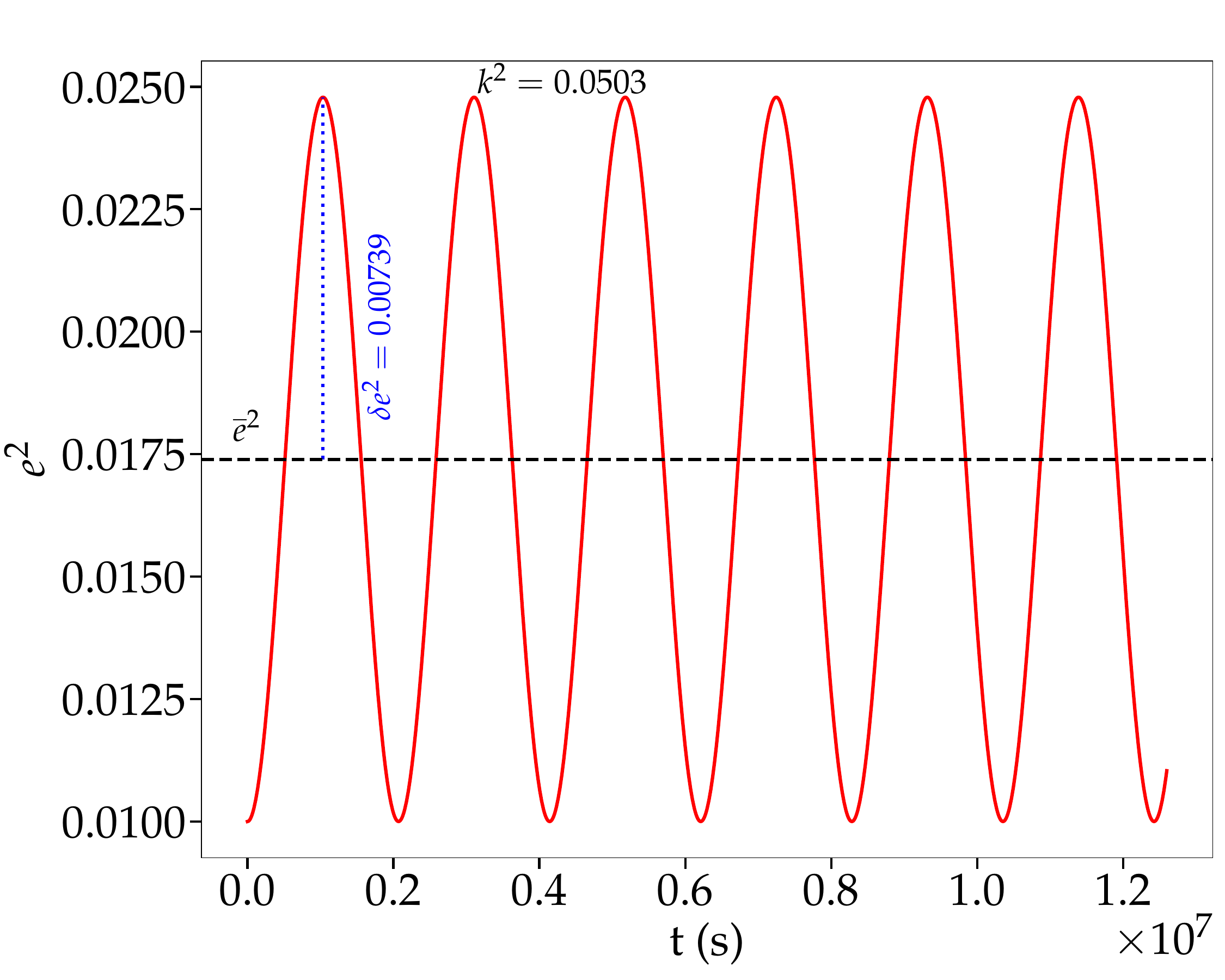}
    \caption{ $\overline{e}^2 =  0.0174$, $\delta e^2 = 0.00734$, $k^2 = 0.0503$.}
    \label{subfig:eKL_iota_30}
    \end{subfigure}
    \hfill
    \begin{subfigure}[t]{0.49\linewidth}
    \centering
    \includegraphics[width=\linewidth]{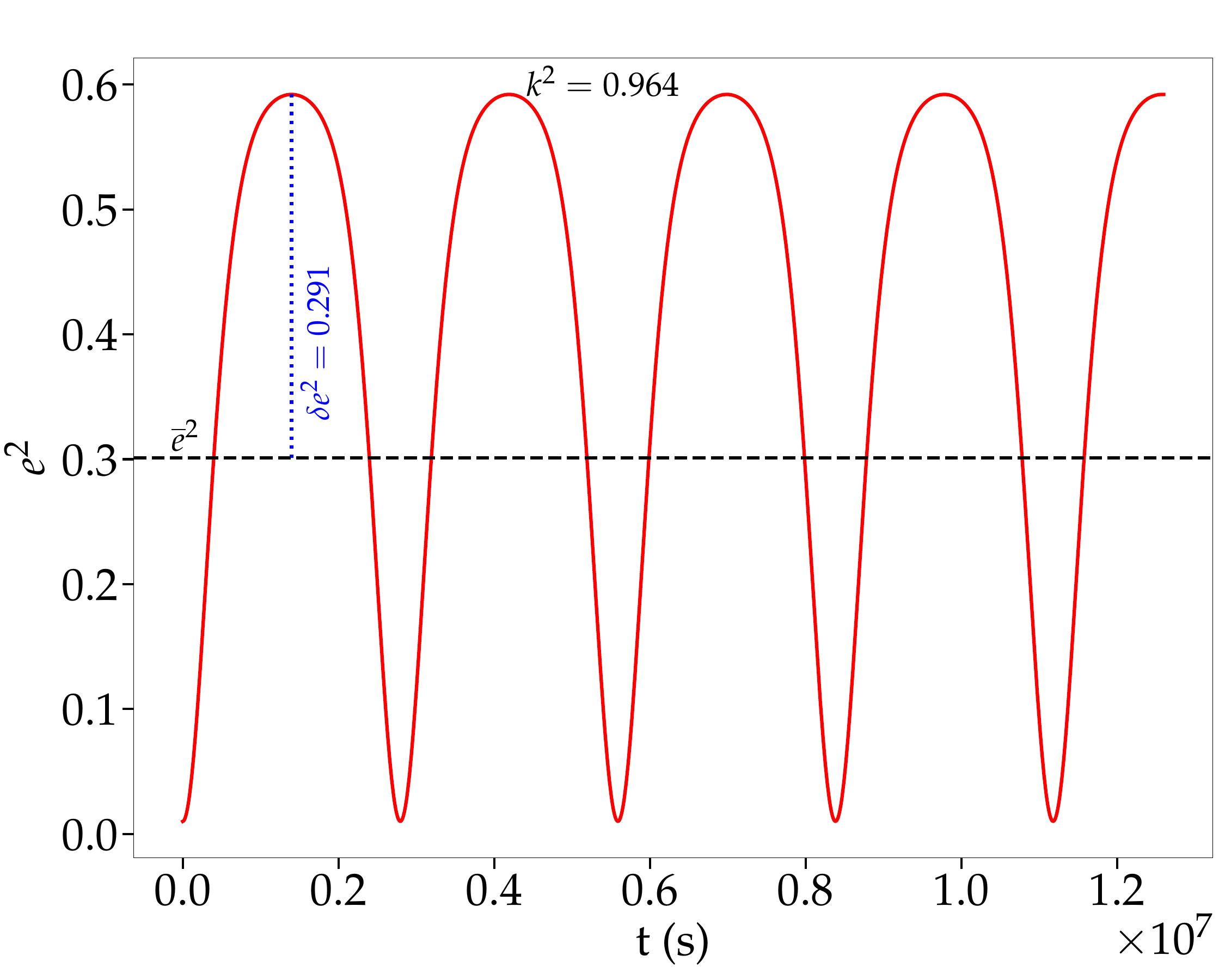}
    \caption{ $\overline{e}^2 =  0.301$, $\delta e^2 = 0.291$, $k^2 = 0.965$.}
    \label{subfig:eKL_iota_60}
    \end{subfigure}
    \caption{\EDIT{Illustration of eccentricity oscillations due to KL effect. We $e^2$ as a function of $t$ in the absence of RR. Here, $m_1=m_2 = 10^4 M_{\odot}, m_3 = 10^6 M_{\odot}, a_0 = 400 m, R=150 m_3, e_0 =0.1$, and $\omega_0 =0$. In~\cref{subfig:eKL_iota_30} $\iota_0 = \pi/6$, while in~\cref{subfig:eKL_iota_60} $\iota_0 = \pi/3$. Observe the clear change in the behavior of $e^2$, along with the values of $\{ \overline{e}^2, \delta e^2, k^2 \}$ between the two cases.}}
    \label{fig:eKL}
\end{figure}

Let us develop a physical interpretation of~\cref{eqn:2timescale_eccentricity_sol} by studying the behavior of some of the key parameters associated with the KL oscillations in the absence of RR. To do so, consider a system with masses $m_1=m_2 = 10^4 M_{\odot}, m_3 = 10^6 M_{\odot}$, initial semimajor axis \EDIT{$a_0 = 400 m$} and third-body separation \EDIT{$R=150 m_3$} such that $\epkl \ll 1$. Additionally let the initial eccentricity $e_0 = 0.1$ and $\omega_0 =0$. We will study two different values of ($\iota_0 = \pi/6$ and $\iota_0 =\pi/3$) which lead to very different KL oscillations, as can be seen from~\cref{fig:eKL}. Observe that although the parameter values chosen are illustrative, they coincide with the set $\{ m_1, m_2, m_3, a_0, R, e_0, \omega_0, \iota_0 = \pi/6 \}$ used in Sec.~\ref{sec:validation} to study several constraints in parameter space and the approximations used in Sec.~\ref{subsec:3timescale,sec:waveforms}.

The KL oscillations of $e$ can be understood through 3 parameters -- $\overline{e}, \delta e$, and $k$, where we introduce $\overline{e}^2 = (\emax^2+\emin^2)/2$ and $\delta e^2 = (\emax^2-\emin^2)/2$ and $k^2$ is defined in~\cref{eqn:params_KL_exact}. Physically, $\overline{e}^2$ represents the average value of the eccentricity oscillation, as can be seen from~\cref{eqn:2timescale_eccentricity_sol}, while $\delta e^2$ presents the amplitude of the oscillation. The parameter $k^2$ controls the shape and time period of the KL oscillations, where the latter property is clearly seen from~\cref{eqn:PKL_exact}.  

\onecolumngrid

\twocolumngrid
Figure~\ref{fig:eKL} shows clearly how the parameters $\{\overline{e}, \delta e, k \}$ capture the physical properties of the KL eccentricity evolution, such as the average value, the amplitude, the time period, and the `shape' of the oscillations. As $\iota_0$ changes from $\pi/6$ to $\pi/3$, the overall magnitude of $e$ changes significantly producing large-amplitude oscillations (note the values of $\delta e^2$ and $\overline{e}^2$) for the latter case. We also see that the ``shape'' of the oscillations is more sinusoidal for the smaller amplitude case of $\iota_0 = \pi/6$, which is consistent with the fact that in the limit $k \rightarrow 0$, $\sn(\theta, k^2) \rightarrow \sin (\theta)$. Also, the time period of oscillations is larger for the case $\iota_0 = \pi/3$, where the amplitude of oscillations is larger, and can be understood from~\cref{eqn:PKL_exact}, since $K(k^2) \rightarrow \infty $ when $k \rightarrow 1$ and that $\sn (\theta, k^2) \rightarrow \tanh (\theta)$ in this limit.

\subsection{Three-timescale analysis: Combining effects of Kozai-Lidov oscillations and Radiation-Reaction} \label{subsec:3timescale}
Under the effect of RR, the orbital separation will shrink, thereby decreasing the orbital time period. Noting that the KL time period scales as $a^{-3/2}$, as the inner binary inspirals, the period of KL oscillations will increase, and the KL effect will continue to get smaller~\cite{Blaes:2002cs} as we can see from~\cref{eqn:KL_equations}. In what follows, we describe how to combine the effects of both the KL oscillations and RR using a 3-timescale MSA analysis. \\ 

When we have both KL and RR perturbations, to capture the long-timescale behavior, we need to introduce two slow timescales given by $\ellkl =\epkl \ell $ and $\ellrr = \eprr \ell$. The total derivative with respect to the true anomaly becomes,
\begin{align}
\dv{}{\ell} = \pdv{}{\ell} + \epkl \pdv{}{\ellkl} + \eprr \pdv{}{\ellrr}.
\end{align}
The general structure of the evolution equations for the orbital elements $\mu_i$ is then given by
\begin{align}
\dv{\mu_i}{\ell} = \epkl g^{(\text{KL})}_i (\mu_{\alpha}) + \eprr g^{(\text{RR})}_i (\mu_{\alpha}), \label{eqn:mu_i_kl_rr}
\end{align}
where the functions $g^{(\text{KL,RR})}_i (\mu_{\alpha})$ result from the perturbations of the third body and RR respectively. The explicit form of these functions can be found in Appendix~\ref{app:perturbations}. To LO in MSA, we can promote the orbital elements to functions of 3 timescales, $\mu_i (\ell) \rightarrow \mu_i (\ell,\ellkl,\ellrr) $. We can also decompose the orbital elements into  
\begin{align}
        \mu_i (\ell,\ellkl,\ellrr) &= \mu_i^{(0)}(\ell,\ellkl,\ellrr) 
        \nonumber \\
        &+ \epkl \, \mu_i^{(1,KL)} (\ell,\ellkl,\ellrr) 
        \nonumber \\
        &+ \eprr \, \mu_i^{(1,RR)} (\ell,\ellkl,\ellrr)    
\end{align}

With this at hand,~\cref{eqn:mu_i_kl_rr} separates into a system of partial differential equations (PDEs) given by
\EDIT{\begin{align}
    \pdv{\mu_i^{(0)}}{\ell} &=0 \label{eqn:3timescale_mu_0}\\
    \pdv{\mu_i^{(0)}}{\ellkl} + \pdv{\mu_i^{(1,\text{KL})}}{\ell} &= g^{(\text{KL})}_i (\mu_{\alpha}^{(0)}), \label{eqn:3timescale_mu_1_kl} \\
    \pdv{\mu_i^{(0)}}{\ellrr} + \pdv{\mu_i^{(1,\text{RR})}}{\ell} &= g^{(\text{RR})}_i (\mu_{\alpha}^{(0)}). \label{eqn:3timescale_mu_1_rr}
\end{align}}
\Cref{eqn:3timescale_mu_0} means that the orbital elements are constant on the orbital timescale as they should be. We can now carry out our orbit averaging over both the inner and outer orbits to obtain
\begin{align}
    \Big ( \pdv{\mu_i^{(0)}}{\ellkl} \Big )_{\text{sec}} & = \langle g^{(\text{KL})}_i (\mu_{\alpha}^{(0)}) \rangle_{\ell, \ellout}, \label{eqn:3timescale_mu0_sec_kl} \\
    \Big ( \pdv{\mu_i^{(0)}}{\ellrr} \Big )_{\text{sec}} &=  \langle g^{(\text{RR})}_i (\mu_{\alpha}^{(0)}) \rangle_{\ell, \ellout}. \label{eqn:3timescale_mu0_sec_rr}
\end{align}
We first drop the ``secular notation'' on the derivatives and also relax the indices $(n)$ on the orbital elements as we had done earlier in Sec.~\ref{subsec:2timescale}, which amounts to
\begin{align}
    \Big ( \pdv{\mu_i^{(0)}}{\ellkl} \Big )_{\text{sec}} & =  \pdv{\mu_i}{\ellmeankl} , \nonumber \\
    \Big ( \pdv{\mu_i^{(0)}}{\ellrr} \Big )_{\text{sec}} &=  \pdv{\mu_i}{\ellmeanrr}. \label{eqn:ell_sec_PDE}
\end{align}
Further, we introduce $d/d \ellmean$ as the `total secular derivative' defined by
\begin{align}
    \dv{}{\ellmean} = \epkl \pdv{}{\ellmeankl} + \eprr \pdv{}{\ellmeanrr},
\end{align}
and then we can recast the PDE system given \EDIT{by~\cref{eqn:3timescale_mu0_sec_kl,eqn:3timescale_mu0_sec_rr}} into an ODE system,
\begin{align}
     \dv{\mu_i}{ \ellmean} = \ellkl \langle g^{(\text{KL})}_i (\mu_{\alpha}) \rangle_{\ell, \ellout} + \ellrr \langle g^{(\text{RR})}_i (\mu_{\alpha}) \rangle_{\ell, \ellout}, \label{eqn:ODE_three_timescales}
\end{align}
which is particularly useful for performing numerical integration, as discussed in Sec.~\ref{sec:validation}. We can also rewrite~\cref{eqn:ODE_three_timescales} as an ODE system with $t$ as the dependent variable by making use of the evolution equation for $\ellmean$, given by 
\EDIT{\begin{align}
    \dv{\ellmean }{t} = \Big (\dfrac{a^3}{m}\Big )^{-1/2}, \label{eqn:mean_anomaly_three_timescales}
\end{align}}
which is more convenient for doing the numerical integration. 

The RHS of~\cref{eqn:3timescale_mu0_sec_kl} is what we found in~\cref{eqn:KL_equations}, while the RHS of~\cref{eqn:3timescale_mu0_sec_rr} is given by the familiar result from~\cite{Peters,Peters-Mathews}. 

Thus, the relevant PDEs over the KL and RR timescales are given by
\begin{align}
\pdv{e}{\ellmeankl} &=  \dfrac{15}{4} \tilde{F}^{-2} e (1-e^2)^{1/2} \sin^2 \iota \sin \omega \cos \omega, \label{eqn:e_ellmeankl} \\
\pdv{e}{\ellmeanrr} & = -\dfrac{304}{15} \eta \tilde{F}^{5/3} e \dfrac{1+\frac{121}{304}e^2}{(1-e^2)^{7/2}}, \label{eqn:3timescale-e-rr} \\
\pdv{\ftilde}{\ellmeankl} &= 0 \label{eqn:ftilde_ellmeankl},\\
\pdv{\ftilde}{\ellmeanrr} &= \dfrac{96}{5} \eta \tilde{F}^{8/3} \dfrac{1+\frac{73}{24}e^2 + \frac{37}{96}e^4}{(1-e^2)^{7/2}}, \label{eqn:3timescale-ftilde-rr}\\
\pdv{\iota}{\ellmeankl} &= - \dfrac{15}{4} \tilde{F}^{-2} e^2 (1-e^2)^{-1/2} \sin \iota \cos \iota \sin \omega \cos \omega, \label{eqn:iota_ellmeankl}\\
\pdv{\iota}{\ellmeanrr} &= 0, \label{eqn:iota_ellmeanrr} \\
\begin{split}
\pdv{\omega}{\ellmeankl} &= \dfrac{3}{4} \tilde{F}^{-2} (1-e^2)^{-1/2} (5 \cos^2 \iota \sin^2 \omega \\ &+ (1-e^2)(5 \cos^2 \omega -3)), \label{eqn:omega_ellmeankl}
\end{split} \\
\pdv{\omega}{\ellmeanrr}&= 0, \label{eqn:omega_ellmeanrr}
\end{align} 
where $\ftilde = F /F_0$ is the scaled orbital frequency, with $F_0$ being the initial value of $F$. When only the KL effect is present, we knew that $\mcgamma$ and $\mczeta$, given by~\cref{eqn:KL_constants}, are constants. Under the influence of RR, they are constant only over the KL timescale, but will slowly vary over the RR timescale. Consequently, they are promoted to functions of $\ellmeanrr$ but have the same functional form as given in~\cref{eqn:KL_constants}. The equations over the KL timescale can be simplified to a single PDE for the quantity $x$, similar to the ODE we had obtained in~\cref{eqn:x_KL_exact}. The solution to the PDE is similar to~\cref{eqn:2timescale_eccentricity_sol} with the difference being that $\emax,\emin$ and $k$ must be promoted to functions of $\ellrr$, while $\theta$ becomes a function of both $\ellkl$ and $\ellrr$. Therefore, the solution to the eccentricity over both timescales can formally be expressed as 
\begin{align}
\begin{split}
        e^2 (\ell_{KL},\ell_{RR}) &= e_{\text{min}}^2 (\ell_{RR}) + (e_{\text{max}}^2 (\ell_{RR}) -e_{\text{min}}^2 (\ell_{RR})) \\
       & \times \text{sn}^2 [ \theta(\ellkl,\ellrr) ,k^2(\ellrr) ]. \label{eqn:3timescale_ecc_ansatz}
\end{split}
\end{align}

To obtain the frequency dependence of the eccentricity $e(F)$, we adopt the following steps:
\begin{enumerate}
\setlength\parskip{-0.05cm}
\setlength\parsep{-0.05cm}
    \item \underline{\textit{Master equation}}: We first use~\cref{eqn:3timescale_ecc_ansatz} with~\cref{eqn:3timescale-e-rr,eqn:e_ellmeankl} to find a master equation that involves the variation of $\emax, \emin$ and $\theta$ over both the KL and RR timescales. 
    \item \underline{\textit{Eccentricity expansion and KL-averaging}}: We then determine the evolution of eccentricity over the orbital frequency $d e^2 / d \ftilde$ by simply dividing $d e^2 / d \ellmean$ by $d\ftilde / d \ellmean$. When the KL effect is absent, the resulting equation can be integrated exactly to find $F(e)$~\cite{Peters}. In the presence of the KL effect, such a simple direct integration is not possible. Instead, we use the $e \ll1$ approximation to systematically expand $d e^2 / d \ftilde$ order by order and obtain evolution equations for $\emax$ and $\emin$.
    \item \underline{\textit{Frequency dependence of $\emax, \emin, \zeta$ and  $\kappa$}}: The resulting equations for $\emax$ and $\emin$ can be averaged over the KL oscillations. Upon averaging, we further expand the resulting Jacobi Elliptic functions in small $k^2$. The magnitude of $k^2$ is comparable to that of $e^2$ in the regime we work in, thus allowing for a Fourier expansion of the Jacobi Elliptic functions which results in a bi-variate expansion in $e,k$. We then solve for $\emax, \emin, \zeta$ and $\kappa$ as functions of the orbital frequency.
    \item \underline{\textit{Frequency dependence of $\theta$}}: We go back to the master equation to determine the frequency evolution of $\theta$. We also simplify the calculation by neglecting the evolution of $\theta$ over the RR timescale and we show the robustness of this simplification by direct comparison with numerical results. Thus, we obtain the solution for the eccentricity as a function of the orbital frequency, given by~\cref{eqn:e2_F_final}, which is our first main result. 
\end{enumerate}

\subsubsection{Master equation}
We first obtain the master equation involving the evolution of $\emax, \emin$ and $\theta$ over both the KL and RR timescales. The evolution of $e^2$ has the form
\begin{align}
\dv{e^2}{\ellmean} &= \epkl \pdv{e^2}{\ellmeankl} + \eprr \pdv{e^2}{\ellmeanrr} \label{eqn:e_ellmean}.
 \end{align}
From~\cref{eqn:3timescale_ecc_ansatz}, the variation over the KL timescale is given by
\begin{align}
    \pdv{e^2}{\ellmeankl} &= 2 \ (e_{\text{max}}^2 -e_{\text{min}}^2 ) \ \sn (\theta,k^2) \ \cn (\theta,k^2) \nonumber \\
    &\times \dn (\theta,k^2) \ \pdv{\theta}{\ellmeankl}, \label{eqn:e2_ansatz_KL_variation}
\end{align}
and this is clearly odd in $\theta$ so it will vanish upon averaging over KL cycles. We now obtain the evolution equation for $\theta$ over the KL timescale from using~\cref{eqn:e2_ansatz_KL_variation} with~\cref{eqn:e_ellmeankl,eqn:omega_ellmeankl,eqn:iota_ellmeankl}, which gives us
\begin{align}
\dfrac{\partial \theta}{\partial \ellmeankl} = \tilde{F}^{-2} \dfrac{3 \sqrt{6}}{8 \pi} \sqrt{\alpha_2-\alpha_0}  \label{eqn:theta_KL}
\end{align}
To calculate the rest of the master equation,~\cref{eqn:e_ellmean}, we consider the variation of~\cref{eqn:3timescale_ecc_ansatz} over the RR timescale, which gives
\begin{align}
    \pdv{e^2}{\ellmeanrr} &= \dv{\emax^2}{\ellmeanrr} \sn^2 (\theta,k^2) + \dv{\emin^2}{\ellmeanrr} \cn^2 ( \theta ,k^2) \nonumber \\
    &+ \pdv{\sn^2(\theta,k^2)}{\ellmeanrr}. \label{eqn:e2_ansatz_RR_Variation}
\end{align}
The third term in~\cref{eqn:e2_ansatz_RR_Variation} is odd in $\theta$, which means that upon averaging over $\theta$, it will vanish. We can use~\cref{eqn:e2_ansatz_RR_Variation,eqn:3timescale-e-rr}, to determine the evolution equation for $\theta$ over the RR timescale, 
\begin{align}
     \pdv{\sn^2(\theta,k^2)}{\ellmeanrr} & = -\dfrac{304}{15} \eta \tilde{F}^{5/3} e^2 \dfrac{1+\frac{121}{304}e^2}{(1-e^2)^{7/2}} \nonumber \\
     &- \dv{\emax^2}{\ellmeanrr} \sn^2 (\theta,k^2) - \dv{\emin^2}{\ellmeanrr} \cn^2 ( \theta ,k^2), \label{eqn:theta_RR_Variation}
 \end{align}
which completes the formal calculation of the master equation. Once we have the frequency dependence of $\emax$ and $\emin$,~\cref{eqn:theta_RR_Variation} can be used to obtain the evolution of $\theta$ over the RR timescale explicitly. However, from a direct comparison with numerical results (detailed in Sec.~\ref{sec:validation}), we find that it is a very good approximation to ignore the dependence of $\theta$ and $k$ on the RR timescale, and therefore we will not be needing~\cref{eqn:theta_RR_Variation} for the rest of the calculation.
 
 \subsubsection{Eccentricity expansion and KL-averaging}

Upon ignoring the variation of $k$ and $\theta$ over the RR timescale, we set $k=k_0$ in~\cref{eqn:3timescale_ecc_ansatz}, and this greatly simplifies our calculation of $e (F)$. Physically, $k_0$ controls the ``shape'' and the KL time period associated with the KL oscillations, as was seen from Sec.~\ref{subsubsec:2timescale_physical}.
 
Since $\emax$ and $\emin$ only vary over the RR timescale, we can average over the shorter KL cycles to compute the dependence on the orbital frequency. We start with the expression for $d e^2 / d \ftilde$, 
\begin{align}
\dv{e^2}{\ftilde} & = \epkl \pdv{e^2}{\ellmeankl} \times \eprr^{-1} \pdv{\ellmeanrr}{\ftilde} +  \pdv{e^2}{\ellmeanrr} \times \pdv{\ellmeanrr}{\ftilde},
\end{align}
and then average over $\theta$ on both sides which gives us
\begin{align}
 \Big \langle \dv{e^2}{\ftilde} \Big \rangle_{\theta} &=  \Big \langle \pdv{e^2}{\ellmeanrr} \times \pdv{\ellmeanrr}{\ftilde} \Big \rangle_{\theta}, \label{eqn:e2_freq}
\end{align}
where the first term on the RHS, being odd in $\theta$, vanishes upon averaging. We introduce $e_{\text{max},0}$ as the initial value of $\emax$ and $e_{\text{min},0}$ as the initial value of $\emin$. The initial conditions for the LO and \textit{next to leading order} (NLO) pieces are given by $e_{\text{max/min,LO}} (0) = e_{\text{max/min},0} $ and $e_{\text{max/min,NLO}} (0) = 0$ respectively. We use a small-eccentricity approximation, motivated by~\cite{NicoBertiArunWill}, to extract the frequency dependence of $e$, and to this end we use the ansatz,
\begin{align}
\begin{split}
e_{\text{max}}^2 (F) &= \ e_{\text{max,LO}}^2 (F) + \ e_{\text{max,NLO}}^2 (F), \\
e_{\text{min}}^2 (F) &= \ e_{\text{min,LO}}^2 (F) + \ e_{\text{min,NLO}}^2 (F),
\end{split}\label{eqn:emaxmin_ansatz} 
\end{align}
where the NLO contributions, given by $e_{\text{max,NLO}}^2$ and $e_{\text{min,NLO}}^2$, are of $\mathcal{O}(e^4)$. We will also be expanding in powers of $k_0$ and for the cases of interest, $k_0$ is comparable in value to $e$ so we adopt a bi-variate expansion in $e,k_0$. We will keep terms up to $\mathcal{O} (e^4, e^2 k_0^2 , k_0^4)$, and ignore higher order terms, which can be systematically obtained.  Using~\cref{eqn:3timescale_ecc_ansatz,eqn:emaxmin_ansatz}, the \EDIT{left-hand side} (LHS) of \cref{eqn:e2_freq} works out to be
\begin{align}
    \begin{split}
       \Big  \langle \dv{e^2}{\ftilde} \Big \rangle_{\theta} = \dv{\emin^2}{\ftilde} \langle \cn^2 (\theta,k_0^2) \rangle_{\theta} + \dv{\emax^2}{\ftilde} \langle \sn^2 (\theta,k_0^2) \rangle_{\theta}.
    \end{split}
\end{align}
We can perform the averaging over $\theta$ for each term, and by plugging in the ansatz for $\emax$ and $\emin$, we obtain the following equation for the LHS of \cref{eqn:e2_freq},
\begin{align}
LHS &= \dfrac{1}{2} \Big ( \dfrac{d e_{\text{max,LO}}^2}{d \tilde{F}} + \dfrac{d e_{\text{min,LO}}^2}{d \tilde{F}} \Big ) \nonumber \\
&+ \dfrac{1}{2} \Big ( \dfrac{d e_{\text{max,NLO}}^2}{d \tilde{F}} + \dfrac{d e_{\text{min,NLO}}^2}{d \tilde{F}} \Big ) \nonumber \\
&+ \dfrac{3}{16} k_0^2 \Big ( \dfrac{d e_{\text{max,LO}}^2}{d \tilde{F}}- \dfrac{d e_{\text{min,LO}}^2}{d \tilde{F}} \Big ).
\end{align}
We can work out the \EDIT{right-hand side} (RHS) in a similar fashion by expanding to a desired order in $e,k_0$, averaging over $\theta$ and using the ansatz for $\emax$ and $\emin$. The RHS of \cref{eqn:e2_freq}, to $\mathcal{O} (e^4)$ is then given by
\begin{align}
RHS &= -\dfrac{19}{9 \tilde{F}} \Bigg [ \dfrac{e_{\text{max,LO}}^2+e_{\text{min,LO}}^2}{2} + \dfrac{e_{\text{max,NLO}}^2+e_{\text{min,NLO}}^2}{2} \nonumber \\
&- \dfrac{3323}{7296} \big ( 3 (e_{\text{max,LO}}^4+e_{\text{min,LO}}^4 +2 e_{\text{max,LO}}^2 e_{\text{min,LO}}^2) \big ) \nonumber \\
&+ \dfrac{1368}{7296} k_0^2 (e_{\text{max,LO}}^2 - e_{\text{min,LO}}^2 ) \Bigg ].
\end{align}

With this at hand, we can now find expressions for the LO and NLO parts of the master equation. At LO, we have
\begin{align}
\begin{split}
    \Big ( \dfrac{d e_{\text{max,LO}}^2}{d \tilde{F}} + \dfrac{19}{9} \dfrac{e_{\text{max,LO}}^2}{\ftilde} \Big ) &= 0, \\
\Big ( \dfrac{d e_{\text{min,LO}}^2}{d \tilde{F}} + \dfrac{19}{9} \dfrac{e_{\text{min,LO}}^2}{\ftilde} \Big ) &= 0 ,
\end{split} \label{eqn:e_LO_evolution}
\end{align}
and at NLO, we have
\begin{widetext}
\begin{align}
\begin{split}
    \Big ( \dfrac{d e_{\text{max,NLO}}^2}{d \tilde{F}} + \dfrac{19}{9} e_{\text{max,NLO}}^2 \Big ) +\Big ( \dfrac{d e_{\text{min,NLO}}^2}{d \tilde{F}} + \dfrac{19}{9} e_{\text{min,NLO}}^2 \Big ) &= -  \dfrac{\EDIT{38}}{9 \tilde{F}} \Bigg [ - \dfrac{3323}{7296} \big (\EDIT{3e_{\text{max,LO}}^4+3e_{\text{min,LO}}^4 +2 e_{\text{max,LO}}^2 e_{\text{min,LO}}^2}\big ) \\
    &+ \dfrac{1368}{7296} k_0^2 (e_{\text{max,LO}}^2 - e_{\text{min,LO}}^2 ) \ \Bigg ] - \dfrac{3}{8} k_0^2 \Big ( \dfrac{d e_{\text{max,LO}}^2}{d \tilde{F}} - \dfrac{d e_{\text{min,LO}}^2}{d \tilde{F}} \Big ).
\end{split}\label{eqn:e_NLO_evolution}
\end{align}
\end{widetext}
\EDIT{To summarize thus far, we have orbit averaged over the KL cycles to determine the behavior of $\emax$ and $\emin$ over the RR timescale. We first obtained the evolution of $\emax, \emin$ to LO in~\cref{eqn:e_LO_evolution}, and subsequently, in~\cref{eqn:e_NLO_evolution}, we found the NLO evolution equation for $\emax, \emin$, which are `sourced' by the LO pieces. We show below how to solve the LO and NLO equations.}
 \subsubsection{Frequency dependence of $\emax, \emin, \zeta$ and $\kappa$}
We first solve the LO evolution equation,~\cref{eqn:e_LO_evolution}, and we obtain 
\begin{align}
    \begin{split}
        e_{\text{max,LO}}^2 (F) =e_{\text{max},0}^2 \ \ftilde^{-19/9}, \label{eqn:emax_LO} \\
        e_{\text{min,LO}}^2 (F) = e_{\text{min},0}^2 \ \ftilde^{-19/9},
    \end{split}
\end{align}
where $e_{\text{max},0}$ and $e_{\text{min},0}$ are the initial values on $\emax$ and $\emin$ respectively. This LO behavior for $\emax$ and  $\emin$ is also what one finds as the LO behavior of the eccentricity for an isolated eccentric binary in the small-eccentricity approximation~\cite{NicoBertiArunWill}. Therefore, to LO, we have that the quantities $\emax$ and $\emin$ behave independently, i.e.~their evolution is decoupled. 

Using~\cref{eqn:emax_LO}, the RHS of the NLO evolution, given by~\cref{eqn:e_NLO_evolution}, simplifies considerably, and we can separate out the evolution of the NLO pieces into
\begin{align}
\begin{split}
    \Big ( \dfrac{d e_{\text{max,NLO}}^2}{d \tilde{F}} + \dfrac{19}{9} e_{\text{max,NLO}}^2 \Big ) & =  \dfrac{3323}{1728 \tilde{F}^{47/9}} e_{\text{max},0}^2 \\
   & \times (3 e_{\text{max},0}^2 + e_{\text{min},0}^2), \label{eqn:emaxmin_NLO_evol} \\
\Big ( \dfrac{d e_{\text{min,NLO}}^2}{d \tilde{F}} + \dfrac{19}{9} e_{\text{min,NLO}}^2 \Big ) & =  \dfrac{3323}{1728 \tilde{F}^{47/9}} e_{\text{min},0}^2 \\
\times & (3 e_{\text{min},0}^2 + e_{\text{max},0}^2).
\end{split}
\end{align}
The above equations obey reflection symmetry, specifically under the exchange of $e_{\text{min},0}^2 \leftrightarrow e_{\text{max},0}^2 $, and so will the solution to the equations. \EDIT{Further, note that in~\cref{eqn:emaxmin_NLO_evol}, the dependence on $k_0^2$ cancels out upon inserting the LO evolution for $\emax$ and $\emin$ in the RHS of~\cref{eqn:e_NLO_evolution}}. Now, the solution for $e_{\text{max,NLO}}^2$ is given by
\begin{align}
\begin{split}
    e_{\text{max,NLO}}^2 &= \dfrac{3323}{3648} ( e_{\text{max},0}^2 \tilde{F}^{-19/9} ) (3e_{\text{max},0}^2 + e_{\text{min},0}^2) \\
    &\times (1-\tilde{F}^{-19/9}),
\end{split}
\end{align}
and the full solution for $e^2_{\text{max}}$ is
\begin{align}
\begin{split}
e^2_{\text{max}} &= e_{\text{max},0}^2 \tilde{F}^{-19/9} \Big (1+ \dfrac{3323}{3648}(3 e_{\text{max},0}^2 + e_{\text{min},0}^2) \\
& \times (1-\tilde{F}^{-19/9}) \Big ).
\end{split}\label{eqn:emax_NLO_sol}
\end{align}
Similarly, the full solution for $e^2_{\text{min}}$ is
\begin{align}
\begin{split}
    e^2_{\text{min}} &= e_{\text{min},0}^2 \tilde{F}^{-19/9} \Big (1+ \dfrac{3323}{3648}(3 e_{\text{min},0}^2 + e_{\text{max},0}^2) \\
    & \times (1-\tilde{F}^{-19/9}) \Big ).
\end{split} \label{eqn:emin_NLO_sol}
\end{align}

Now we turn towards determining the frequency dependence of the quantities $\zeta, \kappa$ and $\sqrt{\alpha_2-\alpha_0}$. We begin by determining how $\zeta$ depends on $F$. Since the inclination angle does not explicitly vary over the RR timescale, differentiating with respect to $\ellmeanrr$, we obtain
\begin{align}
\begin{split}
    \dfrac{d \zeta}{d \ellmeanrr} &= -2 e \dfrac{\partial e}{\partial \ellmeanrr} \cos^2 \iota  \\
& \simeq  \dfrac{608}{15} \eta e^2 \zeta \ \tilde{F}^{5/3},
\end{split}
\end{align}
where we have truncated $\partial e/ \partial \ellmeanrr $ and $\zeta/(1-e^2)$to LO in $e^2$. Therefore, we can compute $d \zeta/d \tilde{F}$ to LO, 
\begin{align}
\dfrac{d \zeta}{d \tilde{F}} \simeq \dfrac{19}{9} \dfrac{e^2 \zeta}{\tilde{F}}.
\end{align}
Since we are only interested in the secular behavior of $\zeta$, we can take liberty in averaging over the KL phase $\theta$, thereby making $e^2 \rightarrow \overline{e}^2$, where
\begin{equation}
\overline{e}^2(F) = \dfrac{e^2_{\text{max,LO},}(F)+e^2_{\text{min,LO}}(F)}{2}  \label{eqn:ebar}
\end{equation}
represents to LO, the average value of the eccentricity oscillation that was introduced in Sec.~\ref{subsubsec:2timescale_physical}. This makes the evolution of $\zeta$ become,
\begin{align}
\dfrac{d \zeta}{d \tilde{F}} &= \dfrac{19}{9} \dfrac{\overline{e}^2 \zeta}{\tilde{F}} \\
\implies \zeta &= \zeta_0 \exp \Big [ \overline{e}_0^2 (1- \tilde{F}^{-19/9}) \Big ] \label{eqn:zeta_freq_unexpanded},
\end{align}
where we have integrated directly as the equation is separable, and introduced $\overline{e}^2_0 \equiv \overline{e}^2 (0)$ as well as $\zeta_0 \equiv \zeta (0)$. Given that the argument of the exponential contains a term that is $\mathcal{O} (e^2)$, we can expand it to LO to obtain,
\begin{align}
\zeta = \zeta_0 \Big [1+ \overline{e}_0^2 (1- \tilde{F}^{-19/9}) \Big ]. \label{eqn:zeta_freq}
\end{align}
Now we turn to the expression for $\kappa$ and we find,
\begin{equation}
\kappa = \dfrac{1}{6} \Big [ 2+8 \overline{e}^2 + 10 \zeta - (1 + 8 \overline{e}^2 + 16 \overline{e}^4 - 10 \zeta + 40 \overline{e}^2 \zeta + 25 \zeta^2)^{1/2} \Big ].
\end{equation}
We can then use~\cref{eqn:zeta_freq_unexpanded} and expand to $\mathcal{O} (e^2)$ to find 
\begin{align}
\begin{split}
    \kappa (F) &= \dfrac{1}{6} \Bigg \{ 3 + 5 \zeta_0 + \bigg [ \tilde{F}^{19/9} (-1 + 5 \zeta_0) \bigg ]^{-1} \\
   & \times \overline{e}_0^2 \bigg [ -12 - 5 (-5 + \tilde{F}^{19/9}) \zeta_0 + 25 (-1 + \tilde{F}^{19/9}) \zeta_0^2 \bigg ] \Bigg \} .
\end{split} \label{eqn:kappa_freq}
\end{align}
Using~\cref{eqn:kappa_freq,eqn:zeta_freq}, we can obtain the frequency dependence of $\gamma$. Further, using~\cref{eqn:kappa_freq}, we obtain an expression for $\sqrt{\alpha_2-\alpha_0}$ as a function of the mean orbital frequency, which is given by
\begin{align}
\begin{split}
    \sqrt{\alpha_2-\alpha_0} & = \dfrac{1}{6 \sqrt{\dfrac{5}{3} \zeta_0 -1}} \Big \{ \big [ -6+5 (2+\overline{e}^2_0)\zeta_0 \big ] \\
    & + \overline{e}^2_0 (12+25 \zeta_0-25 \zeta_0^2) \tilde{F}^{-19/9}  \Big \}. \label{eqn:root_diff_freq}
\end{split}
\end{align}

 \subsubsection{Frequency dependence of $\theta$}
Finally, we now show how to obtain the frequency dependence of the KL phase $\theta$. Recall from~\cref{eqn:theta_KL}, that the evolution of $\theta$ depends on the quantity $\sqrt{\alpha_2-\alpha_0}$, given by~\cref{eqn:root_diff_freq}. From the KL averaged form of~\cref{eqn:3timescale-ftilde-rr}, computed to order $\mathcal{O} (e^2)$, we integrate~\cref{eqn:theta_KL} with respect to $F$. Doing so, we find
\begin{align}
\begin{split}
\theta (F) &= - \dfrac{\epkl}{\eprr} \dfrac{15}{36608 \sqrt{2} \pi} \Big [ (5 \zeta_0 -3 )^{-1/2} \Big \{ 52 \tilde{F}^{-11/3}\\
& \times (-6+5\zeta_0 (2+\overline{e}^2_0) ) + 33 \tilde{F}^{-52/9} \ \overline{e}^2_0 (12+25 \zeta_0 \\
& - 25 \zeta_0^2) (5 \zeta_0-1)^{-1} \Big \} \Big ] +\mathcal{I} _{\theta_0}, \label{eqn:theta_F}
\end{split}
\end{align}
where
\begin{align}
\begin{split}
 \mathcal{I}_{\theta_0} &= \theta_0 + \dfrac{\epkl}{\eprr} \dfrac{15}{36608 \sqrt{2} \pi} \Big [ (5 \zeta_0 -3 )^{-1/2} \Big \{ 52 (-6+5\zeta_0 \\
 & \times (2+\overline{e}^2_0) ) + 33 \overline{e}^2_0 (12+25 \zeta_0-25 \zeta_0^2)(5 \zeta_0-1)^{-1} \Big \} \Big ].
\end{split} \label{eqn:thetaVSfreqMSA}
\end{align}
Using~\cref{eqn:emax_NLO_sol,eqn:emin_NLO_sol,eqn:theta_F}, we write down the evolution of the eccentricity as a function of orbital frequency, as
\begin{equation}
    e^2 (F) = \emin^2 (F) + [ \emax^2 (F) - \emin^2 (F) ] \sn^2 [ \theta (F), k_0^2] \label{eqn:e2_F_final}.
\end{equation}

Let us now summarize what we have done so far. First, we have obtained the orbital evolution over the KL and RR timescales using MSA. Then, using both the small-eccentricity approximation and averaging over KL cycles, we have also determined the frequency dependence of $e, \iota$ and $\omega$ to $\mathcal{O} (e^4)$, which can be systematically extended to higher order. \Cref{eqn:emin_NLO_sol,eqn:emax_NLO_sol,eqn:theta_F,eqn:thetaVSfreqMSA} determine $e (F)$, while~\cref{eqn:e2_F_final,eqn:KL_constants} determine $\omega (F)$ and $\iota(F)$. 

\section{Gravitational waveform modeling} \label{sec:waveforms}
\setlength{\parskip}{0mm}

In this section, we show how to compute the GWs from the inner binary that is undergoing KL oscillations. We will then apply the analytic results for the evolution of the orbital elements and finally obtain the amplitude and phase of the GW polarizations in the Fourier domain. The method for computing the GWs is as follows:
\begin{enumerate}
\setlength\parskip{-0.1cm}
\setlength\parsep{-0.1cm}
    \item We compute the waveform polarizations observed by a GW detector placed on the $Z-$axis of the FF and show how to transform to a more general detector frame ($DF$). We use the \textit{quadrupole formula}~\cite{Wahlquist:1987rx} to determine the polarizations in the time domain.  
    \item We then expand the time domain polarizations, as a Fourier series just as was done in~\cite{NicoBertiArunWill,blake_2018,blake_2019}. We provide explicit expressions for the coefficients of this Fourier series.
    \item We proceed to compute the Fourier transform of the polarizations and make use of the SPA in evaluating the integrals. We show how the stationary-phase condition can be adapted to accommodate the KL oscillations.
    \item We then evaluate the amplitude and phase of the polarizations at a given harmonic. For the phase, we use a small-eccentricity approximation in order to perform the integrals over the frequency domain.
\end{enumerate}
\subsection{Choice of Reference Frame}
The GWs emitted by the inner binary propagate in the direction of $\Vec{L}_{\text{in}}$. To compute the polarizations of the GWs, the coordinates of the propagation unit vector $\hat{L}_{\text{in}}$ must be specified in a certain reference frame. Since the outer orbit is taken to be stationary, it is suitable to compute the GW polarizations in the FF by placing a GW detector on the $Z-$axis, which is the direction of $\Vec{L}_{\text{out}}$. Explicitly we can write the coordinates of the detector in the FF as $\Vec{D} = D_L [0,0,1]_{FF} $, where $D_L$ is the luminosity distance to the source. The propagation unit vector in the FF becomes $\hat{L}_{\text{in}} = [\sin \iota \cos \beta , \sin \iota \cos \beta , \cos \iota ]_{FF}$, where $\beta = \pi/2 - \omega$. Furthermore, we use the traceless, transverse (TT) gauge for computing the GW polarizations, and once they are computed in the FF, we can transform the polarization tensor $h_{ij}^{\text{TT}}$ to a more general $DF$. To accomplish this transformation, we note that the unit vector $\hat{L}_{\text{out}}$ in the DF is given by $\hat{L}_{\text{out}} = [\sin \iota_{\text{out}} \cos \beta_{\text{out}} , \sin \iota_{\text{out}} \beta_{\text{out}} , \cos \iota_{\text{out}} ]_{DF}$, where the angles $\iota_{\text{out}}$ and $\beta_{\text{out}}$ are parameters associated with the outer orbit that would have to be measured by the detector. The difference between the two reference frames is a result of rotating $\Vec{D}$ or $\hat{L}_{\text{out}}$ through a set of two rotations. The transformation of the GW polarizations is then explicitly given by 
\begin{equation}
 \textbf{h}^{\text{TT,DF}} = \textbf{R}_{FF \rightarrow DF} \ \textbf{h}^{\text{TT,FF}} \textbf{R}^{-1}_{FF \rightarrow DF}
\end{equation}
where the rotation matrix is given by
\begin{align}
\begin{split}
    \textbf{R}_{FF \rightarrow DF} = 
    \begin{pmatrix} 
    \cos \beta_{\text{out}} & -\sin \beta_{\text{out}} & 0\\
    \sin \beta_{\text{out}}& \cos \beta_{\text{out}} & 0 \\
    0 & 0 & 1
    \end{pmatrix} \\ 
    \times 
      \begin{pmatrix} 
    \cos \iota_{\text{out}} &  0 & \sin \iota_{\text{out}}\\
    -0 & 1 & 0\\
    -\sin \iota_{\text{out}} & 0 & \cos \iota_{\text{out}}
    \end{pmatrix}  
\end{split} \label{eqn:Dectector_transform}
\end{align}
We drop the notation of the reference frame keeping in mind that we are working in the FF, until otherwise stated.  Following~\cite{NicoBertiArunWill,martel-poisson,moreno-garrido,poisson_will_2014}, and making use of the quadrupole formula, we obtain for the time-domain polarizations,
\begin{align}
\begin{aligned}
\begin{split}
    h_{+} &= -\dfrac{\mu}{p D_L} \Big [ \Big ( 2 \cos (2\phi - 2 \beta) + \dfrac{5 e}{2} \cos (\phi -2\beta) \\
    &+\dfrac{e}{2} \cos (3\phi - 2\beta) + e^2 \cos 2 \beta \Big )(1+\cos^2\iota) \\
    &+ (e\cos \phi + e^2)\sin^2 \iota \Big ], \\
    h_{\times} & = -\dfrac{\mu}{p D_L} \Big [ 4 \sin (2\phi - 2\beta) + 5e \sin (\phi - 2\beta) \\
    &+ e \sin (3 \phi -2\beta) -2e^2 \sin 2 \beta \Big ] \cos \iota.
\end{split}
\end{aligned}
\end{align}

\begin{figure}[H]
    \centering
    \begin{subfigure}[t]{0.49\linewidth}
    \centering
    \includegraphics[width=\linewidth]{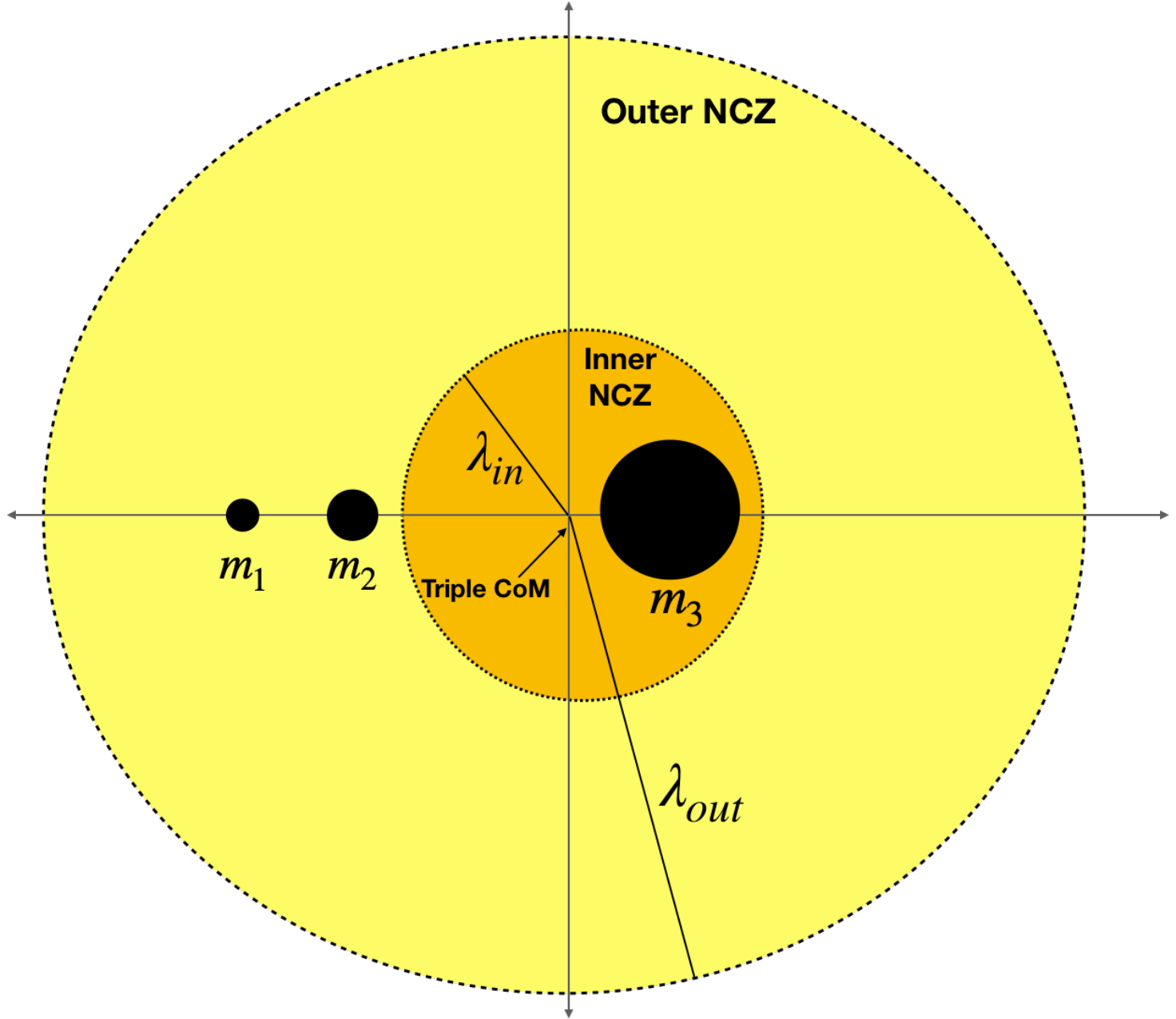}
    \caption{}
    \label{fig:NCZ_a}
    \end{subfigure}
    \hfill
    \begin{subfigure}[t]{0.49\linewidth}
    \centering
    \includegraphics[width=\linewidth]{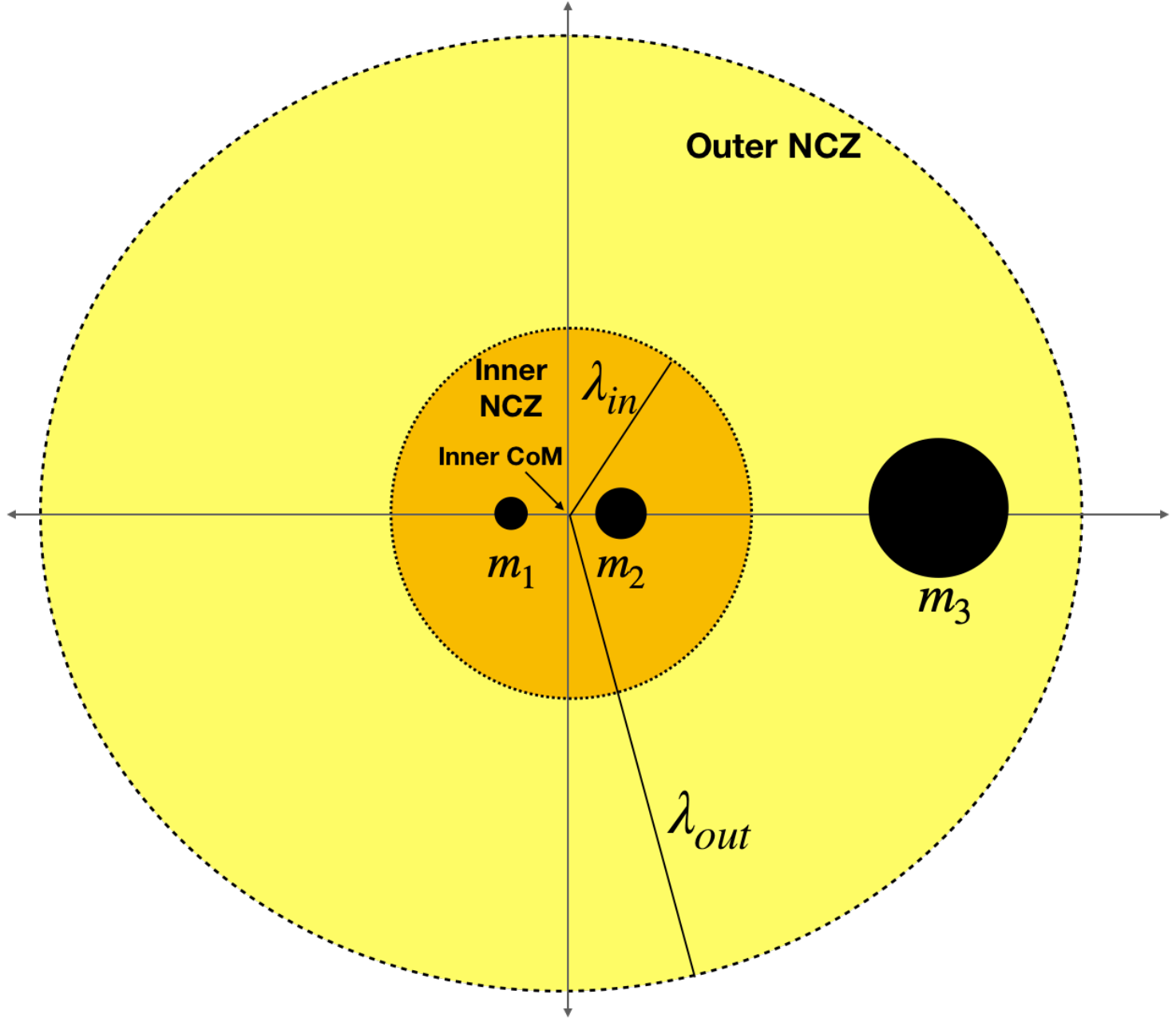}
    \caption{}
    \label{fig:NCZ_b}
    \end{subfigure}
    \caption{Cartoon representation (not to scale) of the NCZs of the triple system for different choices of the origin of the coordinate system. In both figures, we have suppressed the third spatial dimension and have assumed co-planar orbits for the inner and outer binaries for the sake of illustration. In panel~\subref{fig:NCZ_a}, the origin of the coordinates is chosen to be the CoM of the triple, while in panel~\subref{fig:NCZ_b}, it is chosen to be that of the CoM of the inner binary. Observe that the inner binary lies within the inner NCZ in the latter, thereby making the use of the quadrupole formula admissible.}
    \label{fig:NCZ}
\end{figure}

We note that there can be ambiguity in applying the quadrupole formula for a hierarchical triple system due to the existence of two `Near Coordinate Zones' (NCZs) -- the `Inner NCZ' and the `Outer NCZ'~\cite{Bonetti:2017hnb}, that are characterized by the gravitational wavelengths $\lambda_{\text{in}} \sim a/v_{\text{in}}$ and $\lambda_{\text{out}} \sim R / v_{\text{out}}$ respectively. If the origin of the coordinate system is chosen to be the center of mass (CoM) of the triple, it can happen that the inner binary lies outside the `Inner NCZ' (shown in~\cref{fig:NCZ}), particularly when the third body is much more massive and is much further away. 

Since, in this work, it is the inner binary that is in the spotlight when it comes to gravitational radiation, care must be taken in applying the quadrupole formula in order to avoid nonphysical signatures in the gravitational waveform. A simple way to do so (shown in Fig.~\ref{fig:NCZ}), as pointed out in~\cite{Bonetti:2017hnb}, is to instead choose the origin of the coordinate system to be the CoM of the inner binary, which is what we have done in our work. Given that $v_{\text{in}} \ll 1$ and $v_{\text{out}} \ll 1$, this would guarantee that the inner and outer binaries lie outside their respective NCZs, thereby admitting the use of the quadrupole formula for the inner binary. 

\onecolumngrid

\setlength\belowcaptionskip{-1ex}

\twocolumngrid

\subsection{Fourier analysis of gravitational wave polarizations} \label{subsec:Fourier_analysis_GW}
\setlength{\parskip}{0mm}

The time domain polarizations can then be analyzed through a Fourier series, which is an extension of the Fourier analysis of the Kepler problem. A thorough review can be found in~\cite{NicoBertiArunWill,blake_2018,blake_2019,Maggiore:1900zz,moreno-garrido} and here we only present the final result. The polarizations take the form
\begin{align}
    h_{+,\times} = \mathcal{A} \sum \limits_{n=1}^{\infty} \Big [ C_{+,\times}^{(n)} \cos (n \tilde{\ell}) + S_{+,\times}^{(n)} \sin (n \tilde{\ell}) \Big ], 
\end{align}
where the amplitude is given by 
\begin{equation}
    \mathcal{A} = - \dfrac{\mathcal{M}}{D_L} (2 \pi \mathcal{M} F)^{2/3},
\end{equation}
where $\mathcal{M} = \mu^{3/5} m^{2/5}$ is the chirp mass. The polarization coefficients at each harmonic are given by $C_{+, \times}^{(j)}$ and $S_{+, \times}^{(j)} $, which are explicitly given below
\begin{align}
\begin{split}
    C_{+}^{(n)} &= \Big [ 2 s_{\iota}^2 J_{n} (n e) + \dfrac{2}{e^2} (1+c_{\iota}^2) c_{2 \beta} \Big ( (e^2-2) J_{n} (ne) \\
    & + n e (1-e^2) (J_{n-1}(ne)-J_{n+1} (ne)) \Big )   \Big ], \\
S_{+}^{(n)} & = -\dfrac{2}{e^2} \sqrt{1-e^2} (1+c_{\iota}^2) s_{2 \beta} \Big [ -2(1-e^2) n J_{n} (ne) \\
&+ e (J_{n-1}(ne)-J_{n+1} (ne)) \Big ], \\
C_{\times}^{(n)} &= -\dfrac{2}{e^2} c_{\iota}s_{2 \beta} \Big [ 2(-e^2+2) J_{n} (ne) \\
&+ 2 ne (1-e^2) \Big ( J_{n-1} (ne) - J_{n+1} (ne) \Big ) \Big ] \\
S_{\times}^{(n)} &= - \dfrac{4}{e^2} \sqrt{1-e^2} c_{\iota} c_{2 \beta} \Big [ -2(1-e^2) n J_{n} (ne) \\
&+ e (J_{n-1}(ne)-J_{n+1} (ne)) \Big ].
\end{split}
\end{align}
These expressions rectify a typo in the paper by~\cite{blake_2018,moreno-garrido} and we can perform a sanity check by taking the small eccentricity limit and comparing it with the expressions given in~\cite{NicoBertiArunWill}. For the $n=1$ harmonic, we have
\begin{align}
\begin{split}
    C_{+}^{(1)} &= (1+c_{\iota}^2) c_{2 \beta} \Big ( -\dfrac{3}{2}e + \dfrac{2}{3} e^3 + \mathcal{O}(e^5) \Big ) \\
&+ s_{\iota}^2 \Big ( e - \dfrac{e^3}{8} + \mathcal{O}(e^5) \Big ), \\
S_{+}^{(1)} & = (1+c_{\iota}^2) s_{2 \beta} \Big (-\dfrac{3}{2}e + \dfrac{23}{24} e^3 + \mathcal{O}(e^5) \Big ), \\
C_{\times}^{(1)} &= c_{\iota}s_{2 \beta} \Big ( 3 e -\dfrac{4}{3}e^3 + \mathcal{O} (e^5) \Big ), \\
S_{\times}^{(1)} &= c_{\iota} c_{2 \beta} ( -3 e +\dfrac{23}{12}e^3 + \mathcal{O} (e^5) \Big ),
\end{split}
\end{align}
which are consistent with~\cite{NicoBertiArunWill}. 

\subsection{Waveform in Fourier-Domain}

\subsubsection{Stationary-phase Approximation}
The waveform in the Fourier domain can be computed using the SPA. The motivation behind this approximation is to simplify the integral in Fourier transforming the time domain waveform. The simplification is possible because the time domain waveform is composed of a rapidly oscillating phase and slowly varying amplitude. In what follows, we go through the key approximations and highlight some caveats when applying it when the waveform evolves over both the KL and RR timescales. \\

First, we review this approximation when the KL effect is absent. The fact that the amplitude is slowly varying can be summarized through the conditions $d \log \mathcal{A} /dt \ll d \ellmean /dt$ and $d^2 \ellmean /dt^2 \ll  (d \ellmean /dt)^2$. Therefore the contribution to the Fourier integral comes from a point $t=t_{sp}$ where the phase is stationary. Expanding the phase using a Taylor series, the integrand can be simplified considerably to yield a Gaussian, which can be trivially integrated. A detailed overview of the SPA can be found in~\cite{bender}, and in the context of GWs, see~\cite{NicoBertiArunWill,blake_2018,blake_2019}. The stationary-phase condition is then defined by $n \dot{\ellmean} (t_{sp}) = 2 \pi f (t_{sp})$, which can be rewritten as a mapping (at $t=t_{sp}$) between the GW frequency and the mean orbital frequency; the latter is given by $f (t_{sp})=n F(t_{sp})$, where $n$ is a positive integer and we identify $F \equiv \dot{\ellmean} /(2 \pi)$. 

The GWs emitted by a coalescing binary have a finite time duration, which then implies that they have power over a finite range of  frequencies. The initial orbital frequency is simply $F_0$, but the upper bound on the final orbital frequency comes from the limitation of using PN theory which typically breaks down at the `Last Stable Orbit'(LSO). For small eccentricities, the LSO can be approximated with the Inner-Most Stable Circular Orbit (ISCO) and we can use $F_{\text{ISCO}}$ for the final orbital frequency. The expression for the $n$th harmonic of the Fourier polarization $\tilde{h}_{+,\times}^{(n)}(f)$, is then obtained as
\begin{equation}
\begin{aligned}
\tilde{h}_{+,\times}^{(n)}(f) & \approx-\frac{\mathcal{M}}{2 D_L} \frac{\left(2 \pi \mathcal{M} F\left(t_{sp}^{*}\right)\right)^{2 / 3}}{\sqrt{n \dot{F}\left(t_{sp}^{*}\right)}}\\
& \times \left[C_{+, \times}^{(n)}\left(t_{sp}^{*}\right)+i S_{+, \times}^{(n)}\left(t_{sp}^{*}\right)\right] e^{-i ( \psi_{n} + \pi/4)} \\
& \times \Theta\left(f-n F_{0}\right) \Theta\left(n F_{\mathrm{ISCO}}-f\right), \label{eqn:hplus_cross_1}
\end{aligned}
\end{equation}
where $\Theta$ is the Heaviside-Theta function.~\Cref{eqn:hplus_cross_1} can be simplified by using the expression for $\dot{F}$ and we obtain,
\begin{align}
\begin{split}
\tilde{h}_{+,\times}^{(n)} (f) &= - \sqrt{\dfrac{5}{384}} \dfrac{\mathcal{M}^{5/6}}{D_l} (n F(t_{\text{sp}}))^{-7/6} \\
\times & \Big [ C_{+,\times}^{(n)} (t_{\text{sp}}) + i S_{+,\times}^{(n)} (t_{\text{sp}}) \Big ] \mathcal{G} (t_{\text{sp}}) e^{-i(\psi_n + \pi/4)} \\
& \Theta\left(f-n F_{0}\right) \Theta\left(n F_{\mathrm{ISCO}}-f\right),
\end{split} \label{eqn:hpluscross_SPA}
\end{align} 
where
\begin{equation}
\mathcal{G} = \dfrac{(1-e^2))^{7/4}}{\sqrt{1+\frac{73}{24}e^2 + \frac{37}{96}e^4}}. \label{eqn:G_func}
\end{equation}

The phase for a specific harmonic given by $\psi_n$ is determined through the following indefinite integral,
\begin{equation}
\psi_n [F(t_{\text{sp}})] = n \phi_n [F(t_{\text{sp}})] - 2\pi f t_n [F(t_{\text{sp}})], \label{eqn:PsiSPA}
\end{equation}
where we have 
\begin{align}
\begin{split}
    \phi_n [F(t_{\text{sp}})] &= \phi_c + 2\pi \bigintsss \limits^{F(t_{\text{sp}})}  \tau'  dF', \\
    t_n [F(t_{\text{sp}})] &= t_c + \bigintsss \limits^{F(t_{\text{sp}})}  \left(\tau' /F' \right) \ dF',
\end{split}\label{eqn:phi_t_SPA}
\end{align}
which are evaluated at the stationary point $F(t_{\text{sp}}) = f/n$. The quantities $\phi_c$ and $t_c$ are constants which can be determined as outlined in~\cite{blake_2018,blake_2019}. In the above equations, we have defined $\tau$ in the following way
\begin{align}
\tau & \equiv \dfrac{F}{\dot{F}}  = \dfrac{5 \mathcal{M}}{96} (2\pi \mathcal{M} F)^{-8/3} \mathcal{G}^2. \label{eqn:tau_func}
\end{align}

In the presence of the KL effect, the overall amplitude of the waveform will vary over both the KL timescale as well as the slower RR timescale. The coefficients $C_{+, \times}^{(j)}$ and $S_{+, \times}^{(j)} $ vary over both the timescales since they depend on $e,\iota$ and $\omega$. Specifically we have the hierarchy $d \log \mathcal{A} /dt \ll d \log C_{+,\times}^{(n)} /dt \sim d \log S_{+,\times}^{(n)} /dt \ll d \ellmean /dt$. Since the behavior over the KL timescale is the oscillation of the eccentricity and inclination angle, one can expect that this will introduce additional frequencies (or beats). Qualitatively, the stationary-phase condition should be modified by these additional frequencies, in a manner similar to what one finds in the case of an isolated spin precessing binary~\cite{Chatziioannou:2017tdw,Chatziioannou:2016ezg,Klein:2013qda}.

Specifically, one can expect that the defining condition of the SPA will be corrected to the expression $n_1 \dot{\ellmean} (t_{sp}) \pm n_2 \dot{\tilde{\theta}}= 2 \pi f (t_{sp})$, where $n_1$ and $n_2$ are integers and $\tilde{\theta} = \theta \pi/K$ (this would show up when the Jacobi Elliptic functions are expanded in a Fourier series) with $K$ being the complete elliptic integral of the first kind. Here, $n_1 \ellmean$ are the `orbital harmonics' that result due to the Fourier expansion of the polarizations. Meanwhile, $ n_2 \tilde{\theta}$ can be thought of as the `KL harmonics' that would result from expanding the coefficients $C_{+, \times}^{(j)}$ and $S_{+, \times}^{(j)} $ using the solutions for $e$, $\iota$ and $\omega$. However, such a decomposition remains elusive at this point and we leave that as a desirable result for future work. 

Instead we approximate the modification introduced by the additional `KL harmonics' within the framework of MSA. Since the term $n_2 \dot{\tilde{\theta}}$ would vary over the KL timescale, it is smaller than $n_1 \dot{\ellmean}$ by a factor of $\epkl$. Also, we do not expect large integer values for $n_1$ or $n_2$ to be significant since we adopt a small-eccentricity approximation. Therefore, to LO, we simply have the stationary-phase condition corresponding to that of an isolated eccentric binary given by $n \dot{\ellmean} (t_{sp}) = 2 \pi f (t_{sp})$, where we dropped the subscript on the integers. The corrections to the mapping of the GW frequency will be induced on the longer KL timescale, and can be expressed in the form $ f = n F + \mathcal{O} (\epkl)$. In our LO adiabatic analysis, we neglect these corrections to the GW frequencies and use equations derived earlier in this section.

\subsubsection{Postcircular Kozai-Lidov GW phase} \label{subsec:PCKL-phase}
To compute the GW phase at a given harmonic, given by~\cref{eqn:PsiSPA}, we first compute $\phi_n (F)$ and $t_n(F)$, which are given by~\cref{eqn:phi_t_SPA}. The improper integrals over the orbital frequency are difficult to do without making certain approximations. We first expand $\tau$ using a small-eccentricity approximation and then average over the KL cycles, similar to~\cref{eqn:e2_freq}, to extract the dependence on orbital frequency. The averaging over $\theta$ is justified because the main effect of the KL oscillations is to leave a cumulative effect on the GW phase. More details on this averaging procedure for the GW phase calculation can be found in~\cref{app:KL_averaging}. 

We start by expanding~\cref{eqn:tau_func} for small eccentricity, to obtain
\begin{align}
\tau &= \dfrac{5 \mathcal{M}}{96} (2\pi \mathcal{M} F)^{-8/3} \Big (1-\dfrac{157}{24}e^2 + \dfrac{13759}{576} e^4 \Big ). \label{eqn:tau_higher_order}
\end{align}
We use~\cref{eqn:e2_F_final} in~\cref{eqn:tau_higher_order}, average over $\theta$, and expand the resulting Jacobi Elliptic functions for $k_0^2 \ll 1$ to obtain
\begin{align}
\langle \tau \rangle_{\theta} (F) &= \dfrac{5 \mathcal{M}}{96} (2\pi \mathcal{M} F)^{-8/3} \Bigg \{ 1 - \tilde{F}^{-19/9} \Big [ \dfrac{157}{24} \overline{e}_0^2 \nonumber \\
&+ \dfrac{521711}{43776} (2 \overline{e}_0^4 + \delta e_0^4) + \dfrac{157}{192} k_0^2 \delta e_0^2  \Big ] \nonumber \\
&+ \tilde{F}^{-38/9} \dfrac{1044553}{43776} (2 \overline{e}_0^4 + \delta e_0^4) \Bigg \} \label{eqn:tau_PCKL},
\end{align}
where $\overline{e}_0$ is described by~\cref{eqn:ebar}, and $\delta e_0^2 = (e_{\text{max},0}^2 - e_{\text{min},0}^2)/2$, which physically captures the initial difference between the average value and the minimum value of the eccentricity oscillations. We can now integrate~\cref{eqn:tau_PCKL} to obtain the PCKL phase, denoted as $\psi_{n}^{PCKL}$,
\begin{align}
    \psi_{n}^{PCKL} (f) &= n \phi_c -2 \pi f t_c  - \dfrac{3}{128 u} \Big ( \dfrac{n}{2} \Big )^{8/3} \nonumber \\
    &\times \Bigg [ 1 -  \dfrac{785}{888896} \chi^{-19/9} (1824 \overline{e}_0^2 + 6646 \overline{e}_0^4 + \delta e_0^4 \nonumber \\
    & + 228 \delta e_0^2 k_0^2) + \dfrac{5222765}{1997888} \chi^{-38/9} (2\overline{e}_0^4+\delta e_0^4) \Bigg ], \label{eqn:PCKL_phase}
\end{align}
where $u = (\pi \mathcal{M} f)^{5/3}$, and $\chi = f / (n F_0)$. The KL effect is manifest through the parameters $\{ \overline{e}_0 , \delta e_0, k_0 \}$. Recall that $\overline{e}_0$ represents the initial average value of the eccentricity oscillations, $ \delta e_0 $ represents the initial difference between the average value and the minimum value of the eccentricity oscillations, $k_0$ controls the `shape' and time period of the eccentricity oscillations, and all 3 are combinations of $\{ e_0, \iota_0, \omega_0 \}$. As a sanity check, we can `turn off' the KL effect by taking the limit $\epkl \rightarrow 0$, which means that there are no KL oscillations and therefore $e_{\text{max,0}} \rightarrow e_0$ and $e_{\text{min},0} \rightarrow e_0$. Consequently we get  $ \overline{e}_0 \rightarrow e_0, \delta e_0 \rightarrow 0$, and $k_0 \rightarrow 0$ as the appropriate limits that correspond to the KL effect being `turned off'. In the limit $\epkl \rightarrow 0$, we obtain for the phase
\begin{align}
            \psi_{n}^{PC} (f) &= n \phi_c -2 \pi f t_c  - \dfrac{3}{128 u} \Big ( \dfrac{n}{2} \Big )^{8/3} \nonumber \\
            &\times \Bigg [ 1 -  \dfrac{785}{888896} \chi^{-19/9} (1824 \overline{e}_0^2 + 6646 \overline{e}_0^4) \nonumber \\
            &+ \dfrac{5222765}{998944} \chi^{-38/9} \overline{e}_0^4  \Bigg ], \label{eqn:PC_phase}
\end{align}
where we identify $\overline{e}_0 \rightarrow e_0$ in this limit, thereby recovering the conventional ``postcircular" result to the prescribed order. Therefore, we can write the PCKL phase as a sum of two sets of terms. One set is identical, in the limit $\epkl \rightarrow 0$, to using the ``postcircular" result, which holds true for an isolated eccentric binary and is given by $\psi_{n}^{PC}$ in~\cref{eqn:PC_phase}. The other set is due to the cumulative effect of the KL oscillations that are induced by the third body, and these terms, denoted by $\psi_{n}^{KL}$, can be obtained by taking the difference between~\cref{eqn:PCKL_phase} and~\cref{eqn:PC_phase}. We can summarize this using the following equation,
\begin{align}
    \psi_{n}^{PCKL} (f) = \psi_{n}^{PC} (f) + \psi_{n}^{KL} (f), \label{eqn:PCKL_decompose}
\end{align}
which allows for a clear understanding of how the KL effect manifests in the GW phase. While~\cref{eqn:PCKL_decompose} was obtained from a calculation done through $\mathcal{O} (e^4)$, we expect it to hold true at higher orders in eccentricity as well.

\subsubsection{Postcircular Kozai-Lidov GW amplitude}

The evaluation of the amplitude of the polarizations, \EDIT{using the stationary phase condition $f=n F(t_{sp})$}, is relatively straightforward as we can simply calculate $| \tilde{h}_{+,\times}^{(n)} (f) | $ from~\cref{eqn:hpluscross_SPA}, to get
\EDIT{\begin{align}
\begin{split}
| \tilde{h}_{+,\times}^{(n)} (f) | &= \mathscr{A} f^{-7/6} \ \Big | C_{+,\times}^{(n)} (f) + i S_{+,\times}^{(n)} (f) \Big | \mathcal{G} (f) \\
& \times \Theta\left(f-n F_{0}\right) \Theta\left(n F_{\mathrm{ISCO}}-f\right),
\end{split} \label{eqn:PCKL_hpluscross_amplitude}
\end{align}}
\EDIT{where $\mathscr{A} = \sqrt{5/384} \mathcal{M}^{5/6} / D_L$, and $\Theta$ is the Heaviside step function}. Note that the Fourier GW amplitude depends on $e $, $\iota $ and $\omega$ through the functions $C_{+, \times}^{(n)}$ and $S_{+, \times}^{(n)}$. Using the stationary-phase condition, and the MSA solution in~\cref{eqn:e2_F_final} to $e(F) , \iota (F)$ and  $\omega (F)$, we can evaluate~\cref{eqn:PCKL_hpluscross_amplitude} \EDIT{and express it as a function of the GW frequency $f$. Since the KL oscillations enter through $e(f)$ and $\iota (f)$, the Fourier GW amplitude (at a given harmonic) contains a \emph{direct} imprint of the oscillations. Details on the evaluation of $| \tilde{h}_{+,\times}^{(n)} (f) |$ using the MSA solution and SPA can be found in the Supplemental Material.~\cite{supp}}
\section{Validation of analytical results using numerical solutions} \label{sec:validation}
In this section, we present the numerical validation of our analytic results by using a candidate system as a case-study. Identifying a potential candidate system requires determining the region of the parameter space in which the KL effect can be probed by LISA. Since we focus on systems with an IMBH inner binary and a SMBH third body, we will pick masses $m_1=m_2= \{ 10^4 M_{\odot} , 10^5 M_{\odot} \}, m_3= \{ 10^6 M_{\odot}, 4\times 10^6 M_{\odot}, 10^7 M_{\odot} \}$ and we will eventually focus on $m_1 = m_2 = 10^4 M_{\odot}, m_3 = 10^6 M_{\odot}$ as a representative case. In Sec.~\ref{subsec:constraint_params}, we discuss the constraints that play a role in determining the feasible region of the parameter space, and present the region of parameter space inside which these constraints are satisfied. The details of the numerical implementation are presented in Sec.~\ref{subsec:IC}, which also describes the initial conditions we use. In Sec.~\ref{subsec:validation_MSA} we validate the solution found using MSA, given by~\cref{eqn:e2_F_final}, with our numerical solution. We then validate our expressions for the GW phase in~\cref{eqn:PCKL_phase} and amplitude in~\cref{eqn:PCKL_hpluscross_amplitude} in Sec.~\ref{subsec:validation_waveform}.

\subsection{Constraints on the Parameter Space} \label{subsec:constraint_params}
There are several constraints that play a role in determining a feasible region in parameter space for probing the KL effect (see~\cite{Naoz:2012bx} for eg.), but we focus on three of them -- (1) \textit{Stability}, (2) \textit{Quenching}, and (3) \textit{Chirping}. In the following, we discuss each of these constraints in more detail.

\subsubsection{Stability}
The stability criterion for the triple is given in~\cite{Blaes:2002cs} and was originally derived in~\cite{Mardling}. Although the derivation is based on mutually co-planar orbits between the inner and outer binary, the fact that mutually inclined orbits are more stable means that this criterion is a conservative one. We rewrite it for the case where the outer binary's orbit is circular and obtain,
\begin{equation}
    \dfrac{R}{a} \gtrsim 2.8 \Big ( \dfrac{m_3}{m} \Big )^{2/5}, \label{eqn:stable}
\end{equation}
which can be recast into
\begin{align}
     R \gtrsim (0.4 \text{AU}) \Big ( \dfrac{a}{ 10^7 \text{km}} \Big ) \Big ( \dfrac{m_3}{10^6 M_{\odot}} \Big )^{2/5} \Big ( \dfrac{m}{10^4 M_{\odot}} \Big )^{-2/5}.
\end{align}

\subsubsection{Quenching}

If precession due to PN effects starts to dominate over the KL effect, the KL oscillations can become quenched, and at best they are suppressed. The quenching effectively happens because the precession due to RR acts to oppose precession due to KL oscillations~\cite{Blaes:2002cs}. The criterion for the quenching of KL oscillations becomes equivalent to the criterion under which precession due to PN effects is faster than that due to KL effects. Thus, for quenching to not happen, we effectively need $P_{\text{KL}} < P_{\text{gr,prec}} $, which gives
\begin{equation}
    \dfrac{R^3}{a^3} \lesssim  \dfrac{m_3}{m} \dfrac{a}{m} (1-e^2)^{3/2}.
\end{equation}
Since $1-e^2<1$ always, we can push the upper bound to \begin{equation}
    \dfrac{R^3}{a^3} <  \dfrac{m_3}{m} \dfrac{a}{m}, \label{eqn:no_quench}
\end{equation}
which can be rewritten as 
\begin{align}
R & \lesssim (2.7\text{AU}) \Big ( \dfrac{a}{10^7 \text{km}} \Big )^{4/3} \Big ( \dfrac{G m /c^2 }{1.5 \times 10^4\text{km}} \Big )^{-1/3} \nonumber \\
& \times \Big ( \dfrac{m_3}{10^6 M_{\odot}} \Big )^{1/3} \Big ( \dfrac{m}{10^4 M_{\odot}} \Big )^{-1/3}.
\end{align}
\subsubsection{Chirping}
\EDIT{Due to RR, the semimajor axis shrinks and the gravitational-wave frequency chirps. For a circular binary, to leading PN-order, 
\begin{align}
    a (t) = \Big [ a_0^4 - \dfrac{256}{5} \eta \ m^3 t \Big ]^{1/4}, \label{eqn:a_circ}
\end{align}
and from this, $f(t)$ (where $f=2F$) can be computed, through Kepler's third law. Since we are interested in observing the GWs produced by the inner binary, we require that it chirps sufficiently to produce a signal-to-noise-ratio (SNR) that is above a certain detection threshold, at a reasonable astrophysical luminosity distance. We determine $a_0$ such that, over a 4yr observation period of LISA, we obtain an $\text{SNR} \gtrsim 10$ (at $D_L = 1$Gpc). The sky-averaged SNR is estimated using
\begin{align}
\text{SNR}^2 \sim \dfrac{64}{5} \eta^2 \dfrac{m^4}{a_0^2 D_L^2} \dfrac{P_{\text{obs}}}{S_{n,0}} 
\label{eq:SNR-const}
\end{align}
where $S_{n,0}$ is a characteristic value of the LISA sensitivity curve during the observation; we approximate $S_{n,0}=S_n (f_0)$~\cite{Cornish:2018dyw}, where $f_0 = 2F_0$. By requiring $\text{SNR} \gtrsim 10$, we can solve for $a_0$, thereby obtaining the constraint on the maximal initial semimajor axis. For slowly radiating systems with fixed $\{ m,\eta, D_L, S_{n,0}, P_{\text{obs}} \}$, we can see that $\rho \propto a_0^{-1}$, reaffirming that wider binaries will chirp less and produce less SNR. Explicitly, we find
\begin{align}
\begin{split}
    a_0 &\lesssim 0.0957\text{AU} \Big ( \dfrac{\eta}{0.25} \Big ) \Big ( \dfrac{m}{2 \times 10^4 M_{\odot}} \Big )^2 \Big ( \dfrac{D_L}{1 \text{Gpc}} \Big )^{-1} \\ 
    & \times \Big ( \dfrac{\text{SNR}}{10} \Big )^{-1} \Big ( \dfrac{P_{\text{obs}}}{4 \text{yr}} \Big )^{1/2} \Big ( \dfrac{S_{n,0}}{4 \times 10^{-36} s} \Big )^{-1/2}, 
\end{split}\label{eqn:inner_chirp}
\end{align}
which is our \emph{chirping inner binary} constraint.

Since we approximate the outer binary to be stationary, we also neglect the effect of RR on the outer orbit. Therefore, we will require that the radius of the outer orbit does not change appreciably over the observable inspiral of the inner binary. To quantify this, we choose an initial $R_0$ for the outer binary such that its fractional change, $\delta R /R_0$, is not more than $1 \%$ over the initial inspiral of the inner binary, where the KL effect is most dominant (e.g.~for 1yr of evolution). Using
\begin{align}
    R (t) = \Big [ R_0^4 - \dfrac{256}{5} \eta_{\text{out}} \ (m+m_3)^3 t \Big ]^{1/4}, \label{eqn:R_circ}
\end{align}
with $\eta_{\text{out}} \equiv m \ m_3 / (m+m_3)^2$, we obtain 
\begin{align}
    R_0 \gtrsim (0.94 \text{AU})\Big [ \Big ( \dfrac{\eta_{\text{out}}}{10^{-2}} \Big ) \Big ( \dfrac{m+m_3}{10^6 M_{\odot}} \Big )^3 \Big ( \dfrac{P_{\text{obs}}}{1 yr} \Big ) \Big ]^{1/4}, \label{eqn:outer_chirp}
\end{align}
which is our \emph{chirping outer binary} constraint. Together,~\cref{eqn:inner_chirp,eqn:outer_chirp} constitute the \emph{chirping} constraints. }

\subsubsection{Parameter space for candidate systems} \label{subsec:candidate}

The constraints in~\cref{eqn:stable,eqn:no_quench,eqn:inner_chirp,eqn:outer_chirp} lead to a two-dimensional parameter space spanned by $R/m_3$ and $a/m$. \EDIT{We consider $m_1 = m_2 \in \{ 10^3 M_{\odot}, 10^4 M_{\odot} \}$ and $m_3 \in \{ 10^6 M_{\odot}, 10^7 M_{\odot} \}$ to be the characteristic masses of the triple.} \EDIT{For each combination of masses,~\cref{fig:regions} shows the region in parameter space resulting from the constraints. \EDIT{In addition to~\cref{eqn:stable,eqn:no_quench,eqn:inner_chirp,eqn:outer_chirp}, we have also included in this figure the constraint resulting from the double averaging approximation (DAA) breaking down, which occurs when $R \lesssim 10 \ a$~\cite{Luo_2016}}. In~\cref{fig:regions_a,fig:regions_c}, there is a large region of parameter space inside which all of our constraints are satisfied. This also suggests there could be many IMBH binaries, with masses in the range $\sim ( 10^3 M_{\odot}$ -- $ 10^4 M_{\odot})$, that live in the feasible region of~\cref{fig:regions_a,fig:regions_b}, and would thus be interesting sources for LISA. However, the feasible region shrinks for $\eta_{\text{out}}$ (larger $m_3$ for a fixed $m$ or smaller $m$ for a fixed $m_3$), so systems with $m_3 \gtrsim 10^7$ are not of interest to us. Since~\cref{fig:regions_c} has the largest feasible region, we will focus on this system for the rest of our work. }

\subsection{Numerical implementation and initial conditions} \label{subsec:IC}

With these preliminaries set up, let us now discuss how to carry out numerical integrations of the orbit-averaged equations given by~\cref{eqn:ODE_three_timescales,eqn:mean_anomaly_three_timescales}. When working numerically, it is more convenient to use $t$ as the dependent variable and so we transform~\cref{eqn:ODE_three_timescales} using~\cref{eqn:mean_anomaly_three_timescales}. We use Mathematica's \texttt{NDSolve} for solving the ODE system with the flags \texttt{PrecisionGoal} $\rightarrow 10$, \texttt{AccuracyGoal} $\rightarrow 10$, and \texttt{MaxSteps} $\rightarrow$ \texttt{Infinity}. The numerical integration is carried out until either $a = a_{\text{ISCO}}$, or $F>0.05$Hz (half the upper bound of LISA's gravitational-wave frequency window). \\

Let us now discuss the initial conditions with which to begin our numerical integrations, all of which we make sure respect the approximations we use throughout this work. Recall that we employ a small-eccentricity approximation, particularly for the computation of the GW phase, and in obtaining the frequency dependence of the orbital elements. For $\iota_0 \gtrsim 40^{\circ}$, the KL effect can induce large-amplitude eccentricity oscillations that lead to $e_{\text{max},0} > 0.3$. Given this, we restrict $\iota_0 \lesssim 40^{\circ}$ and $e_0 \lesssim 0.1$. Note that once $e_0$ and $\iota_0$ have been restricted, $\omega_0$ can be chosen arbitrarily from the range $[0,2\pi]$. Different values of $\omega_0$ simply alter $e_{\text{max},0}$ and $e_{\text{min},0}$ relative to $e_0$. In Appendix~\ref{app:roots_KL_problem}, we provide more details on the behavior of $e_{\text{max},0}$ and $e_{\text{min},0}$ as a function of the initial conditions, as well as other well-known phase-space behavior of the KL problem. \\

\begin{figure}[H]
	\centering
\begin{subfigure}[b]{0.49\linewidth}
\centering
	\includegraphics[width=\linewidth]{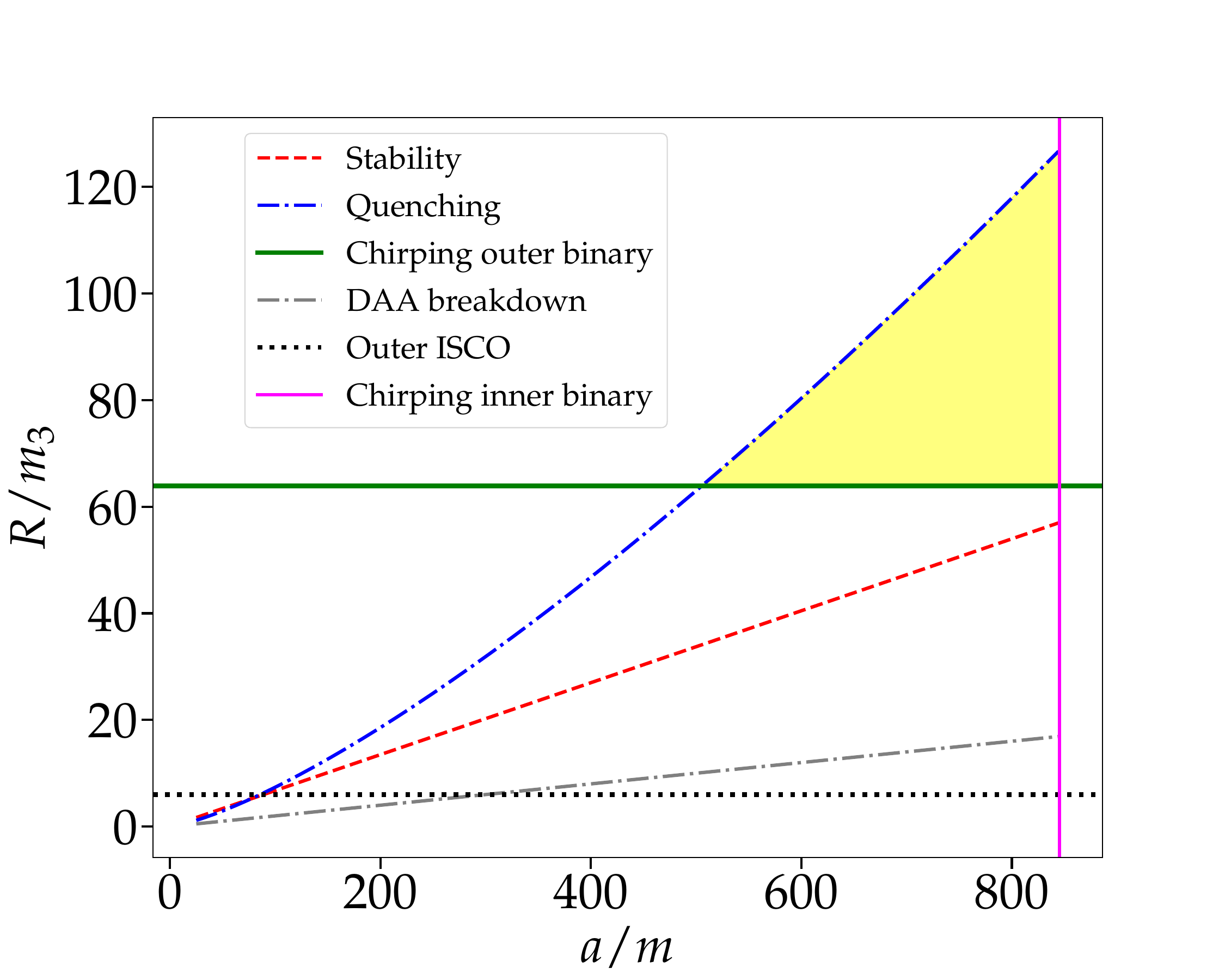}
	\caption{$m_1=m_2=10^3 M_{\odot}, \ m_3 = 10^6 M_{\odot}$}
	\label{fig:regions_a}
\end{subfigure}
\hspace{0.4em}%
\begin{subfigure}[b]{0.49\linewidth}
\centering
	\includegraphics[width=\linewidth]{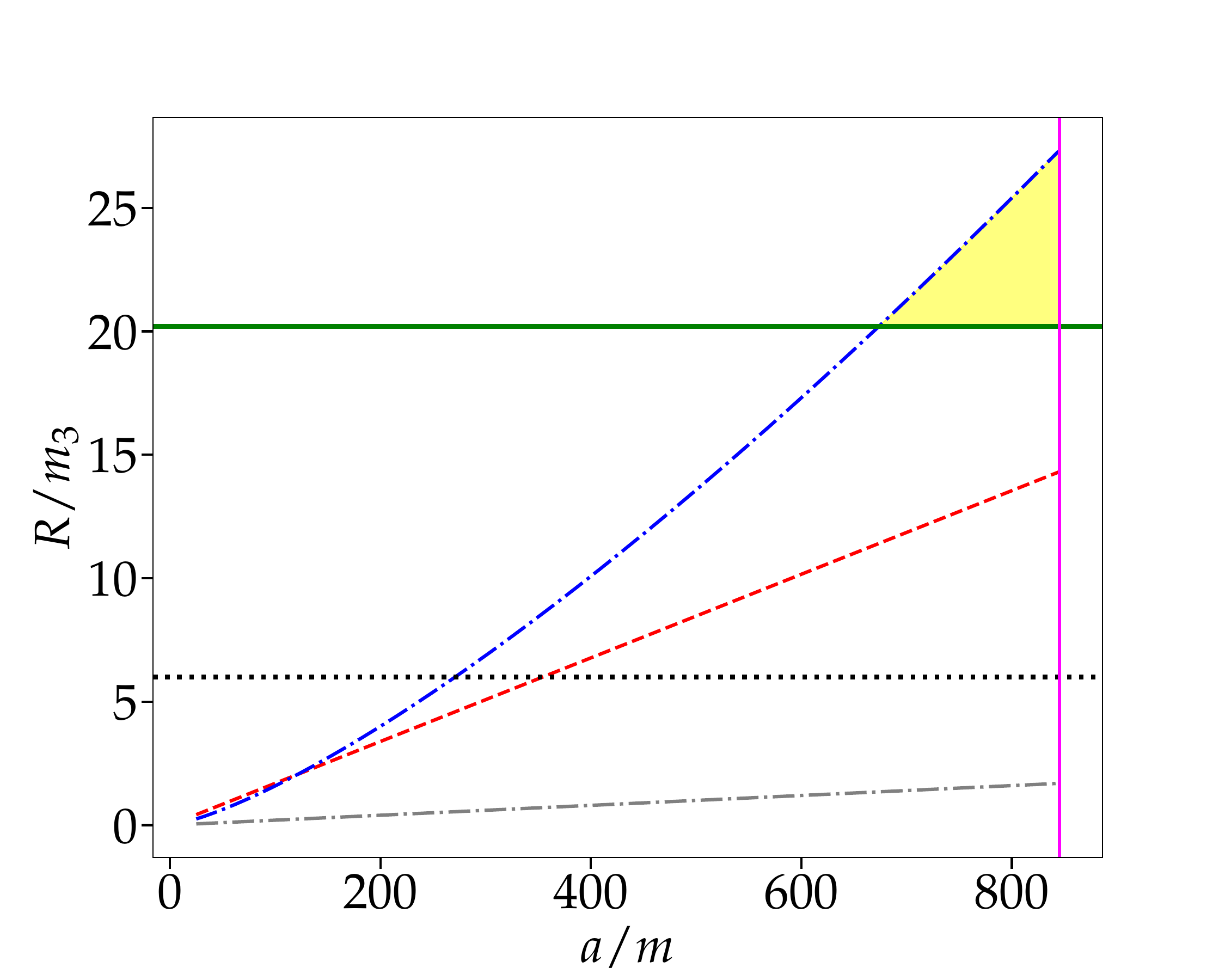}
	\caption{$m_1=m_2=10^3 M_{\odot}, \ m_3 = 10^7 M_{\odot}$}
	\label{fig:regions_b}
\end{subfigure}
\hspace{0.4em}%
\begin{subfigure}[b]{0.49\linewidth}
\centering
	\includegraphics[width=\linewidth]{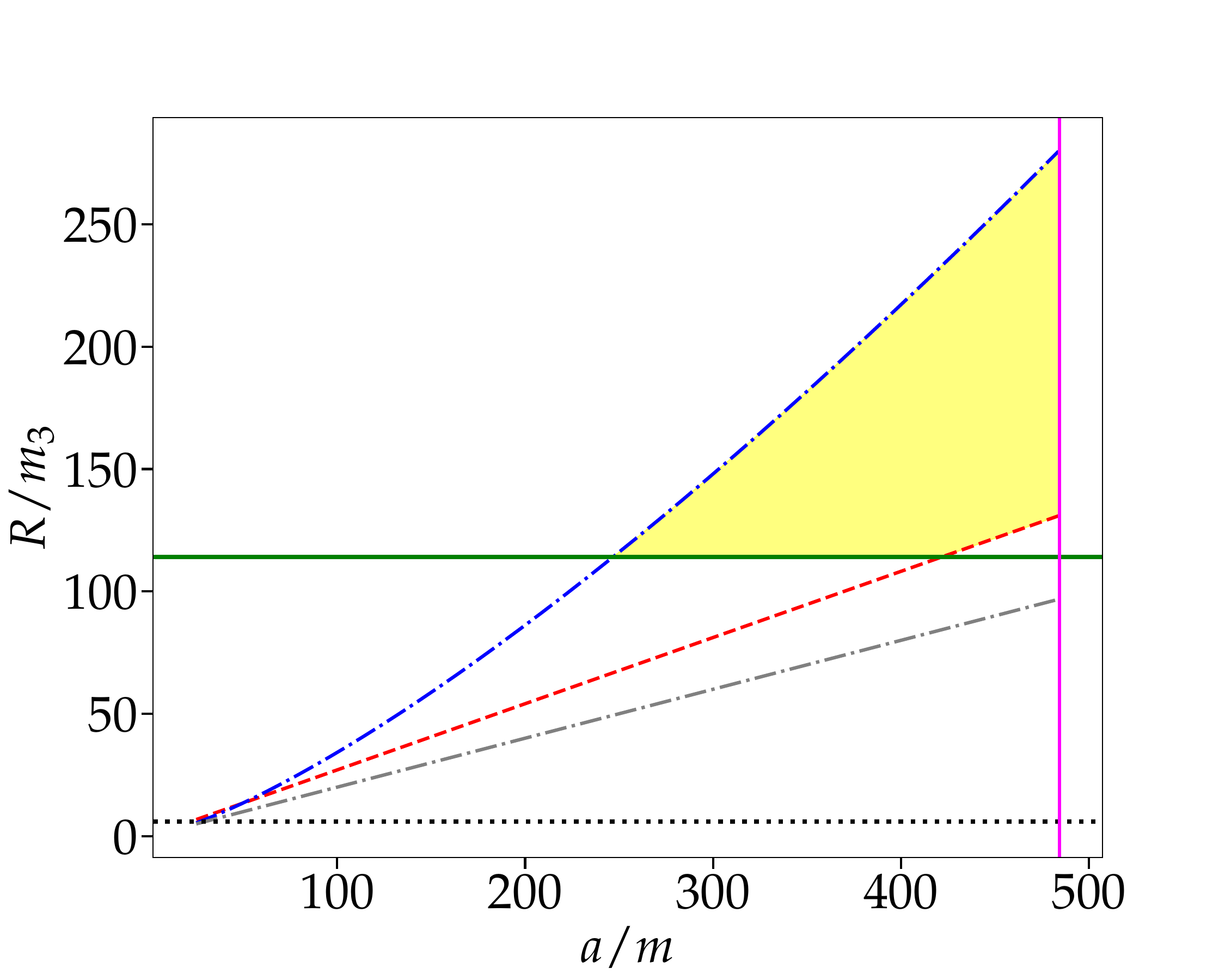}
	\caption{$m_1=m_2=10^4 M_{\odot}, \ m_3 = 10^6 M_{\odot}$}
	\label{fig:regions_c}
\end{subfigure}
\hspace{0.4em}%
\begin{subfigure}[b]{0.49\linewidth}
\centering
	\includegraphics[width=\linewidth]{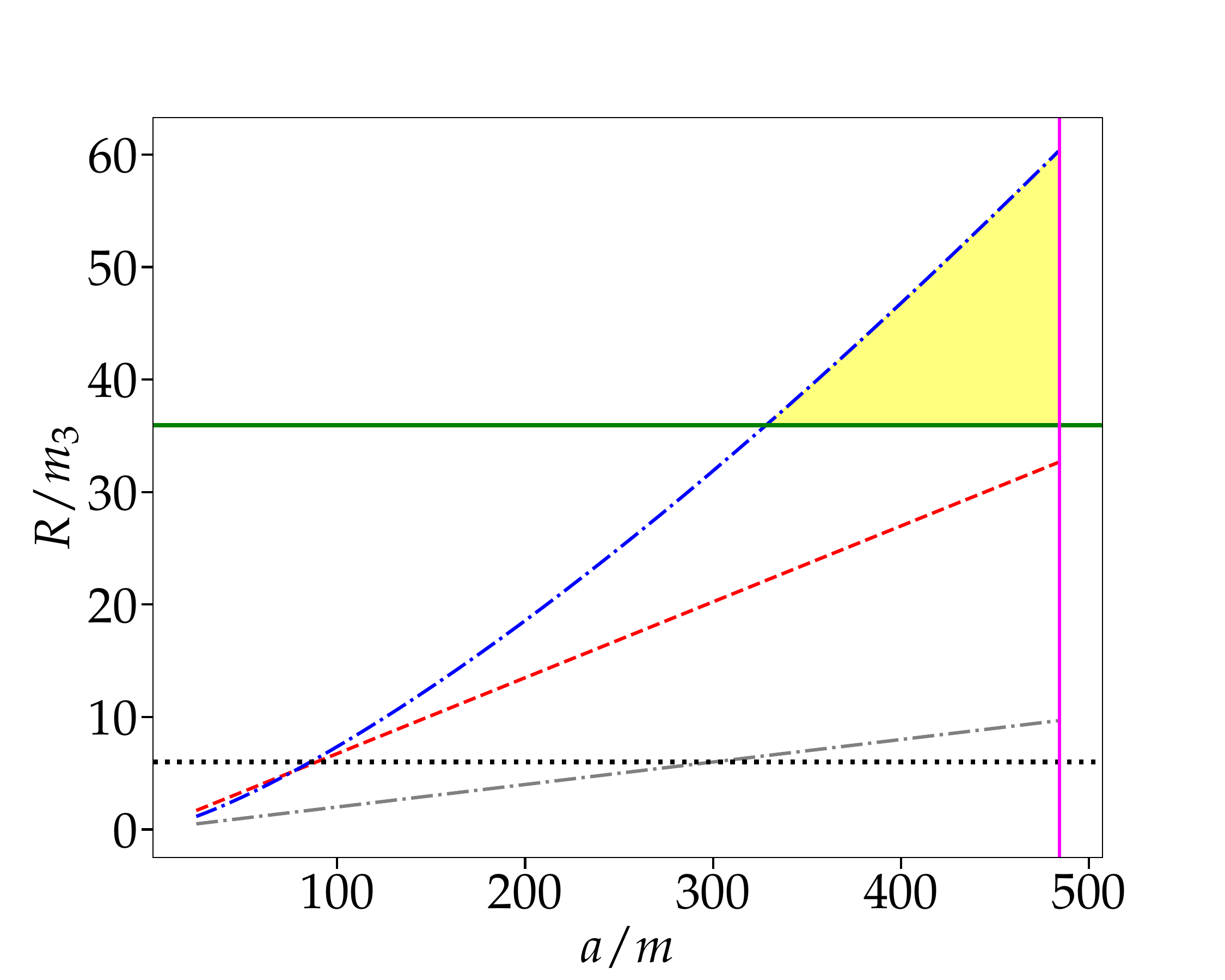}
	\caption{$m_1=m_2=10^4 M_{\odot}, \ m_3 = 10^7 M_{\odot}$}
	\label{fig:regions_d}
\end{subfigure}
	\caption{\EDIT{Two-dimensional parameter space of $a/m$ and $R/m_3$. The lines represent different constraints in the triple system. In the region above the blue (dash-dotted) line, the KL oscillations are quenched. Below the red (dashed) line, hierarchical triples are unstable. Below the green (solid) line, the outer binary chirps sufficiently to break the stationary outer orbit approximation. To the right of the magenta (solid) line, the inner binary does not chirp sufficiently to produce a loud enough SNR. Below the gray (dash-dot) line, the double-averaged approximation breaks down. The yellow shaded region thus indicates the feasible region of parameter space that satisfies all of our constraints. Notice that as the ratio of $m/m_3$ increases, the feasible (yellow) region shrinks, and that the system with $m_1=m_2=10^4 M_{\odot}, \ m_3 = 10^6 M_{\odot}$ has the largest feasible region.}}
	\label{fig:regions}
\end{figure}

For our candidate system, with masses $m_1=m_2=10^4 M_{\odot}$ and $m_3=10^6 M_{\odot}$, keeping in mind the small-eccentricity approximation, we chose the initial values $e_0 = 0.1 ,\iota_0 = \pi/6$, and $\omega_0 =0$. Based on~\cref{fig:regions_c}, we choose \EDIT{$a_0 = 400 m$} and \EDIT{$R = 150 m_3$}. Given these initial conditions, the inner binary inspirals for about \EDIT{$3.89$yr} before the 1PN precession dominates and the KL effect is quenched. Consequently, at least for the first year of the inspiral starting at $a_0$, there are many KL cycles that leave a cumulative effect on the GW phase. We therefore use the first year of inspiral for validating our analytic results. 
\onecolumngrid

\twocolumngrid

\subsection{Validation of the orbital dynamics description through multiple-scale analysis}

\label{subsec:validation_MSA}

\subsubsection{Three-timescale analysis}\label{subsubsec:validation_MSA_threetimescale}
We have made several approximations and simplifications in Sec.~\ref{subsec:3timescale} when computing the analytic MSA solution for the orbital elements. To validate our analytic calculations, we compare the evolution of $e(F)$, given by~\cref{eqn:e2_F_final} with its numerical counterpart, obtained from \texttt{NDSolve}. We focus on $e(F)$ because $ \iota (F)$ and $\omega (F)$ are related to it through~\cref{eqn:KL_constants}. 

\onecolumngrid

\setlength\belowcaptionskip{-1.5ex}
\begin{figure}[H]
    \centering
    \begin{subfigure}[t]{0.49\linewidth}
    \centering
    \includegraphics[width=\linewidth]{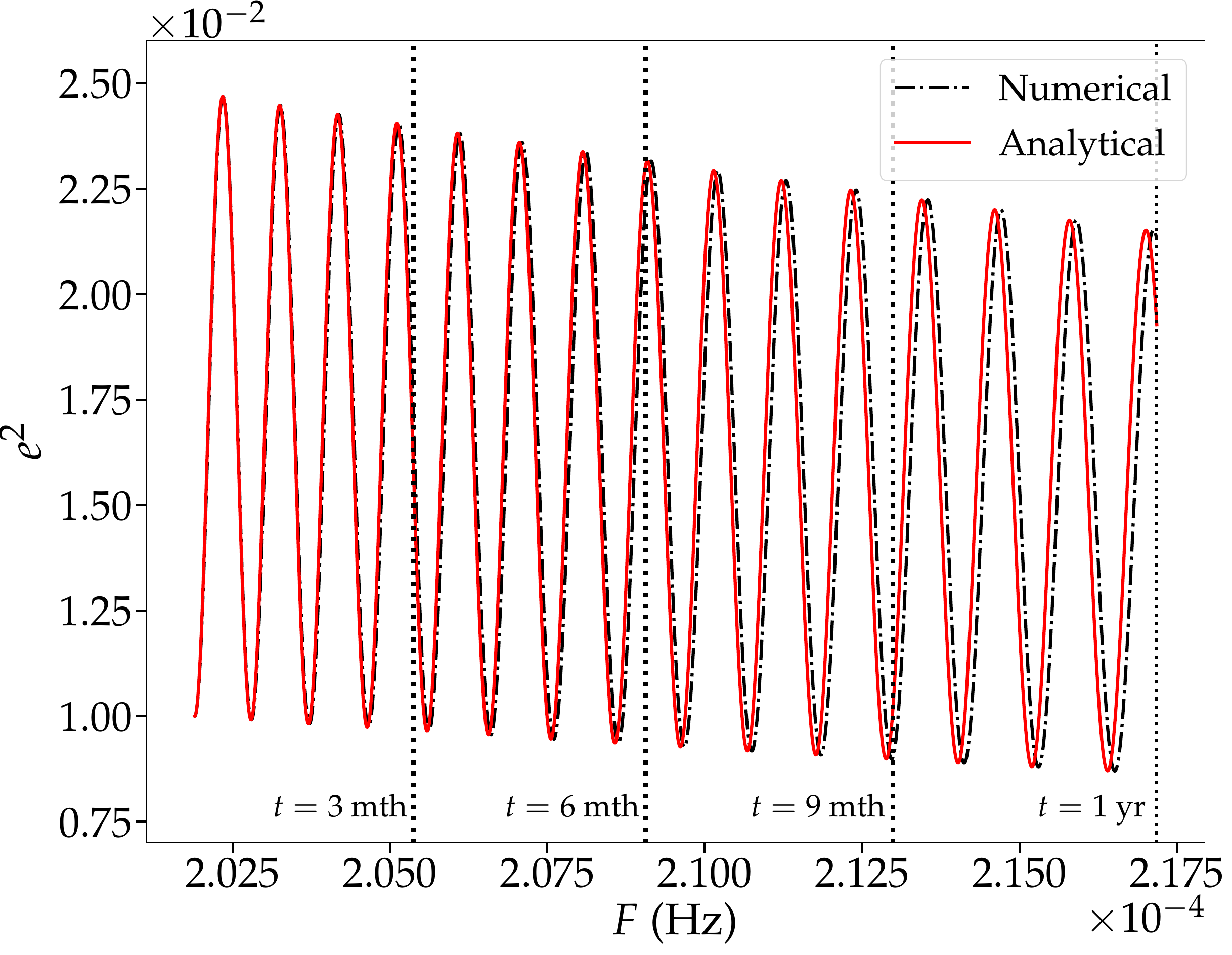}
    \caption{}
    \label{fig:eMSA_Num_compare_a}
    \end{subfigure}
    \hfill%
        \begin{subfigure}[t]{0.49\linewidth}
    \centering
    \includegraphics[width=\linewidth]{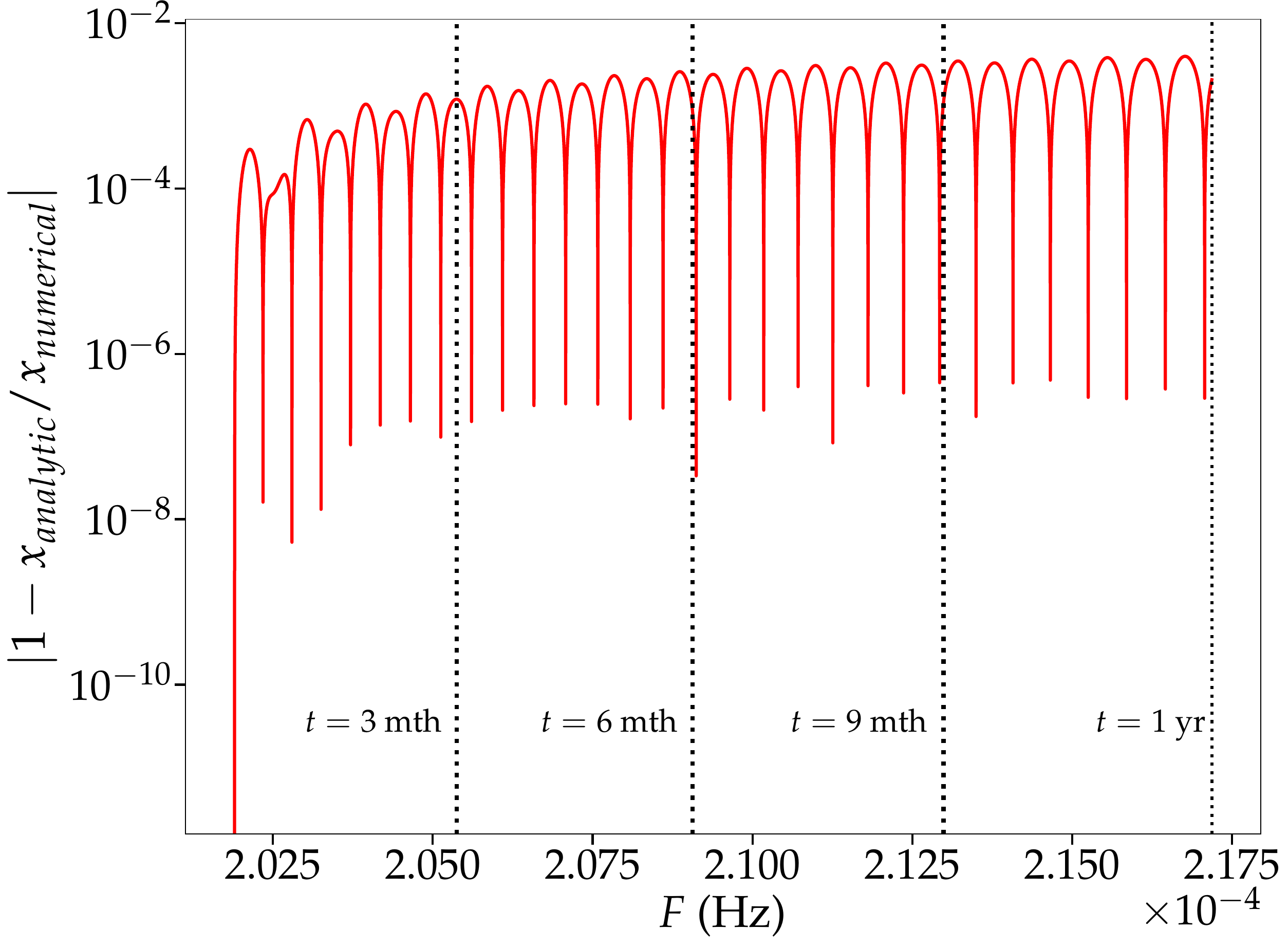}
    \caption{}
    \label{fig:eMSA_Num_compare_b}
    \end{subfigure}
\caption{\EDIT{Comparison of the analytic MSA representation (black solid) and the numerical (red dashed) representation of $e^2$, as a function of mean orbital frequency $F$. In~\subref{fig:eMSA_Num_compare_a}, we show the analytic and numerical representations plotted on top of each other, while in~\subref{fig:eMSA_Num_compare_b}, we show the fractional error between the analytic and numerical representations of $x$. Also for reference, we show the three month time intervals over the total one year evolution. Observe the excellent agreement between the analytic and the numerical results, as well as the characteristic KL oscillations ($\sim 15$ KL-cycles).} }
    \label{fig:eMSA_Num_compare}
\end{figure}

\twocolumngrid

\Cref{fig:eMSA_Num_compare} shows that our analytic MSA solution agrees really well with the numerical solution. In~\cref{eqn:e2_F_final}, the calculation was done to $\mathcal{O} (e^4)$, so the error between the numerical and analytic solutions is of $\mathcal{O} (e^6)$. We can improve on our analytic approximation by simply extending~\cref{eqn:e2_F_final} to higher order in eccentricity. Observe also the distinct behavior of the eccentricity over the KL and RR timescales, which is what we expected from the LO MSA treatment. Over the the KL timescale, the characteristic oscillations are observed clearly while over the RR timescale, the characteristic circularization effect (as expected from the results of~\cite{Peters}) is seen. There are $\sim 40$ KL-cycles during this time interval, where $P_{\text{KL}}$ starts at \EDIT{$\sim 7.55$} days and stretches to \EDIT{$\sim 8.13$} days owing to the RR effect that we discussed in Sec.~\ref{subsec:3timescale}. Further, this means that $P_{\text{KL}} / P_{\text{obs}} \sim \EDIT{10^{-2}}$ which is why we can average over the KL-cycles during the observation window. Therefore, our analytic MSA result is extremely good when compared to numerical integrations, and all of the approximations we made in obtaining~\cref{eqn:e2_F_final} are validated. We have here presented results only for the representative system described earlier, but we have validated our analytic solutions for a wide range of systems inside the region of parameter space where the constraints of Sec.~\ref{subsec:constraint_params} are satisfied.

\subsection{Validation of the Analytic, Fourier-domain Waveform Model} \label{subsec:validation_waveform}

We now validate our analytic, Fourier-domain PCKL waveform model, described by~\cref{eqn:PCKL_hpluscross_amplitude,eqn:PCKL_phase} using numerics. Our goal is to determine if our model is accurate and if the errors introduced by our approximations in the model are well controlled. The key approximations in the model are the small-eccentricity approximation, the LO/adiabatic approximation, and the averaging over KL-cycles. Since we assume $e \ll 1$ throughout this work, we restrict the validation of the waveform to the dominant $n=2$ harmonic. Consequently, for the GW phase, we consider $\psi_2$. For the GW amplitude, we only focus on the ``$+$"-mode polarization $\tilde{h}_{+}^{(2)}$, since the analysis is similar for the ``$\times$"-mode. While it is typical to use a Discrete Fourier Transform (DFT), we instead use a numerical SPA for validating our results. A DFT would introduce windowing and binning errors which are extraneous to the errors that arise from our approximations in the model, which we want to focus on. A numerical SPA is obtained by evaluating~\cref{eqn:hpluscross_SPA,eqn:PsiSPA} using the numerical solutions to the orbital elements. In Secs.~\ref{subsubsec:Validation_amplitude} and~\ref{subsubsec:Validation_phase}, we explain in more detail how the numerical GW amplitude and phase are obtained, along with how we evaluate the analytic results given by~\cref{eqn:PCKL_hpluscross_amplitude,eqn:PCKL_phase}. We then discuss the comparison between the analytic and numerical calculations.


\subsubsection{Validation of the Fourier Amplitude of the Waveform}
\label{subsubsec:Validation_amplitude}

We now discuss the validation of the Fourier GW amplitude of our PCKL waveform model. As explained earlier, we will be focusing on $\tilde{h}^{(2)}_{+} $. We first discuss how the analytic $\tilde{h}^{(2)}_{+} $ is evaluated, and then explain how the numerical counterpart is obtained. 

The analytic $\tilde{h}^{(2)}_{+} $, given by~\cref{eqn:PCKL_hpluscross_amplitude} for $n=2$, is evaluated using the MSA solution to the orbital elements, which are already functions of $F$. Specifically, we insert~\cref{eqn:e2_F_final,eqn:KL_constants} in~\cref{eqn:PCKL_hpluscross_amplitude} and we evaluate it over a binned GW frequency $f$ (corresponding to $n=2$) domain of $[2 F_0, 2 F_{1 yr}]$, where $F_{1 yr}$ is the orbital frequency at $t=1 yr$. The values of $F$ are obtained by tabulating the corresponding numerical solution from \texttt{NDSolve}, and we chose $\Delta t = 100$ for the bin size. We compute the absolute value, $| \tilde{h}^{(2)}_{+} |$ and scale it by its initial value $| \tilde{h}^{(2)}_{+,0} |$. 

The numerical $\tilde{h}^{(2)}_{+} $ is obtained by once again evaluating~\cref{eqn:PCKL_hpluscross_amplitude} for $n=2$, but this time using the numerical solutions to the orbital elements. We obtained the orbital evolution using \texttt{NDSolve}, and each orbital element is given as an \texttt{InterpolatingFunction} of $t$. We make use of the stationary-phase condition $f=2F$ to parametrically evaluate $\tilde{h}^{(2)}_{+} $. We do so by tabulating the orbital elements $e, \omega$, and $\iota$ over the time domain of $1 yr$ with the bin size $\Delta t = 100$. We then use the previously (for the analytic calculation) binned values of $F$ to parametrically evaluate $\tilde{h}^{(2)}_{+} $ as a function of $f$. Once again, we compute the scaled quantity $| \tilde{h}^{(2)}_{+} / \tilde{h}^{(2)}_{+,0} | $. \setlength{\parskip}{0mm}

\onecolumngrid

\setlength\belowcaptionskip{-2ex}
\begin{figure}[H]
	\centering
\begin{subfigure}[t]{0.49\linewidth}
\centering
	\includegraphics[width=\linewidth]{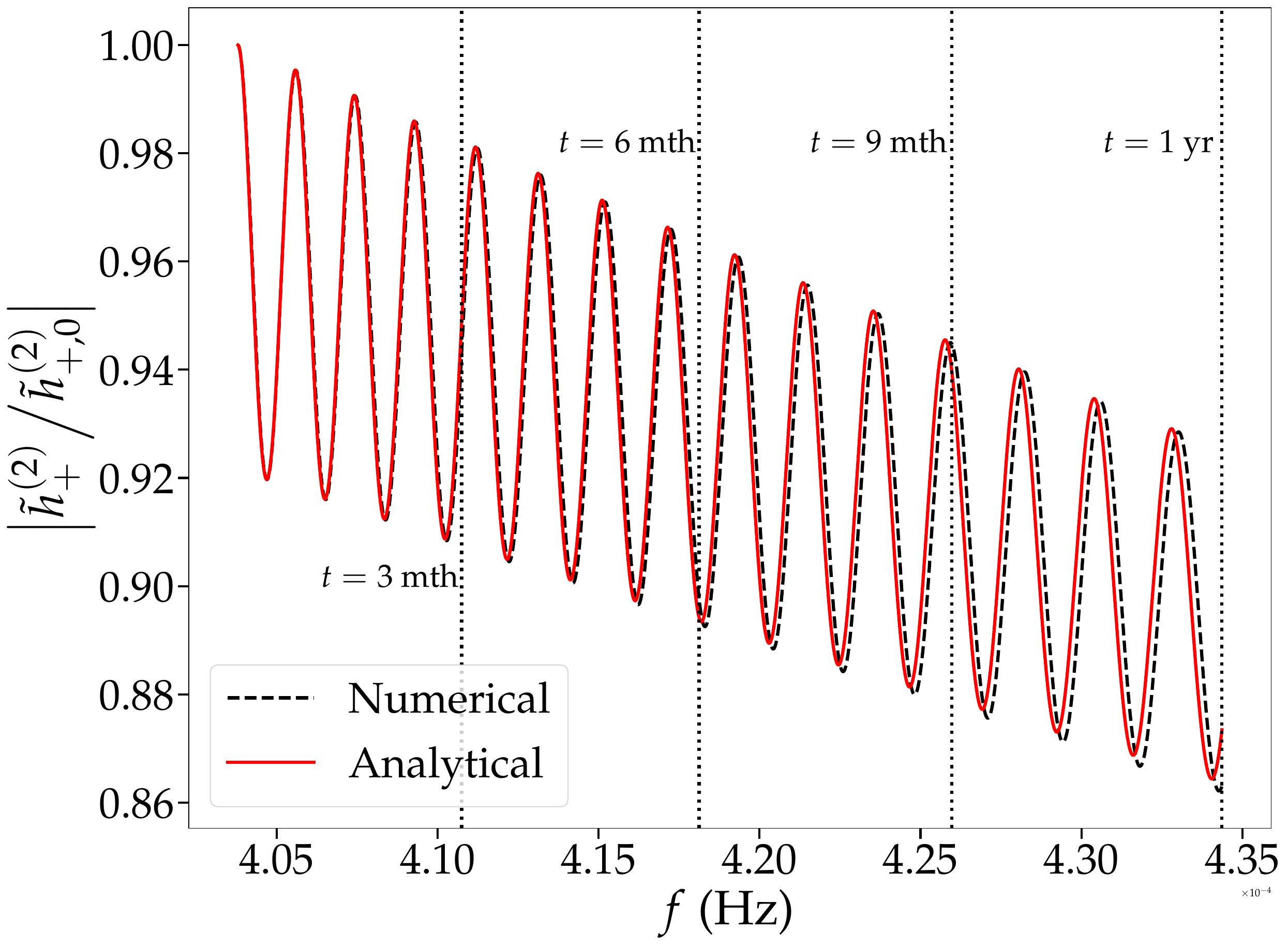}
	\caption{}
	\label{fig:hplusl2_num_analytic_a}
\end{subfigure}
\hfill
\begin{subfigure}[t]{0.49\linewidth}
\centering
	\includegraphics[width=\linewidth]{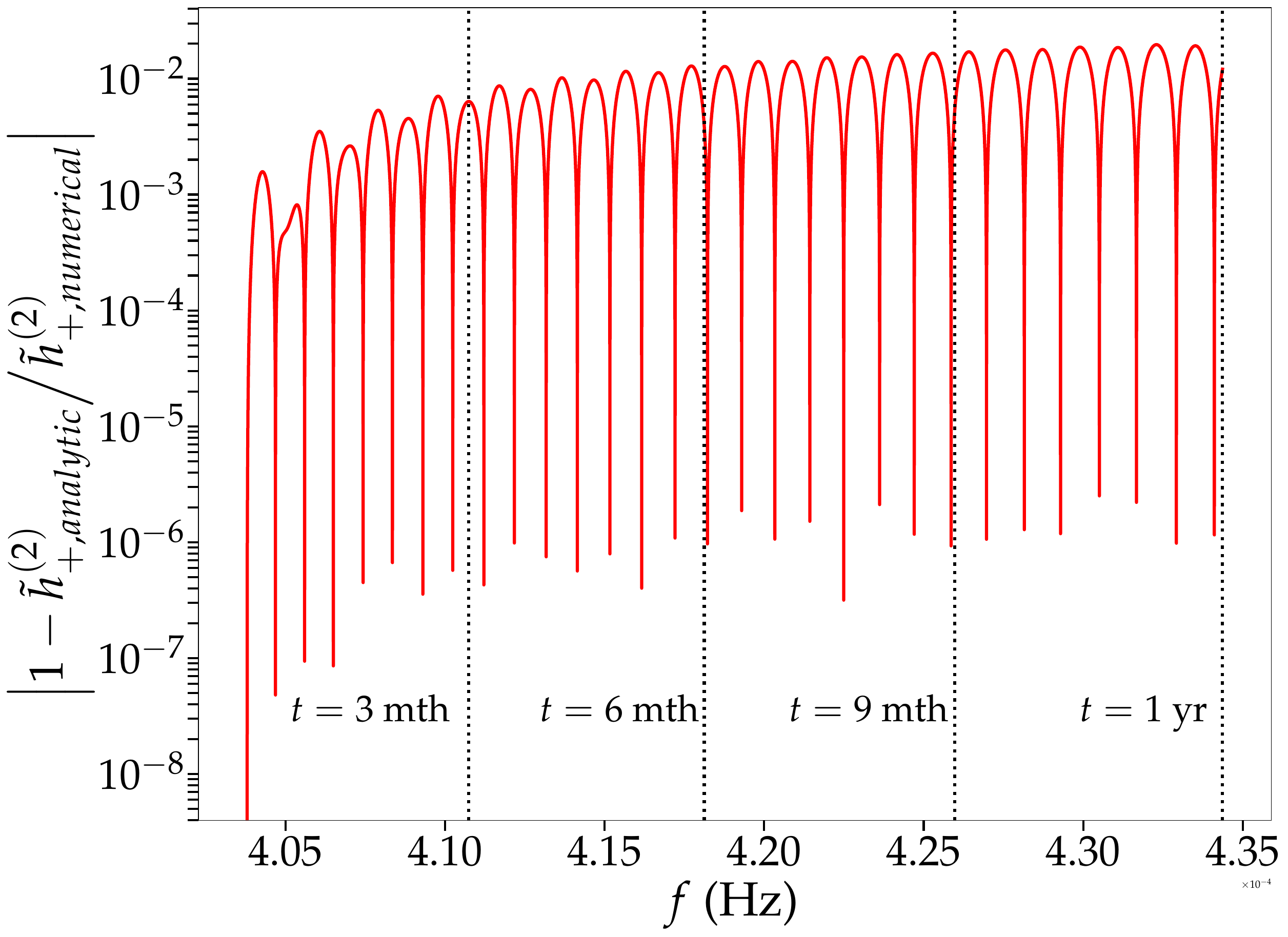}
	\caption{}
	\label{fig:hplusl2_num_analytic_b}
\end{subfigure}
	\caption{\EDIT{Numerical validation of the Fourier waveform amplitude of the $n=2$ harmonic. In~\subref{fig:hplusl2_num_analytic_a}, we show the numerical and analytic scaled amplitude $| \tilde{h}^{(2)}_{+} / \tilde{h}^{(2)}_{+,0} |$ plotted on top of each other. In~\subref{fig:hplusl2_num_analytic_b}, we show the fractional error between the analytic and numerical calculations. We also indicate for reference, the three month time intervals over the total period of one year of evolution. Observe the excellent agreement between the analytic and numerical results, and the clear signature of the KL oscillations.} }
	\label{fig:hplusl2_num_analytic}
\end{figure}

\twocolumngrid
\Cref{fig:hplusl2_num_analytic} shows the analytic and the numerical representation of the Fourier waveform amplitude. Observe that the two representations agree very well with each other, which is consistent with what we saw when validating the MSA representation of the orbital dynamics. Furthermore, we can see the prominent oscillatory features of the KL effect on the amplitude, with an overall $f^{-7/6}$ scaling induced by RR. The oscillations in the amplitude carry information regarding $m_3$ and $R$, which are crucial parameters associated with the third body. This suggests that, in principle, these parameters could be extracted from the data, given a sufficiently loud signal. 

\subsubsection{Validation of the Fourier Phase of the Waveform}
\label{subsubsec:Validation_phase}
\setlength{\parskip}{0mm}

We now discuss the validation of the Fourier GW phase of our PCKL waveform model. Our goal is to validate both the small-eccentricity approximation, as well as the averaging over KL-cycles that were used in computing~\cref{eqn:PCKL_phase}. As explained earlier, we will be focusing on $\psi_2$. We first discuss how the analytic $\psi_2 $ is evaluated, and then explain how the numerical counterpart is obtained. 

To evaluate $\psi_{2,\text{analytic}} $ and $\psi_{2,\text{numerical}} $, we need to specify $\phi_c$ and $t_c$. As outlined in~\cite{blake_2018,blake_2019}, this is done by maximizing the match between the analytic and numerical GW phases. In this work, we do not compute a match, and our goal is simply to validate the approximations we undertook in obtaining~\cref{eqn:PCKL_phase}. Thus, in order to compare  $\psi_{2,\text{analytic}} $ and $\psi_{2,\text{numerical}} $, we simply pick $\phi_c$ and $t_c$ to ensure that $\psi_2 (2F_0) = 0$ for both $\psi_{2,\text{analytic}} $ and $\psi_{2,\text{numerical}} $. We ensure the condition $\psi_2 (2F_0) = 0$ is satisfied by making the choice $\phi_2 (F_0) = 0$ and $t_2 (F_0) = 0$, which fix the values of $t_c$ and $\phi_c$. Since it is the absolute dephasing $| \psi_{2,\text{analytic}} - \psi_{2,\text{numerical}} |$ that captures the accuracy of the analytic result, terms involving $\phi_c$ and $t_c$ will cancel out identically for every value of $f \in [ 2F_0, 2F_{\text{1yr}} ]$. The values of $F$ are once again obtained by tabulating the corresponding numerical solution from \texttt{NDSolve}, and we chose $\Delta t = 100$ for the bin size.

\begin{figure}[H]
	\centering
	    \begin{subfigure}[t]{0.49\linewidth}
    \centering
    \includegraphics[width=\linewidth]{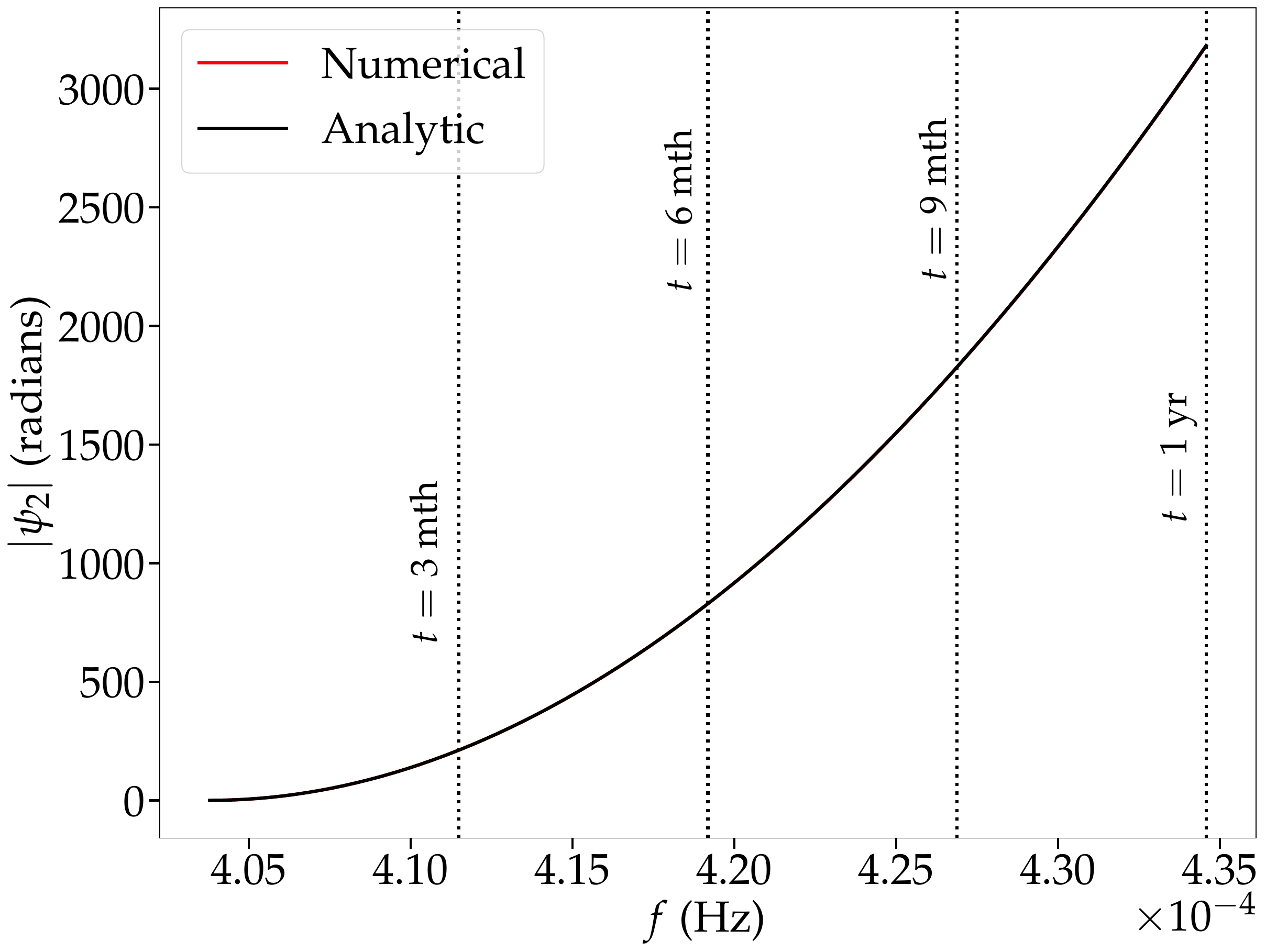}
    \caption{}
	\label{fig:psi_numeric_anlaytic_a}
    \end{subfigure}
    \hfill
	\begin{subfigure}[t]{0.49\linewidth}
	\centering
	\includegraphics[width=\linewidth]{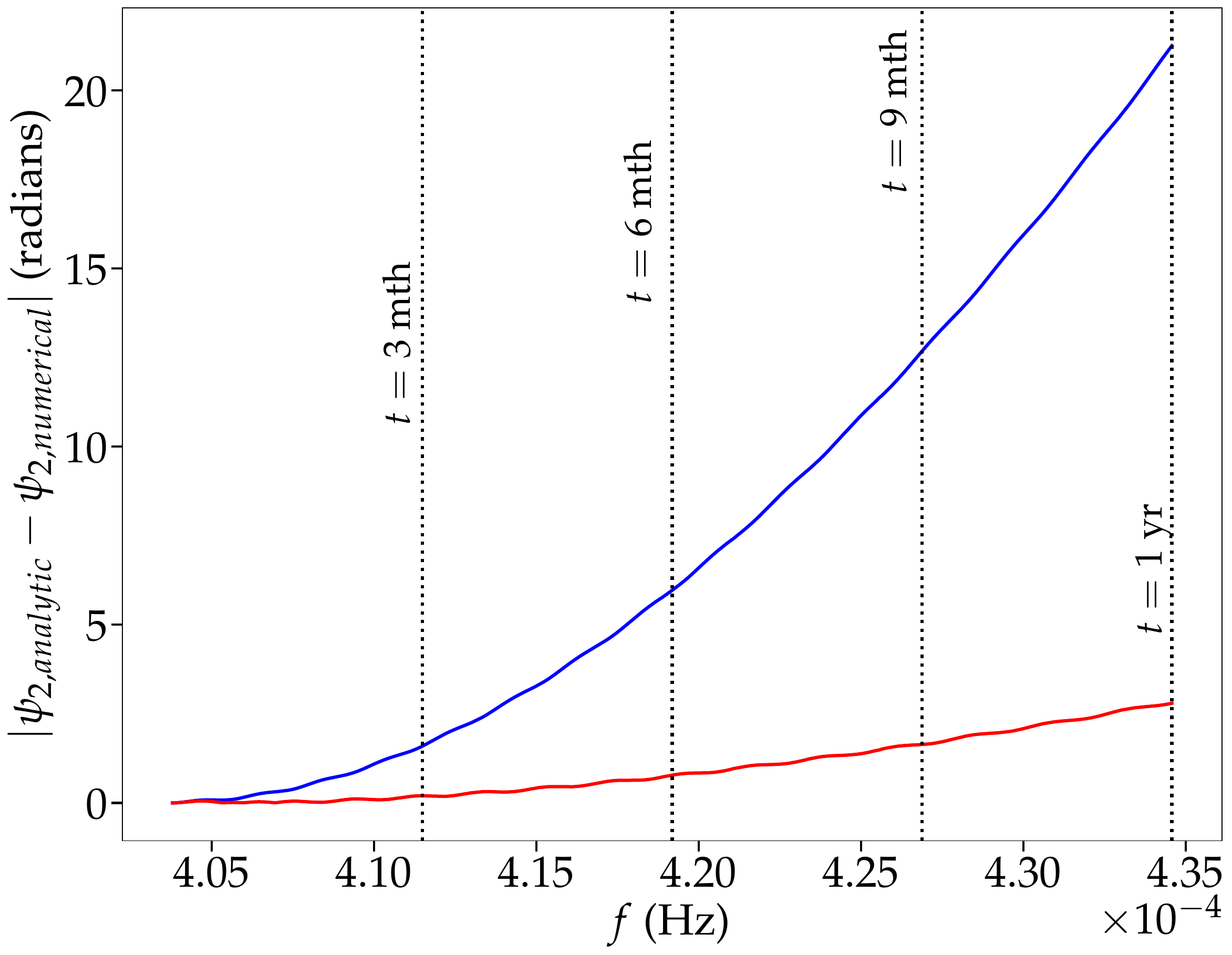}
	\caption{}
	\label{fig:psi_numeric_anlaytic_b}
	\end{subfigure}
	\caption{\EDIT{Numerical validation for the GW phase $\psi_2$, corresponding to the $n=2$ harmonic. In~\subref{fig:psi_numeric_anlaytic_a}, we show the numerical and analytic $\psi_2$, computed up to $\mathcal{O} (e^4)$, plotted on top of one another. In~\subref{fig:psi_numeric_anlaytic_b}, we show the dephasing $ | \psi_{2,\text{analytic}} - \psi_{2,\text{numerical}} | $ computed at $\mathcal{O}(e^2)$ (blue line) and at $\mathcal{O} (e^4)$ (red line). Note that the dephasing decreases as we go to higher order in eccentricity. Also note that the number of cycles of dephasing ($\sim \EDIT{0.39}$GW-cycles) at $\mathcal{O}(e^4)$ is much smaller it is much smaller compared to the total number of GW cycles accumulated, which is $\sim \mathcal{O}(10^3)$ GW-cycles.} }
	\label{fig:psi_numeric_analytic}
\end{figure}

For $\psi_{2,\text{analytic}} $, we tabulate~\cref{eqn:PCKL_phase} using the binned values of the GW frequency $f=2F$, with the appropriate values of $\phi_c$ and $t_c$ as described above. For $\psi_{2,\text{numerical}} $, we use \texttt{NDSolve} on the differential form of~\cref{eqn:phi_t_SPA} to obtain numerical solutions to $\phi_2 (F)$ and $t_2 (F)$ with the initial conditions listed above. We then tabulate $\psi_{2,\text{numerical}} $ using the binned values of the GW frequency. We compute $\psi_{2,\text{analytic}} $ and $\psi_{2,\text{numerical}} $ at $\mathcal{O}(e^2)$ and at $\mathcal{O}(e^4)$. In~\cref{fig:psi_numeric_analytic}, we show in~\subref{fig:psi_numeric_anlaytic_a}, both $| \psi_{2,\text{analytic}} |$ and $| \psi_{2,\text{numerical}} |$ computed at $\mathcal{O} (e^4)$ and plotted on top of each other, while in~\subref{fig:psi_numeric_anlaytic_b}, we show the dephasing $| \psi_{2,\text{analytic}} - \psi_{2,\text{numerical}} |$ computed at $\mathcal{O}(e^2)$ and at $\mathcal{O}(e^4)$.

\onecolumngrid

\setlength\belowcaptionskip{-2ex}

\twocolumngrid
From~\cref{fig:psi_numeric_analytic}, we observe that by going to higher order in eccentricity (from  $\mathcal{O} (e^2)$ to  $\mathcal{O} (e^4)$), the dephasing is reduced by \EDIT{nearly an order of magnitude}. Thus, an $\mathcal{O}(e^2)$ approximation in our model would mean that we miss the true (numerical) waveform signal by $\sim \EDIT{3}$ GW-cycles, while an order $\mathcal{O}(e^4)$ is significantly better. We can systematically extend our calculation to higher orders in eccentricity and we expect that the dephasing will continue to get smaller. We also indicate the absolute value of the numerical phase denoted by $|\psi_{2,\text{numerical}}|$ at the three month time intervals, which allows for a comparison of the dephasing at those intervals with $|\psi_{2,\text{numerical}}|$. 

We infer that at both orders in $e$, the dephasing is much smaller than the total number of cycles accumulated, which is about $\mathcal{O}(10^4)$ GW-cycles owing to the long inspiral time of the source. At the end of $1$yr of inspiral, the dephasing at both orders in $e$ is about 4 orders of magnitude smaller than $|\psi_{2,\text{numerical}}|$.  Moreover, for the $\mathcal{O}(e^4)$ result, the dephasing is only about $\lesssim \EDIT{2.5}$ radians after 1 yr. Consequently, our model would be able to capture the waveform's phase to within the last GW-cycle (at the end of 1 yr), which is significantly less than the total number of GW-cycles accumulated. Therefore for a loud enough signal, it is in principle possible to extract information from the GW phase regarding $\{ \overline{e}_0, \delta e_0, k_0\}$, which are parameters introduced by the KL effect, \EDIT{and are in} turn combinations of $\{ e_0, \omega_0, \iota_0 \}$. \\

Having shown that our PCKL model is capable of capturing the phase up to the last GW cycle, we now address how it compares to the PC model. Specifically we partially address the question of how well a PC waveform (corresponding to an isolated binary with constant eccentricity) would agree/disagree from the PCKL waveform we have constructed in this paper. However, we leave a more detailed and rigorous \emph{match} calculation using Markov Chain Monte Carlo (MCMC) methods/Fisher analysis to future work, and perform a simple dephasing calculation here instead. 

Consider then the absolute difference between $\psi_2^{PCKL} (f) $, given by~\cref{eqn:PCKL_phase}, and $\psi_2^{PC} (f)$, given by~\cref{eqn:PC_phase}, i.e., we compute $|\psi^{KL}_2 (f)|$. With this in hand, let us evaluate $ \mathcal{N} (f) = | \psi_2^{KL} (f)| /(2 \pi)$, which qualitatively gives an estimate of the number of cycles of dephasing due to the KL effect, at a particular frequency $f$. We will compute $\mathcal{N}$ at $f=2F_{\text{orb,1 yr}}$ (which corresponds to the cycles of dephasing accumulated after 1 yr of inspiral) over a region of parameter space corresponding to $e_0 \in [0.003,0.2]$ and $\iota_0 \in [4^{\circ}, 39^{\circ}]$, while keeping a fixed value of $\omega_0 = 0$. When $\mathcal{N} \geq 1$, we can characterize the PC model as being \emph{insufficient} in capturing effects introduced by the KL effect. We emphasize that the dephasing calculation we perform here provides only conservative estimates since we don't rigorously account for correlation and degeneracies between parameters, which can be done with an MCMC/Fisher analysis. 
\setlength\belowcaptionskip{-2ex}
\begin{figure}[H]
    \centering
    \includegraphics[width=\linewidth]{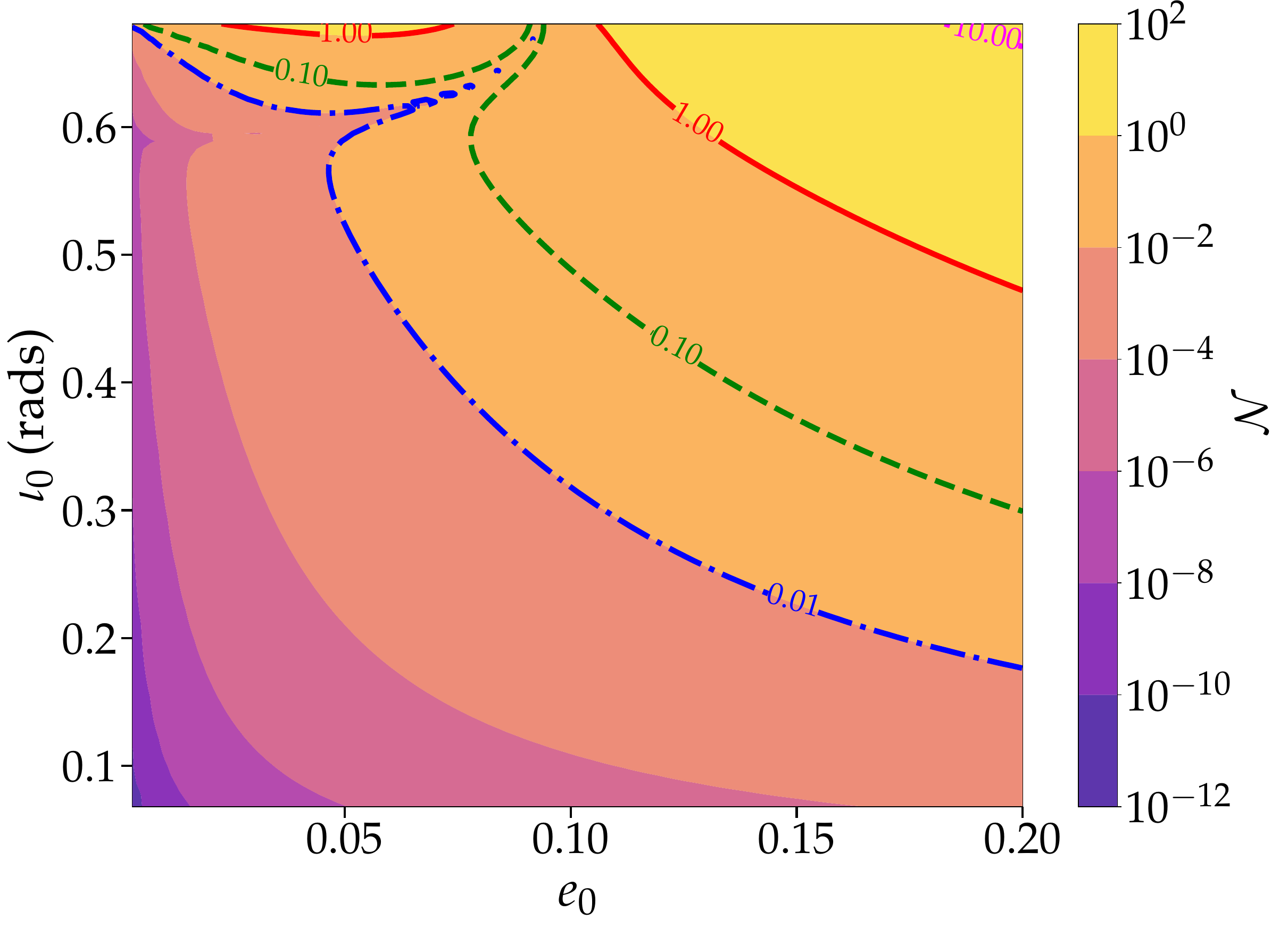}
    \caption{\EDIT{Number of cycles of dephasing due to the KL effect when compared to the PC model. Note the contour levels shown, and that for $e_0>0.1$ there is a wider region of parameter space for which the PC model is insufficient for capturing the KL effect that is manifest in the PCKL model.}}
    \label{fig:contour}
\end{figure}

\EDIT{\Cref{fig:contour}} shows a contour plot of $\mathcal{N}$ for various values of $\iota_0$ and $e_0$. Observe that for small eccentricities, $e_0 < 0.1$, the PC model is insufficient for larger inclinations, $\iota_0 \sim 40^{\circ}$. For larger eccentricities, $e_0 \gtrsim 0.15$, there is a wider region of parameter space (admitting smaller values of $\iota_0$), for which the PC model becomes insufficient. Therefore, the PCKL model is most relevant for regions of parameter space corresponding to larger eccentricities and inclinations, which is further motivation for extending our model to those regimes. However, we note that the corresponding amplitudes of the PCKL and PC models would be easy to distinguish for most of the parameter space shown here, owing to the absence of KL oscillations in the amplitude of the PC model. Therefore, the information from both the amplitude and the phase will indicate better where the PC model becomes insufficient, and we will explore this in more depth by doing \emph{match} calculations in the future.

\section{Conclusion and Future Work} \label{sec:conclusion}

\begin{figure}[H]
    \centering
    \includegraphics[width=0.9\linewidth]{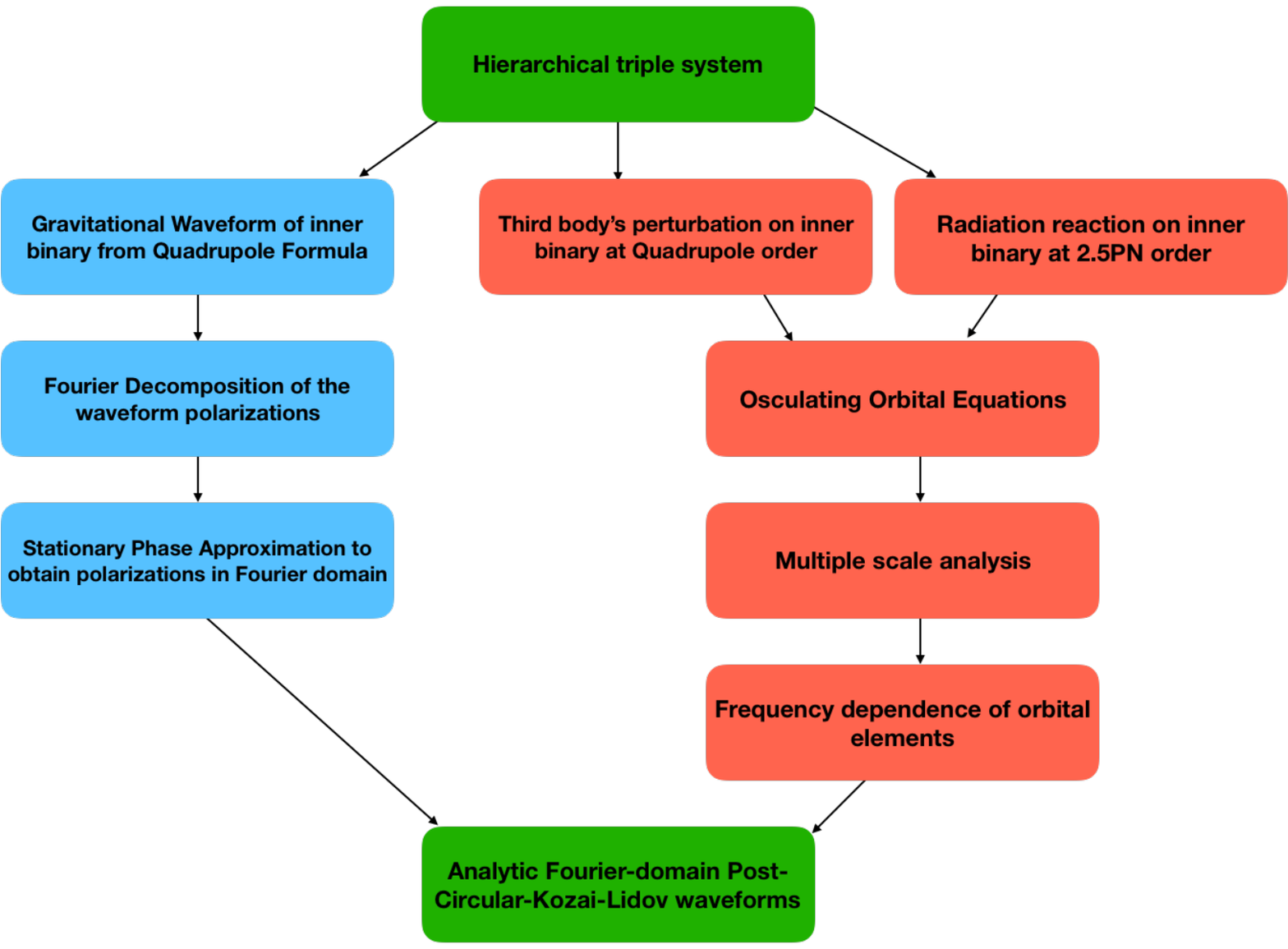}
    \caption{Schematic workflow for the creation of an analytic model for the orbital dynamics and the GWs in the Fourier domain emitted by an inspiraling IMBH binary with KL perturbations induced by a SMBH third-body.}
    \label{fig:schematic}
\end{figure}

In this work, we created an analytic model for the GWs emitted during the inspiral of an IMBH binary that undergoes KL oscillations induced by a SMBH third body. Using the osculating orbit formalism and MSA, we found an analytic representation of the orbital dynamics, and in particular a representation of $e$ and $\iota$ and $\omega$ as a function of orbital frequency that is valid for small eccentricities. We used this analytic representation to then obtain the Fourier amplitude and Fourier phase of the GWs emitted under the SPA. Figure~\ref{fig:schematic} presents a schematic of the workflow that allowed the construction of this analyltic model.

We found that there is a clear signature of the KL oscillations in the Fourier amplitude. We also found that the GW phase can be written as a sum of two sets of terms -- one that is obtained under the postcircular approximation applied to an isolated eccentric binary, and another that contains corrections induced by the cumulative effect of the KL oscillations. We validated our analytic results with numerics and found that our calculations are robust. Although we only went up to $\mathcal{O}(e^4)$ in our calculations, the results can be systematically extended to higher orders in eccentricity. \\

For the purpose of data analysis and parameter estimation, the properties of the source are best inferred from the GW phase. With our result for the phase given in~\cref{eqn:PCKL_phase}, we can perform some Fermi estimates regarding how well we can measure the parameters that enter the phase. Let us define $\delta s$ as the error in measuring a parameter $s$, where e.g.~$s \in \{ e_0, \omega_0, \iota_0 \}$, and $\rho \in \{ 10,20 \}$ as the SNR. A Fermi estimate of the accuracy to which a parameter $s$ can be measured is then $\delta \mathit{s} \sim {\rho}^{-1} ({ \partial \psi_n^{PCKL} }/{\partial s})^{-1}$. Using the $n=2$ harmonic at a GW frequency corresponding to half a year of evolution, $f_{1/2 yr} = 2 F_{1/2 yr}$, we find that, on average, we may be able to estimate $e_0$ to $10^{-4}$, $\iota_0$ to $10^{-3}$ and $\omega_0$ to $10^{-2}$ for an event with a SNR of 20. What this primitive Fermi analysis shows is that there is an in-principle \textit{measurable} imprint on the phase due to the KL effect. Such an imprint suggests that one should consider a more rigorous data analysis study, which we will undertake upon extending our model to higher eccentricity and higher PN-order. 

\onecolumngrid

\setlength\belowcaptionskip{-2ex}

\twocolumngrid


In addition to $\mathcal{M}$, the GW phase in~\cref{eqn:PCKL_phase} depends on the parameters $\overline{e}_0 , \delta e_0$, and $k_0$, which are induced due to the KL effect and are combinations of the parameters $e_0, \iota_0$, and $\omega_0$. There is no dependence on the parameters $m_3$ and $R$, which is due to a limitation of our model. Under the approximation of treating the outer orbit as stationary, at the quadrupolar level, the maximum eccentricity and inclination angle of the KL oscillations are \textit{independent} of the set of parameters $\{ m,\eta, a_0,m_3,R \}$; in fact, they only depend on the initial conditions $e_0,\iota_0$, and $\omega_0$. The KL phase $\theta$ depends on $\{ m, a_0,m_3,R \}$ and since we are averaging over $\theta$ in computing the GW phase (coupled with the stationary outer orbit approximation), the dependence of the phase on $\{ m, a_0,m_3,R \}$ is lost (note that there is still a dependence on $\eta$ through $\mathcal{M}$). However, the GW amplitude contains a strong signature of the oscillations as it still depends on $\theta$, and hence it contains information regarding $ m_3$ and $R$, that is complementary to the GW phase. An important prediction of this is that in extending the family of waveform parameters by $\{ m_3, R, \overline{e}_0, \delta e_0, k_0 \}$, we can capture the KL effect through the waveform. Therefore, for a strong SNR source, one could in principle extract information regarding $\{ m_3, R, \overline{e}_0, \delta e_0, k_0 \}$. \EDIT{We point out that although other methods exist (such as measuring the Doppler shift~\cite{Yunes:2010sm,Inayoshi:2017hgw,Randall:2018lnh}) for obtaining similar information regarding the third body, our approach using MSA provides direct information on how the third body affects the waveform over long-timescales. It therefore complements existing approaches that are aforementioned.}

\EDIT{An important caveat in our work is the assumption of a stationary outer orbit. The variation of the outer orbit due to the KL effect occurs on a longer timescale, and the motion of the inner binary around the third body also leads to PN effects tied to the outer orbit (such as de-Sitter precession~\cite{Yu:2020dlm} and radiation-reaction). The effects of the variation of the outer orbit can be incorporated using MSA and this can be further investigated as part of a future study.}

We used the adiabatic approximation throughout our work, even in obtaining the stationary-phase condition. As we have pointed out, there are corrections to the condition that scale with $\mathcal{O} (\epkl)$, and which are a part of the \textit{postadiabatic} corrections. There are other postadiabatic effects that simply arise from applying MSA to higher order (see~\cite{Loutrel:2018ydu} for an example of such effects). postadiabatic corrections could play a role depending on the regime of parameter space that is being explored. Therefore, we leave a study of such postadiabatic effects for future work. 

Finally, a key limitation in our work is the use of the small-eccentricity approximation in obtaining the frequency dependence of the eccentricity, which was also a limitation of the ``postcircular" approximation~\cite{NicoBertiArunWill}. The work done by~\cite{blake_2019}, where the eccentric waveform model was computed to 3PN-order, bypasses that limitation. In the immediate future, we will be following the methods used by~\cite{blake_2019} to not only extend our model to higher eccentricities, but also to higher PN-order; such an analysis is different from what is being pursued by~\cite{Gupta:2019unn, Kuntz:2021ohi}, as we would be exploring a different region of parameter space. We will also be using our results with those of~\cite{blake_2019} to construct inspiral-merger-ringdown models with the KL effect included.

\begin{acknowledgments}
We thank Coleman Miller, Hector O. Silva, Alejandro C\'ardenas-Avenda\~no, and Scott Perkins for discussions. N.Y. acknowledges financial support through NASA ATP Grants No. 17-ATP17-0225, No. NNX16AB98G and No. 80NSSC17M0041.

\end{acknowledgments}

\appendix

\section{Components of perturbing force} \label{app:perturbations}
In Sec.~\ref{subsec:3timescale}, we introduced both the quadrupolar perturbation due to the third body as well as the perturbation due to the RR force. In this section we provide details on the components of the perturbing forces due to both perturbations. 
\subsection{Kozai-Lidov perturbation}

Following~\cite{poisson_will_2014}, the components of the third-body's perturbing force are
\begin{align}
\begin{split}
    \begin{split}
\mathcal{R}_{\text{3b}}& = -\dfrac{G m_3 r}{R^3} \Big [1- 3 \big (\cos (\omega+\ell) \cos \ellout \\
& + \cos \iota \sin (\omega+\ell)\sin \ellout \big )^2 \Big ], 
\end{split}\\
\begin{split}
\mathcal{S}_{\text{3b}}& = \dfrac{3 G m_3 r}{R^3} \big [\cos (\omega+\ell) \cos \ellout + \cos \iota \sin (\omega+\ell)\sin \ellout \big ] \\
& \times \big [ -\sin (\omega+\ell) \cos \ellout + \cos \iota \cos (\omega+\ell) \sin \ellout \big ],
\end{split}\\
\begin{split}
   \mathcal{W}_{\text{3b}}& = \dfrac{3 G m_3 r}{R^3} \big [\cos (\omega+\ell) \cos \ellout + \cos \iota \sin (\omega+\ell) \sin \ellout \big ] \\
& \times \sin \iota \sin \ellout, 
\end{split}
\end{split}
\end{align}
where $r = p/(1+e\cos \ell)$, with $p$ being the semilatus rectum.

\subsection{Radiation-reaction perturbation}
Following~\cite{poisson_will_2014}, the components of the perturbing force due to RR are given by
\begin{align}
\begin{split}
     \mathcal{R}_{\text{RR}} &= \Vec{a}_{\text{RR,orb}} \cdot \Vec{n}_{\text{orb}}, \\
     \mathcal{S}_{\text{RR}} &= \Vec{a}_{\text{RR,orb}} \cdot \Vec{\lambda}_{\text{orb}},
\end{split}
\end{align}
where the vectors $\Vec{n}_{\text{orb}},\Vec{\lambda}_{\text{orb}}$ are the tangent vectors to the orbital plane of the inner binary and are specified in the orbital reference frame~\cite{poisson_will_2014} ($\Vec{e}_x , \Vec{e}_y, \Vec{e}_z$)and are given by 
\begin{align}
    \Vec{n}_{\text{orb}} &= \cos (\omega+\ell) \Vec{e}_x +  \sin (\omega+\ell) \Vec{e}_y, \\
    \Vec{\lambda}_{\text{orb}} &= -\sin (\omega+\ell) \Vec{e}_x + \cos (\omega+\ell) \Vec{e}_y,
\end{align}
and $\Vec{a}_{\text{RR,orb}}$ is the acceleration due to RR and in the orbital reference frame it is given by
\begin{align}
\begin{split}
        \Vec{a}_{\text{RR,orb}} &= \dfrac{8}{5} \eta \dfrac{m^{7/2}}{p^{9/2}} (1+e\cos \ell)^{3} [ e \sin \ell Q_1 \Vec{n}_{\text{orb}}\\
        &-(1+e \cos \ell) Q_2 \Vec{\lambda}_{\text{orb}} ],
\end{split}
\end{align}
with the coefficients $Q_1, Q_2$ given by
\begin{equation}
\begin{aligned}
Q_1:=& \frac{44}{3}+\frac{35}{3} q_1-5 q_2+e\left(\frac{80}{3}+\frac{125}{6} q_1-\frac{55}{4} q_2\right) \cos \ell \\
&+e^{2}\left[2+\frac{5}{3} q_1+\frac{5}{2} q_2+\left(10+\frac{15}{2} q_1-\frac{45}{4} q_2\right) \cos ^{2} \ell \right], \\
Q_2:=& 4+e\left(10+\frac{35}{6} q_1-\frac{5}{2} q_2\right) \cos \ell \\
&-e^{2}\left[9+\frac{35}{3} q_1-5 q_2-\left(15+\frac{35}{2} q_1-\frac{15}{2} q_2\right) \cos ^{2} \ell \right] .
\end{aligned}
\end{equation}
The parameters $q_1,q_2$ are fixed with an appropriate gauge. At the LO in MSA, as far as the secular evolution is concerned, that is upon orbit averaging, all the terms involving $q_1,q_2$ vanish and we point the reader to~\cite{poisson_will_2014} for more details. Note that even though the unit vectors and acceleration are specified in the orbital reference frame, $\mathcal{R}_{\text{RR}},\mathcal{S}_{\text{RR}}$ are invariant under spatial coordinate transformations since they are dot products. Hence it is sufficient to provide expressions for them in one frame.

\section{Validation of averaging over the Kozai-Lidov cycles} \label{app:KL_averaging}
In Sec.~\ref{subsec:PCKL-phase} we averaged over the KL-cycles to obtain expressions for $\psi_n (f)$ given by~\cref{eqn:PCKL_phase}. The GW phase can be expressed as $\psi_n (f) = \langle \psi_n (f) \rangle_{\theta} + \psi_{n,\text{osc}}(f)$, where $\langle \psi_n (f) \rangle_{\theta}$ is what we computed in Sec.~\ref{subsec:PCKL-phase} and is given by~\cref{eqn:PCKL_phase}, while $\psi_{n,\text{osc}}(f)$ is the oscillatory contribution that we neglected in our calculation. \\

We now restrict to $n=2$, compute $\psi_{2,\text{osc}}(f)$ using numerics and we show that it is justifiable to ignore this correction in our work. We do so by numerically solving for $\phi_{2,\text{osc}}$ and $t_{2,\text{osc}}$ using the differential form of~\cref{eqn:phi_t_SPA} as we had done in Sec.~\ref{subsubsec:Validation_phase}. 
\begin{figure}[H]
	\centering
	\includegraphics[width=\linewidth]{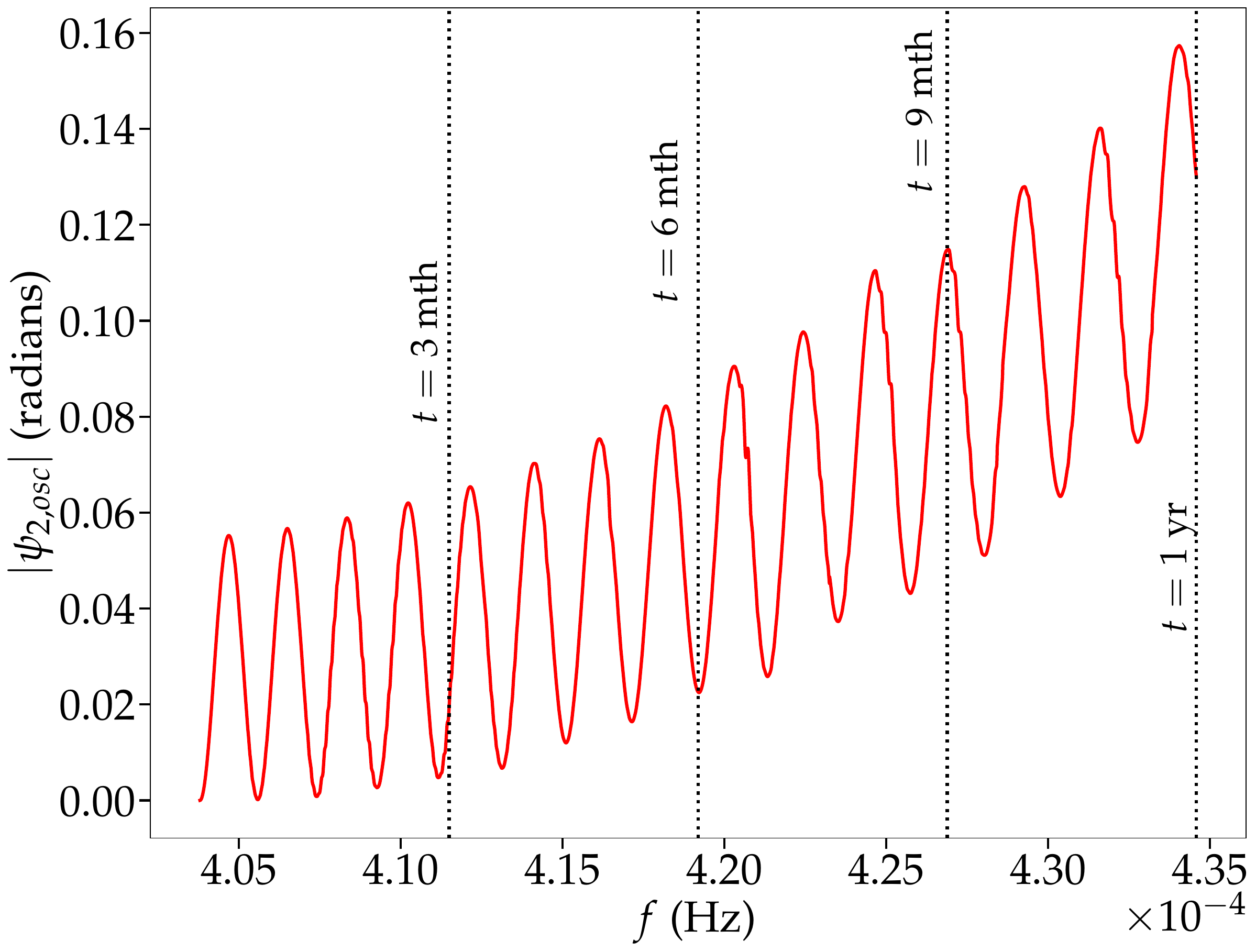}
	\caption{\EDIT{Oscillatory contribution to the phase $| \psi_{2,\text{osc}}| $, plotted against the GW frequency $f=2F$. Note that the maximum value of $| \psi_{2,\text{osc}}| $ is about $0.16$ radians, which is an order of magnitude smaller than the dephasing in~\cref{fig:psi_numeric_analytic}}}
	\label{fig:psiosc}
\end{figure}
The important point to note is that in the defining equation for $\phi_{2,\text{osc}}$ and $t_{2,\text{osc}}$, we ought to use $\tau_{\text{osc}} = \tau - \langle \tau \rangle_{\theta}$, computed to a desired order in eccentricity. We perform this calculation to $\mathcal{O} (e^4)$, and we use~\cref{eqn:tau_PCKL} for $\langle \tau \rangle_{\theta}$. We can then use \texttt{NDSolve} (with the same flags as described in Sec.~\ref{sec:validation}) to integrate the differential equations for  $\phi_{2,\text{osc}}$ and $t_{2,\text{osc}}$ with the initial conditions $\phi_{2,\text{osc}} (F_0) =0$ and $t_{2,\text{osc}} (F_0) =0$. We then tabulate $\psi_{2,\text{osc}}$ using binned values of the GW frequency $f = 2F$ where values for $F$ are obtained from the numerical solution to the orbital elements.\\

From~\cref{fig:psiosc}, we see that $|\psi_{2,\text{osc}} |$ has the characteristic oscillatory behavior, but it also contains a secular growth. Since we are doing a bivariate expansion in $e,k_0$ when we obtain~\cref{eqn:PCKL_phase}, the secular behavior is a feature of the truncation in both $e$ as well as $k_0$. The secular behavior would decrease provided we go to higher order in $e$ and $k$ when obtaining~\cref{eqn:PCKL_phase}. Also, note that  if added as a correction, $\psi_{2,\text{osc}} $ only contributes at most \EDIT{$0.16$} radians and hence can be neglected in our calculation of~\cref{eqn:PCKL_phase}. 

\section{Non-linear aspects and phase-space study of the Kozai-Lidov problem} \label{app:roots_KL_problem}
In Sec.~\ref{subsec:2timescale}, we reviewed the KL problem along with the exact solution in the absence of RR. Since we use a small-eccentricity approximation in our work, we restricted the initial conditions to $\iota_0 \lesssim 40^{\circ}$ and $e_0 \lesssim 0.1$ so that the small-eccentricity approximation is not violated. We now show how we obtained the constraints on $\iota_0$ by reviewing the non-linear properties of the KL oscillations in the absence of RR. We study the dynamical phase-space behavior as well as the behavior of $\emax$ and $\emin$ on $\iota_0$ and $\omega_0$ for a given $e_0$. 
\begin{figure}[H]
    \centering
    \includegraphics[width=\linewidth]{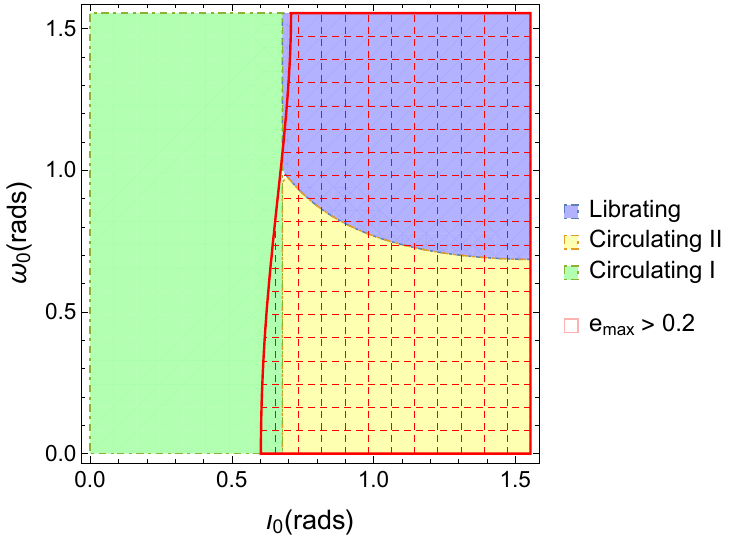} 
    \caption{Dynamical phase-space behavior for the KL problem in the absence of radiation-reaction. Note the demarcated regions based on their dynamical properties -- \emph{Librating} (blue), \emph{Circulating I} (green), \emph{Circulating II} (yellow). Also shown by the red, meshed region is where $\emax > 0.2$, thereby indicating that the small eccentricity regime corresponds to that of  \emph{Circulating I}.}
        \label{fig:circLibRegions}
\end{figure}
There are broadly two types of phase-space trajectories -- \emph{circulating} and \emph{librating}, which are determined by the values of $\zeta$ and $\gamma$. There is a separatrix that characterizes the two types of phase-space trajectories and it is defined by $\gamma_{sp} = \gamma (e_0=0)$ and can be expressed as
\begin{equation}
\gamma_{sp} = 2 (3 \cos^2 \iota_0 -1).
\end{equation} 
We summarize the behavior of the trajectories (see~\cite{Kinoshita:2007,kozai,lidov}) below
\setlength\parskip{-0.2cm}
\setlength\parsep{-0.2cm}
\begin{itemize}
\item \underline{\emph{Circulating I}}: If $\zeta>3/5$, then the motion is circulating and there is no resonance region in phase-space (small amplitude oscillations in eccentricity and inclination, while the pericenter advances). Looking at this, we see that this reduces to the requirement that $c_{i0}^2 > 3/(5 x_0)$, where $c_{i0} \equiv \cos^2 \iota_0$. Since $x_0>0$ always, we have $c_{i0}^2 > 3/5$.
\item \underline{\emph{Circulating II}}: If $\zeta<3/5$ and $\gamma> \gamma_{sp}$, then the motion is again circulating, but there is a resonance region (typically close to $\omega_0 = \pi /2$) in phase-space, leading to large-amplitude of oscillations in eccentricity and inclination, while the pericenter advances.
\item \underline{\emph{Librating}}: If $\zeta<3/5$ and $\gamma<\gamma_{sp}$, then the motion is librating, meaning that the pericenter, inclination, and eccentricity oscillate about the resonance point. This typically occurs around $\omega_0 = \pi/2$ (can lead to large-amplitude oscillations). 
\end{itemize}

In \emph{Circulating I}, the behavior of the roots $\{ x_0^*, x_1^*, x_2^*\}$ is such that $x_1^*<x_0^*<x_2^*$ (see~\cite{Kinoshita:2007,kozai,lidov}) which implies that $\alpha_0 =x_1^*, \alpha_1 = x_0^*$, and $\alpha_2 = x_2^*$. We use this particular hierarchy of the roots when we restrict to small eccentricities. 

In~\cref{fig:circLibRegions}, we show the dynamical behavior for $e_0 =0.1$ over the two-dimensional plane of $\iota_0 - \omega_0$. We demarcate the regions according to the dynamical behavior and also indicate the regime where $\emax > 0.2$, which occurs around $\iota_0 \sim 40^{\circ}$. We see that by restricting to $\iota \lesssim 40^{\circ}$, the dynamical behavior is that of \emph{Circulating I} and this is the regime that we use in our work.\\ 

\begin{figure}[H]
    \centering
    \begin{subfigure}[t]{0.49\linewidth}
    \centering
    \includegraphics[width=\linewidth]{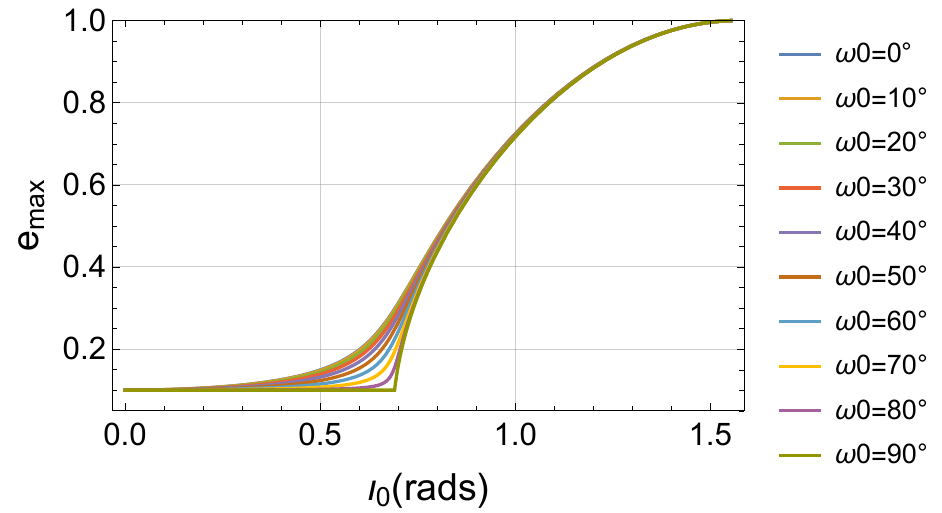}
    \caption{}
    \label{fig:MaxMinEccPlot-a}
    \end{subfigure}
    \hfill
     \begin{subfigure}[t]{0.49\linewidth}
    \centering
    \includegraphics[width=\linewidth]{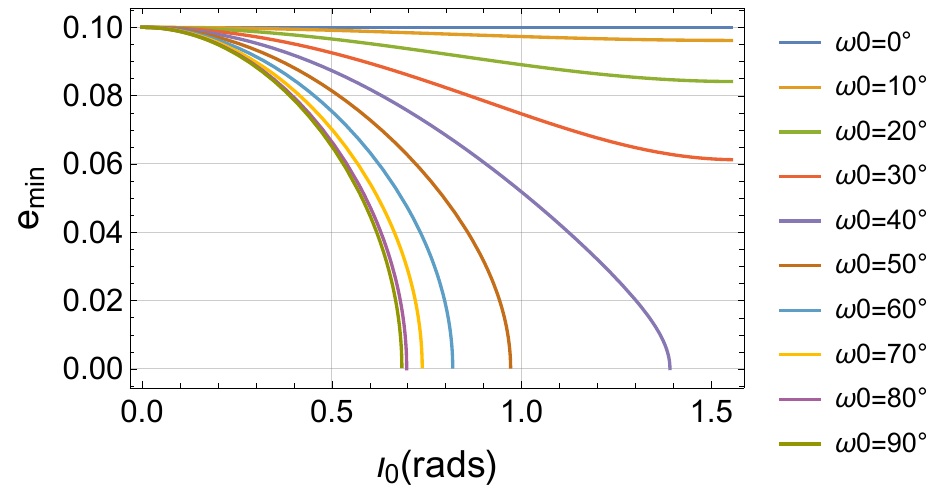}
    \caption{}
    \label{fig:MaxMinEccPlot-b}
    \end{subfigure}
    \label{fig:MaxMinEccPlot}
    \caption{Behavior of $\emax$ and $\emin$ against $\iota_0$ (with $e_0=0.1$)for various values of $\omega_0$ shown in~\subref{fig:MaxMinEccPlot-a} and~\subref{fig:MaxMinEccPlot-b} respectively.}
\end{figure}

\EDIT{We now turn to the behavior of $\emax$ and $\emin$. In~\cref{fig:MaxMinEccPlot}, we show how $\emax$ and $\emin$ vary with $\iota_0$ and $\omega_0$ for $e_0 = 0.1$. Note that $\emax$ depends strongly on $\iota_0$, while $\omega_0$ controls $\emin$ more strongly. We also see that for $\iota_0 > \cos^{-1} \sqrt{3/5}$ (also called the Kozai angle~\cite{smadar_review})}, the small-eccentricity approximation is no longer valid, consistent with what we observed in~\cref{fig:circLibRegions}. In order to obey the small-eccentricity approximation, it therefore suffices to limit $\emax$ and that is how we obtain the constraint $\iota_0 \lesssim 40^{\circ}$, which corresponds to \emph{Circulating I} as we saw earlier.\\

\begin{figure}[H]
    \centering
    \begin{subfigure}[t]{0.32\linewidth}
    \centering
    \includegraphics[width=\linewidth]{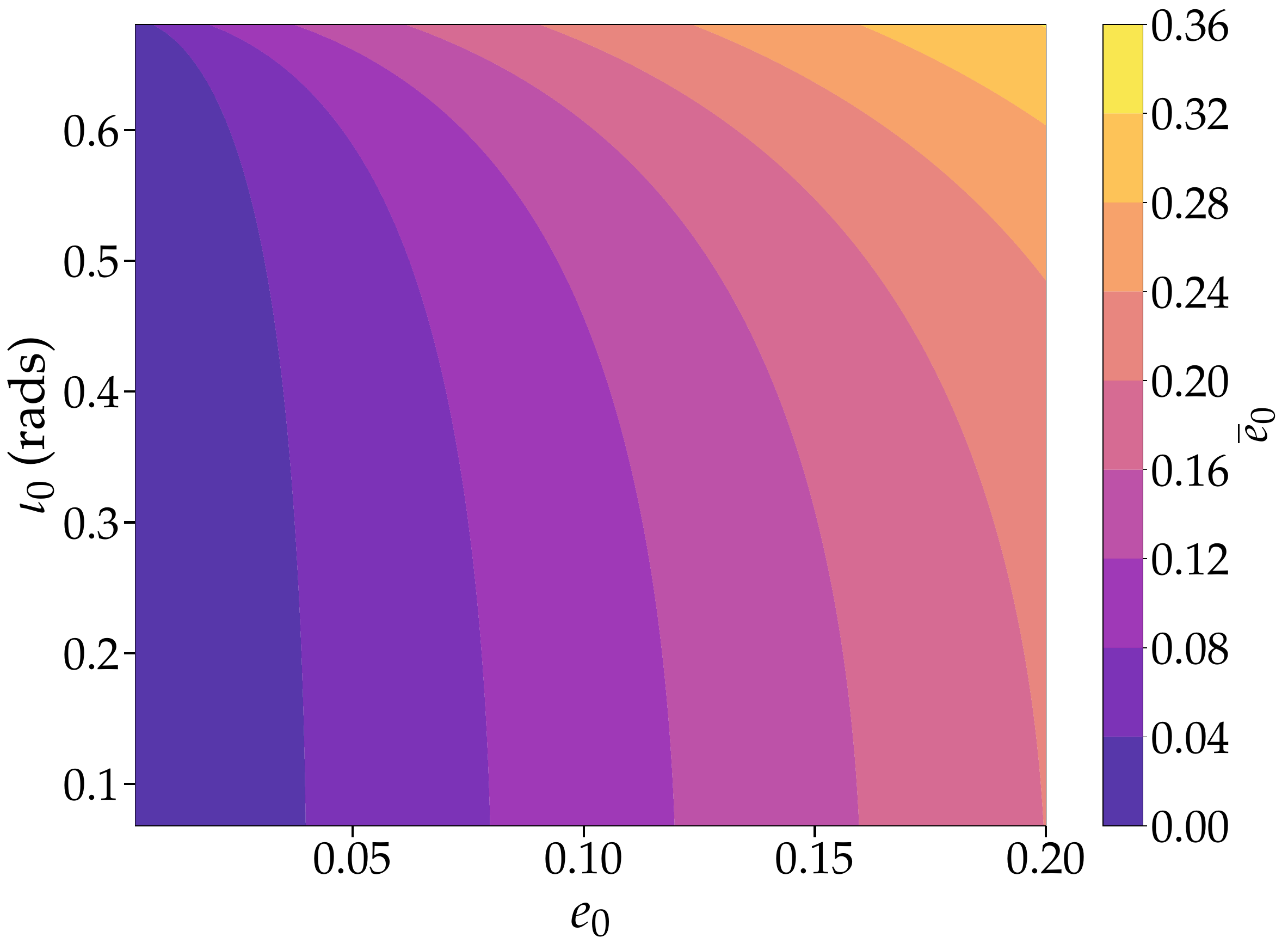}
    \caption{}
    \label{fig:ebardeltaek-a}
    \end{subfigure}
    \hfill
     \begin{subfigure}[t]{0.32\linewidth}
    \centering
    \includegraphics[width=\linewidth]{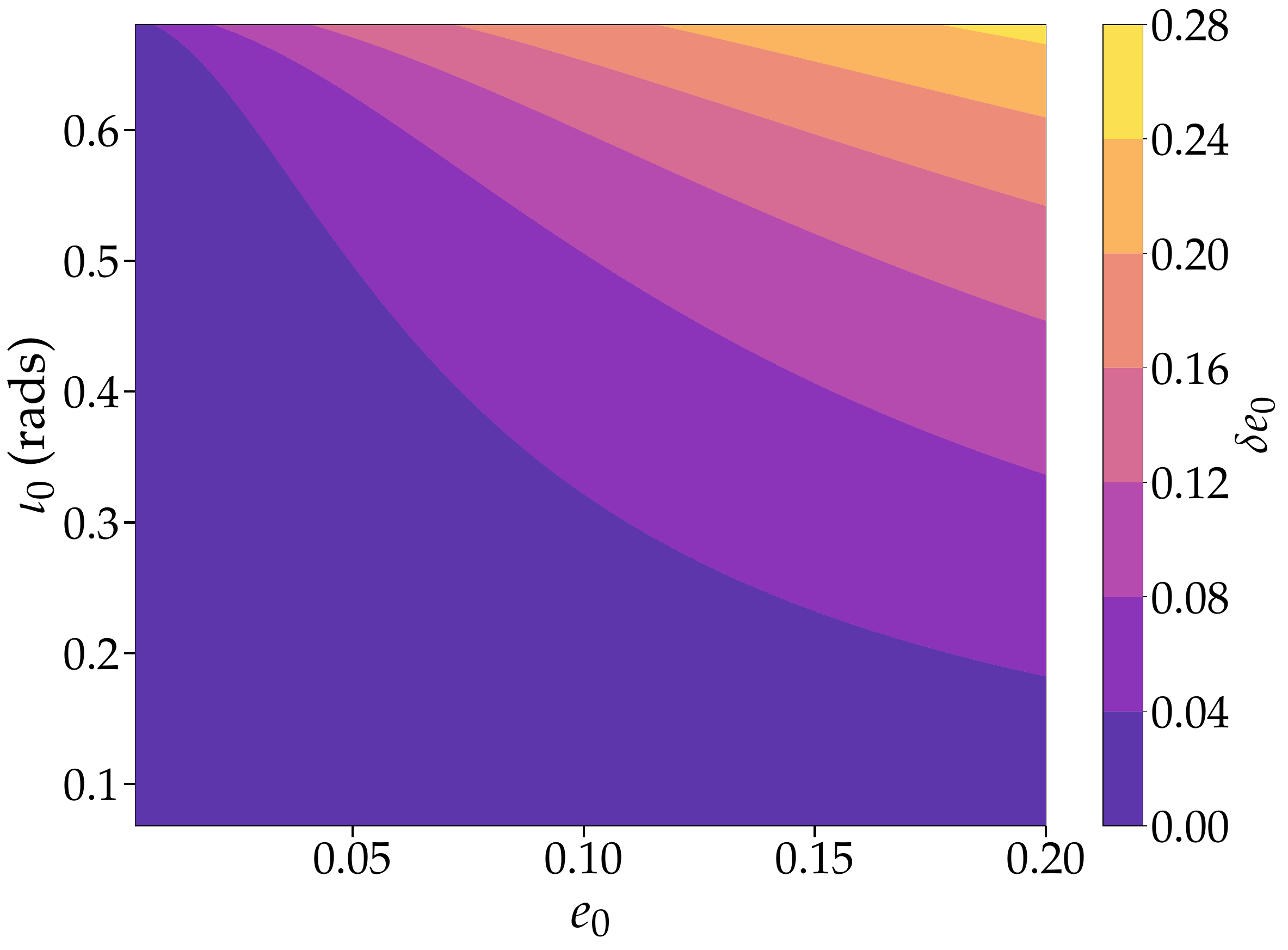}
    \caption{}
    \label{fig:ebardeltaek-b}
    \end{subfigure}
        \hfill
     \begin{subfigure}[t]{0.32\linewidth}
    \centering
    \includegraphics[width=\linewidth]{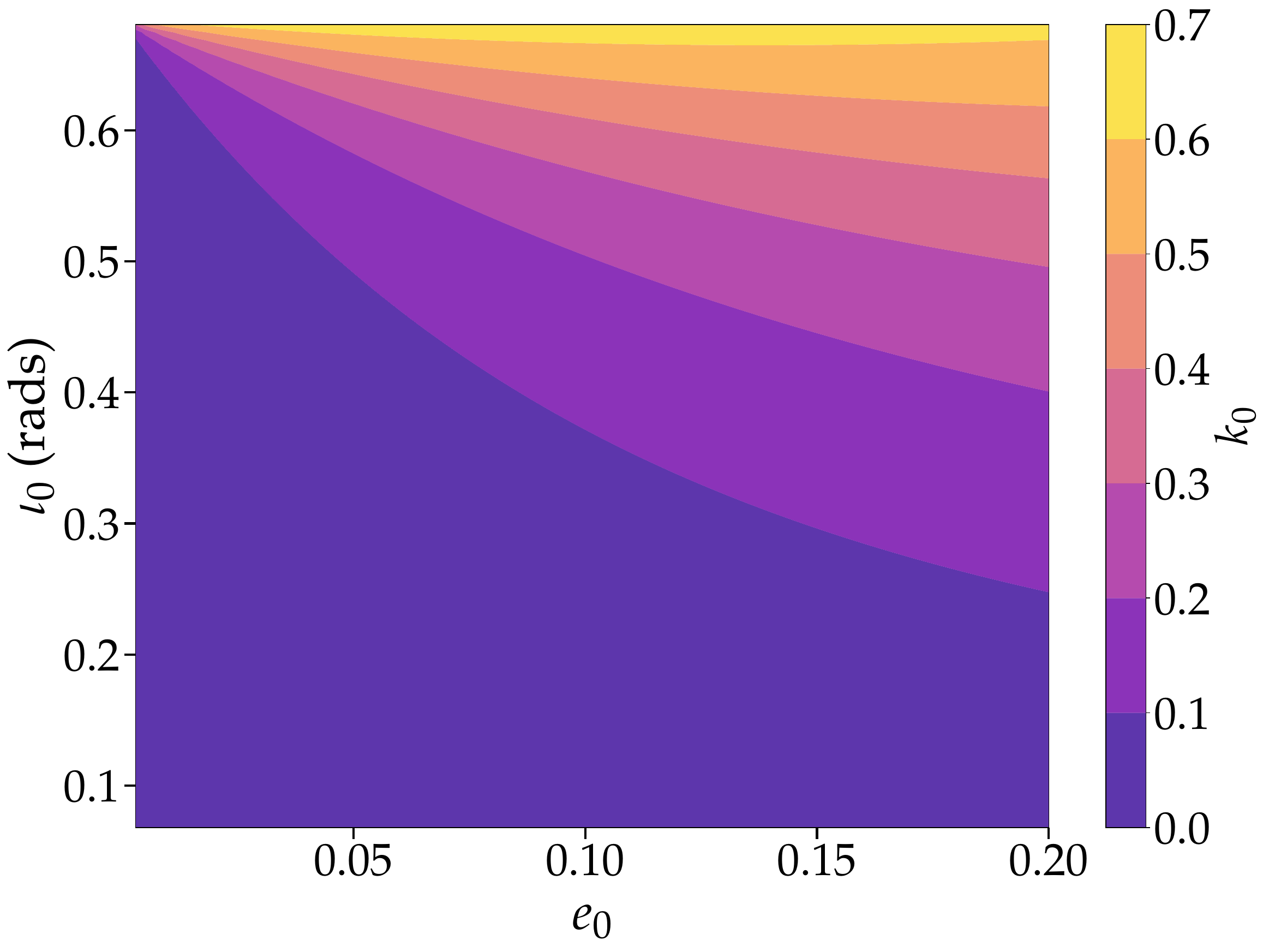}
    \caption{}
    \label{fig:ebardeltaek-c}
    \end{subfigure}
    \label{fig:ebardeltaek}
    \caption{\EDIT{Behavior of PCKL-waveform parameters $\overline{e}_0$, $\delta e_0$, and $k_0$ respectively. The parameter range shown here is the same as in~\cref{fig:contour}.} }
\end{figure}

\EDIT{Finally, we show the behavior of the parameters that enter the GW phase in~\cref{eqn:PCKL_phase} -- $\overline{e}_0$, $\delta e_0$, and $k_0$. In~\cref{fig:ebardeltaek}, we show contour plots for $\{ \overline{e}_0, \delta e_0, k_0 \}$ for the parameter range considered in~\cref{fig:contour} i.e., $e_0 \in [0.03, 0.2], \iota_0 \in [4^{\circ}, 39^{\circ}], \omega_0 =0$. The dephasing seen in~\cref{fig:contour}, together with~\cref{fig:ebardeltaek} provides a more complete understanding of how these parameters affect the GW phase. }

\onecolumngrid

\twocolumngrid

\bibliography{bibFile}

\end{document}